\documentclass[11pt]{article}
\pdfoutput=1
\usepackage{draft}
\usepackage{comment}
\usepackage[normalem]{ulem}
\usepackage{hyperref}
\hypersetup{
 colorlinks=true,
 urlcolor=blue,
 anchorcolor=blue,
 citecolor=blue,
 filecolor=blue,
 linkcolor=blue,
 menucolor=blue,
 pagecolor=blue,
 linktocpage=true,}
\usepackage{graphicx,color,subfig}
\usepackage{cite}
\usepackage{empheq}
\usepackage{bbold}
\usepackage[usenames,dvipsnames,table]{xcolor}
\graphicspath{{./figures/}}
\usepackage{tikz}
\usepackage{braket}

\newcommand{\RR}{\mathbb{R}}

\newcommand{\modS}{\mathbb{S}}	 
\newcommand{\fusion}{\mathbb{F}}	 
\newcommand{\id}{\mathbb{1}}	         

\newcommand{\op}{\mathcal{O}}

\newcommand\half{{1\over 2}}
\newcommand{\sbmatrix}[1]{
{\tiny\arraycolsep=0.3\arraycolsep\ensuremath{\begin{bmatrix}#1\end{bmatrix}}}}

\newcommand{\f}{\frac}
\newcommand{\s}{\sqrt}
\def\ba#1\ea{\begin{align}#1\end{align}}
\def\be#1\ee{\begin{equation}#1\end{equation}}
\begin{document}

\unitlength = .8mm

 \begin{titlepage}
 \begin{center}

 \hfill \\
 \hfill \\

\title{Universal Dynamics of Heavy Operators in Boundary CFT$_2$}

\author{Tokiro Numasawa$^{\alpha}$, Ioannis Tsiares$^{\beta,b}$}
 \address{\text{}\\
  $^{\alpha}$
Institute for Solid State Physics, University of Tokyo, Kashiwa 277-8581, Japan \\
 $^{\beta}$Department of Physics, McGill University,
 Montr\'eal, QC H3A 2T8, Canada
 \\
  $^{b}$Universit\'e Paris-Saclay, CNRS, CEA, Institut de Physique Th\'eorique, \\91191, Gif-sur-Yvette, France
 }

 \email{numasawa@issp.u-tokyo.ac.jp, ioannis.tsiares@ipht.fr}

 \end{center}

 \abstract{We derive a universal asymptotic formula for generic boundary conditions for the average value of the bulk-to-boundary and boundary Operator Product Expansion coefficients of any unitary, compact two-dimensional Boundary CFT (BCFT) with $c>1$. The asymptotic limit consists of taking one or more boundary primary operators -- which transform under a single copy of the Virasoro algebra -- to have parametrically large conformal dimension for fixed central charge. In particular, we find a \textit{single} universal expression that interpolates between distinct heavy regimes, exactly as in the case of bulk OPE asymptotics\cite{Collier:2019weq}. The expression depends universally on the boundary entropy and the central charge, and not on any other details of the theory. We derive these asymptotics by studying crossing symmetry of various correlation functions on higher genus Riemann surfaces with open boundaries. Essential in the derivation is the use of the irrational versions of the crossing kernels that relate holomorphic Virasoro blocks in different channels. Our results strongly suggest an extended version of the Eigenstate Thermalization Hypothesis for boundary OPE coefficients, where the hierarchy between the diagonal and non-diagonal term in the ansatz is further controlled by the boundary entropy. We finally comment on the applications of our results in the context of $\text{AdS}_3/\text{BCFT}_2$, as well as on the recent relation of BCFTs with lower dimensional models of evaporating black holes.
 }

 \vfill

 \end{titlepage}

\eject

\begingroup
\hypersetup{linkcolor=black}

\renewcommand{\baselinestretch}{0.93}\normalsize
\tableofcontents
\renewcommand{\baselinestretch}{1.0}\normalsize

\endgroup

\section{Introduction}\label{sec:intro}
The analysis of boundary conditions is a natural problem in physics. All realistic statistical systems possess boundaries and hence their full theoretical understanding clearly requires a good control of boundary conditions. Additionally, probing the system with a boundary can sometimes be proven fruitful to constrain the original bulk system itself. In the case of conformal field theories (CFTs), the study of boundaries has a long and eminent history. Boundary CFT (BCFT) has by now established its position as one of the main theoretical techniques within the already rich framework of CFTs, and find diverse applications in modern physics. They describe surface phenomena in systems near criticality, with surface critical exponents related to the conformal dimensions of the boundary operators\cite{Cardy:1984bb}, as well as systems with quantum impurities in condensed matter physics\cite{Affleck:1990zd,Affleck:1990iv}. In string theory, two-dimensional worldsheet BCFTs are famously interpreted as D-branes\cite{Callan:1987px,Polchinski:1987tu}. The study of
BCFTs is also an interesting subject in the context of the AdS/CFT correspondence \cite{Maldacena:1997re}.
In the AdS/BCFT proposal \cite{Takayanagi:2011zk,Fujita:2011fp}, the gravity dual of BCFT is described by an end-of-the-world (ETW) brane in the bulk of  spacetime \cite{Karch:2000ct,Karch:2000gx}.  
Recent works have further employed BCFTs as toy models of lower dimensional gravitational systems coupled to an auxiliary CFT which serves as a reservoir for the gravitational system, e.g. \cite{Almheiri:2019hni,Rozali:2019day,Akal:2020twv}. 
Despite these numerous applications to important problems in various branches of physics, the landscape of explicit constructions of BCFTs remains largely unexplored to date, except in a few special cases such as in minimal models or rational conformal field theories (RCFTs) \cite{Cardy:1989ir,Cardy:1991tv,Runkel:1998he,Runkel:1999dz,Behrend:1999bn}, special cases in free CFTs\cite{Callan:1993mw,Callan:1994ub,Affleck:2000ws,Gaberdiel:2001zq}, or Liouville theory \cite{Hosomichi:2001xc,Ponsot:2001ng,Ponsot:2003ss}.

In this work we will study some new universal aspects of unitary, irrational, compact\footnote{By compact here we formally mean two dimensional theories with discrete (bulk and boundary) energy spectra and unique $\mathfrak{sl}(2,\mathbb{C})$ and $\mathfrak{sl}(2,\mathbb{R})$ invariant vacuum states in the bulk and boundary respectively.} two dimensional BCFTs. In two-dimensions, the study of boundaries was pioneered by Cardy in a series of papers, in particular \cite{Cardy:1984bb,Cardy:1986gw,Cardy:1989ir}. The presence of the infinite-dimensional Virasoro symmetry imposes strong non-perturbative constraints on the spectrum and
interactions of such theories and constrains the correlation functions of local operators. Away from any boundaries, a bulk two dimensional CFT is in principle completely specified by its left and right central charges $(c_{L},c_{R})$, the
conformal dimensions of its primary fields $(h_i,\bar{h}_i)$, and the structure constants $C_{ij}^k$ appearing as coefficients of the primary fields $\op_k$ in the Operator Product Expansion (OPE) of primary fields $\op_i$ and $\op_j$. These data uniquely determine the correlation functions of the theory in flat space as well as on an arbitrary compact surface.
Ideally one could in principle solve the constraints of unitarity and conformal invariance (usually referred to as the \textit{conformal bootstrap program}) to determine the possible allowed values of the set $\{h_i, {\bar h}_i, C_{ijk}\}$, and hence completely classify two dimensional CFTs. 

In the case where a conformal boundary $s$ is added to the bulk CFT -- by which we mean that boundary conditions labelled by $s$ along the boundary preserve the conformal symmetry -- two dimensional BCFTs are enriched with an additional set of CFT data.
In this case the conformal boundary inherits a diagonal Virasoro subalgebra from the bulk theory\footnote{In $d$ Euclidean dimensions the presence of a boundary preserves an $SO(d, 1)$ subgroup of the original $SO(d + 1, 1)$ conformal symmetry. For analytic studies of BCFTs in higher dimensions see e.g. \cite{Liendo:2012hy,Hogervorst:2017kbj,Mazac:2018biw}.} which necessarily sets $c_{L}=c_{R}\equiv c$. Besides, there is a novel spectrum of primary operators living solely on the boundary with conformal weights $h^{B}_i$, and two new sets of boundary structure constants: the \textit{bulk-to-boundary structure constants} $C^{(s)i}_{\alpha}$ appearing as coefficients  of the boundary primary fields $\Psi_i$ when we expand a bulk primary field $\op_\alpha$ on the conformal boundary $s$, and the \textit{boundary structure constants} $C_{ij}^{(abc)k}$ which are the coefficients in the OPE expansion of purely boundary primary operators joining three distinct boundary conditions labelled by $(abc)$. We can summarize the CFT data in the presence of a boundary in the following set:
\begin{equation}\label{eq:bcftdata}
\begin{aligned} 
\left\{ \mathfrak{s}_a; \ h^{B}_i, C^{(s)i}_\alpha, C^{(abc)k}_{ij}  \right\}
\end{aligned}
\end{equation}
There is an additional distinguished universal quantity in BCFT$_2$, which we called $\mathfrak{s}_a$, and captures the vacuum amplitude on the disk (i.e. the disk partition function) with conformal boundary labelled by $a$. 
It is commonly known as the \textit{boundary entropy} or the brane tension in string theory applications. This quantity provides a measure of the boundary degrees of freedom and we can think of it as an analogue of the central charge for the boundary theory. As we will see, it will play an important role in our discussion.

The main focus of this paper is to understand universal aspects of the data (\ref{eq:bcftdata}). We will show that in any compact, unitary BCFT$_2$ with finite central charge $c>1$ and finite boundary entropies the bulk-to-boundary and boundary OPE coefficients satisfy the following universal asymptotic behaviour in the high energy regime\footnote{We use the notation $a\sim b$ to denote that $a/b\to1$ in the limit of interest. We will also use the notation $a\approx b$ to denote that $a$ and $b$ have the same leading scaling in the limit of interest.}: 
\begin{equation}\label{UnivFormulaBcft}
\begin{aligned}
\overline{\left|C^{(s)i}_{\alpha }\right|^2} &\sim e^{-\mathfrak{s}_s/2}  \ C_0\left(P_{\alpha},\bar{P}_\alpha,P_i\right)\\
\overline{\left|C_{ij}^{(abc)k}\right|^2} &\sim e^{-(\mathfrak{s}_a+\mathfrak{s}_b+\mathfrak{s}_c)/2} \ C_0\left(P_i,P_j,P_k\right)
\end{aligned}
\end{equation}
where 
\begin{equation} \label{C0}
C_0(h_i,h_j,h_k) \equiv
\frac{1}{\sqrt{2}}{\Gamma_b(2Q) \over \Gamma_b(Q)^3} {\prod_{\pm\pm\pm}\Gamma_b\left({Q\over 2} \pm i P_i \pm i P_j\pm i P_k\right)  \over \prod_{a\in\{i,j,k\}} 
\Gamma_b(Q +2iP_a) \Gamma_b(Q -2iP_a)}~.
\end{equation}
Here $\prod_{\pm}$ denotes a product of eight terms with all possible sign permutations. In addition, instead of using the central charge $c$ and dimensions $h$ and $\bar h$ to express our formula, we have used the ``Liouville parameters":
\begin{equation}\label{params}
c = 1+ 6 Q^2 = 1 + 6 (b+b^{-1})^2,~~~ h=\alpha (Q-\alpha), ~~~\alpha = {Q\over 2} + i P~.
\end{equation}
The function $\Gamma_b$ is the $b$-deformed gamma function, which is meromorphic, with no zeros, and with poles at argument $-mb-nb^{-1}$ for nonnegative integers $m,n$ (similarly to the usual gamma function, which has poles at nonpositive integers). 

The asymptotic formulas (\ref{UnivFormulaBcft}) are true in any (compact) BCFT$_2$ with $c>1$, and are universal in the sense that they depend only on the central charge $c$ and the boundary entropies $\mathfrak{s}_i$ and not on any other details of the theory. We will now make a few comments on the precise interpretation of our formulas. 

First, the \textit{square} of the structure constants denotes the fully contracted quantity with respect to the boundary operator metric (i.e. the normalization of the two-point functions on the boundary), assuming a canonical normalization for any bulk operators on the sphere. To be extremely clear\footnote{Throughout this paper we will be referring to the quantities $C^{(s)i}_{\alpha },C_{ij}^{(abc)k}$ as the boundary ``structure constants'', whereas to the quantities $C^{(s)}_{\alpha i },C_{ijk}^{(abc)}$ -- i.e. the structure constants with all operator indices lowered -- as the ``bulk-to-boundary two-point functions'' and ``boundary three-point functions'' respectively.}, our notation means
$
\left|C^{(s)i}_{\alpha }\right|^2\equiv C^{(s)i}_{\alpha}C^{(s)}_{\alpha i}$, and
$
\left|C_{ij}^{(abc)k}\right|^2\equiv C_{ij}^{(abc)k}C_k^{(cba)ji}
$.
This fact actually implies that our asymptotic results (\ref{UnivFormulaBcft}) are \textit{independent} of the choice of normalization for boundary operators. Furthermore, since the two asymptotic formulas for the bulk-to-boundary structure constants and the boundary OPE coefficients are essentially the same up to factors independent of the conformal dimensions, this seems to suggest that there should be a natural normalization of the boundary operators where we could capture both structure constants by exactly the same formula. Indeed, we confirm the existence of such normalization and we write it in (\ref{normalization}). We will subsequently show in detail why in this particular normalization one lands in a unified asymptotic formula for both the bulk-to-boundary two-point functions and the boundary three-point functions.

Second, (\ref{UnivFormulaBcft}) is an expression for the \textit{average} boundary structure constants, with the heavy operator weight(s) averaged over all Virasoro (either bulk or boundary) primary operators, which is valid for any finite $c>1$. Although we have only written one formula for each structure constant, equations (\ref{UnivFormulaBcft}) are secretly \textit{three} different formulas for \textit{each} boundary structure constant hiding in one: for the bulk-to-boundary structure constants the formula holds in three distinct asymptotic heavy regimes, namely light bulk-heavy boundary, heavy bulk-light boundary or heavy bulk-heavy boundary. Similarly for the boundary OPE coefficients this result holds in the heavy-light-light, heavy-heavy-light or heavy-heavy-heavy regimes of boundary operator dimensions\footnote{As we will explain in detail below, ``heavy" in this context means that for a bulk operator we can either take $h$ or $\bar h$ to be much larger than both the central charge or the boundary entropies -- similarly, for the boundary operator we only take $h^B$ to be large -- and the dimensions of the other operators are held fixed.  For this reason the three different regimes described above are \textit{distinct}, and there is a-priori no reason to expect to get the same result in each regime.}. In each case, the averaging taken in equation (\ref{UnivFormulaBcft}) should be understood as an average over the heavy operator(s), but not over the other operators which are held fixed (which is what we mean by \textit{light}). The amusing result is that we obtain exactly the same formula either for the bulk-to-boundary or for the boundary structure constants in each of these cases.

Recently, a similar crucial observation was made for the case of bulk OPE coefficients \cite{Collier:2019weq}, where the authors showed that there is a single universal formula (consisting of \textit{two} copies of $C_0$) capturing the heavy asymptotics of the OPE coefficients squared in three distinct heavy regimes. We are now in a position to confirm that the same qualitative picture is true in the case of BCFT for the corresponding boundary structure constants where, in addition, our asymptotic formulas depend universally on the boundary entropies. In fact, it is quite surprising that the asymptotic formula for the boundary OPE coefficients in (\ref{UnivFormulaBcft}) is not only symmetric under the exchange of the dimensions of primary operators (since $C_0$ is a symmetric function of its arguments) but it is also symmetric in the three types of boundary entropies that enter the formula. 

The strategy that will lead us to our main results will follow closely the methods of \cite{Collier:2019weq}, where we will now study \textit{sewing constraints} for Riemann surfaces with open boundaries\cite{Cardy:1989ir,Cardy:1991tv,Lewellen:1991tb} and implement the use of the so-called crossing kernels to derive our asymptotic results. For the case of non-orientable Riemann surfaces analogous methods and asymptotic results were derived in \cite{Tsiares:2020ewp}. The present work completes the basic study of Cardy-like results in 2d CFTs (in Euclidean signature) which can be obtained with the leverage of the crossing kernels. A natural generalization from this point is to compute asymptotics of higher moments of the CFT data by studying crossing symmetry (and assembling the appropriate crossing kernels) on more involved Riemann surfaces with boundaries or crosscaps. For the case of bulk OPE coefficients (in compact orientable Riemann surfaces), this extension was recently pursued in \cite{Belin:2021ryy,Anous:2021caj}.

\par The outline of the paper is as follows:
in section \ref{sec:review}, we review the basic properties of BCFT$_2$ on the upper half-plane and the disk, and we carefully define the correlation functions and OPEs that involve boundary operators.
In section \ref{sec:crossingeqn}, we revisit the sewing constraints on a general Riemann surface with conformal boundaries.
As a warm up, we derive the analogue of the Cardy formula for the boundary spectrum illustrating our basic technique which involves the use of a suitable \textit{crossing kernel} that relates different ``dissections'' of the same surface.
As a simple extension to that problem, we also derive the BCFT counterpart of the Kraus-Maloney formula \cite{Kraus:2016nwo} at finite central charge for the diagonal heavy-heavy-light boundary OPE coefficients.
In section \ref{sec:Bbasymptotics} we give a detailed derivation of our main universal asymptotic formula for the square of the bulk-to-boundary structure constants.
Similarly, in section \ref{sec:basymptotics} we derive the universal asymptotic formula for the square of the boundary structure constants.
In section \ref{sec:bcftETH}, we initiate a study of the Eigenstate Thermalization Hypothesis (ETH) for BCFTs based on our asymptotic results. Our findings suggest a novel hierarchy between the various terms in the ETH ansatz which is controlled by suitable boundary entropy factors.
In section \ref{sec:semiclassical}, we discuss the large central charge limits of our main formulas and discuss their holographic interpretation. In appendices \ref{app:explicitForms} and \ref{sec:asymptcrosskernnls}, we summarize some basic properties of the elementary crossing kernels, namely the fusion and modular kernel in the irrational case ($c>1$).
In appendix \ref{sec:CykKerDouble}, we provide details on the construction of the crossing kernel for the cylinder two-point functions that we study in section \ref{sec:Bbasymptotics}.
Finally, in appendix \ref{sec:ZZmodularK} we describe a novel relation between the irrational version of the modular kernel and the bulk-to-boundary structure constant in Liouville theory.\\
\par \textbf{Note added:} While this work was in preparation, the paper \cite{Kusuki:2021gpt} appeared which investigates similar asymptotic formulas for the case of identical boundary conditions.

\section{Review of BCFT$_2$}\label{sec:review}

We start by offering a lightning review of the basic ingredients of boundary two dimensional CFTs on the upper half-plane and the disk. Along the way we fix our notations and state our basic assumptions which will be relevant for the rest of the paper. 
\subsection{Natural variables for Virasoro representation theory}

We first introduce a parametrization of the CFT data that is becoming increasingly useful recently especially in manipulations involving the representation theory of the Virasoro algebra. The central charge $c$ can be written in terms of a ``background charge'' $Q$ or ``Liouville coupling'' $b$ as
\begin{equation}
	c = 1+6Q^2 = 1+6(b+b^{-1})^2. 
\end{equation}
In the present work we will be interested in two dimensional CFTs with $c >1$. With that in mind, we will fix the choice of $b$ by taking $0 < b < 1$ if $c > 25$, and by taking $b$ to lie on the unit circle in the first quadrant if $1 \leq c \leq 25$. To label Virasoro representations we use the ``Liouville momentum'' variable $P$, or sometimes the equivalent $\alpha=\frac{Q}{2}+iP$, which is related to the more common conformal weight by
\begin{equation}
	h = \left(\tfrac{Q}{2}\right)^2+P^2 =\alpha(Q-\alpha),
\end{equation}
and similarly $\bar{P}$ or $\bar{\alpha}$ in place for the anti-holomorphic dimension $\bar{h}$. We notice that this representation for conformal dimensions is redundant since it is invariant under reflections $P\rightarrow -P$ (or $\alpha\to Q-\alpha$). In addition, it naturally splits unitary values of the weights ($h\geq 0$) into two distinct regimes: $h\ge\frac{c-1}{24}$ corresponds to real $P$ (or $\alpha\in \frac{Q}{2}+i\mathbb{R}$), and $0\leq h<\frac{c-1}{24}$, which corresponds to imaginary $P$ (or $\alpha\in(0,{Q\over 2})$). Following \cite{Collier:2018exn}, we will refer to these as the \textit{continuous regime} and the \textit{discrete regime} of conformal dimensions respectively\footnote{This terminology stems from the analytic structure of the fusion kernel of the Virasoro four-point blocks which we will describe in more detail below. The T-channel Virasoro blocks have in general a discrete support (i.e. sum over residues on a finite set of poles) on S-channel blocks for $h < \frac{c-1}{24}$ , plus a continuous support for $h \geq \frac{c-1}{24}$ as discussed extensively in \cite{Collier:2018exn,Kusuki:2018wpa}. This terminology is also relevant in the AdS$_3$/CFT$_2$ context. In Liouville theory, the above regimes correspond to dimensions of non-normalisable and normalisable vertex operators respectively.}.

\subsection{Boundary states}
\par We will be considering Euclidean correlation functions of two-dimensional conformal field theories of central charge $c>1$ on Riemann surfaces with boundaries with prescribed conformal boundary conditions on each boundary\footnote{We will be labelling the boundary conditions on conformal boundaries with latin letters such as $a,b,c,\cdots$.}. The simplest example of such surface is a strip $\mathcal{S}$: $\mathbb{R}\times[0,\pi]$ with no operator insertions and boundary conditions $a,b$ at the two ends of the strip.
Via conformal transformation, it is convenient to map the strip to the upper half-plane (UHP) $\mathbb{H}_{+}=\{z\in\mathbb{C}; \text{Im}(z)\geq0\}$. Once we understand the CFT on the UHP, we can also calculate correlators in more general geometries. This is accomplished simply by a conformal mapping back to the UHP.
 
\par As it was explained originally by Cardy \cite{Cardy:1984bb,Cardy:1986gw,Cardy:1989ir} (for a review see \cite{Cardy:2004hm}) on the UHP with the real axis as the boundary, imposing conformal boundary conditions and in particular requiring the bulk stress tensor to remain traceless implies the following condition:
\begin{equation}\label{bdycond}
\left.T(z)\right|_{\text{Im}(z)=0}=\left.\bar{T}(\bar{z}) \right|_{\text{Im}(z)=0}.
\end{equation}
This has the immediate consequence that correlators of $T$ are those of $\bar{T}$, analytically continued into the lower half plane. In other words, in a boundary CFT only the diagonal Virasoro algebra is preserved by the boundary which further implies that the left and right central charges of the theory are identified $c_L=c_R\equiv c$. It will therefore be important to distinguish between the Hilbert space of states in the bulk and the Hilbert space of states on the boundary. We will denote these Hilbert spaces as
\begin{equation}
\begin{aligned}
\mathcal{H}_{\text{closed}}&=\int_{\mathcal{S}\times\bar{\mathcal{S}}}d\alpha d\bar{\alpha} \  \mathcal{M}_{\alpha,\bar{\alpha}}\otimes \mathcal{V}_{\alpha}\otimes\bar{\mathcal{V}}_{\bar{\alpha}} \ , \ \ \ \ \ \mathcal{S}\times\bar{\mathcal{S}}=\text{spectrum on the circle}.\\
\mathcal{H}^{a,b}_{\text{open}}&=\int_{\mathcal{S}_{a,b}}d\beta \ \mathcal{M}^{a,b}_{\beta}\otimes\mathcal{V}^{a,b}_{\beta} \ , \ \ \ \ \ \ \ \ \ \ \ \ \ \ \   \mathcal{S}_{a,b}=\text{spectrum on the line with bdy conditions $a,b$}.
\end{aligned}
\end{equation}
where $\mathcal{M}_{\alpha,\bar{\alpha}}$ and $\mathcal{M}^{a,b}_{\beta}$ are the multiplicity spaces (that transform trivially under $Vir_c\times \overline{Vir_{c}}$ and $Vir_c$ respectively), and $\mathcal{V}_{i}(\bar{\mathcal{V}}_{\bar{i}})$ denotes an irreducible representation of the holomorphic (anti-holomorphic) copy of the Virasoro algebra. Note that in the boundary sector we only have a single copy $\mathcal{V}^{a,b}_{\beta}$. The corresponding primary operator content of the theory naturally splits into two types of fields:  
\begin{itemize}
\item \textit{bulk} fields $\mathcal{O}_\alpha(z)$ with conformal dimensions $(h_{\alpha},\bar{h}_{\alpha})$ defined on $z\in\mathbb{H}_+$, which transform under conformal transformations in the bulk as:
\begin{equation}
\begin{aligned}
[L_n,\mathcal{O}_{\alpha}(z)]=z^n\left(z\partial_z+h_{\alpha}(n+1)\right)\mathcal{O}_{\alpha}(z)+\bar{z}^n\left(\bar{z}\partial_{\bar{z}}+\bar{h}_{\alpha}(n+1)\right)\mathcal{O}_{\alpha}(z) , \ \ \ \ \ \ n\in\mathbb{Z},
\end{aligned}
\end{equation}
\item \textit{boundary} fields $\Psi^{ab}_i(x)$ with conformal dimension $h_{i}$ defined on $x\in\mathbb{R}$ and joining the boundaries $a$ and $b$ (with $a\neq b$ in general), which transform under conformal mappings in the boundary as:
\begin{equation}
\begin{aligned}
[L_n,\Psi^{ab}_i(x)]=x^n\left(x\partial_x+h_{i}(n+1)\right)\Psi^{ab}_i(x) , \ \ \ \ \ \ n\in\mathbb{Z}.
\end{aligned}
\end{equation}
\end{itemize}
For bulk operators, we henceforth choose a canonical normalization for their two-point function on the sphere, namely $
\langle\mathcal{O}_i(0)\mathcal{O}_{j}(1)\rangle_{S^2}=\delta_{ij}
$. For boundary operators we could in principle proceed in the same way by canonically normalizing their two-point functions on the line. However, as we will see in more detail later, this is not the most natural normalization one could choose since the one-point function of the identity operator is non-trivial in BCFT for different boundary conditions. 
 We will discuss a particular choice of normalization which we think is convenient and natural in this set-up and, crucially, differs from the canonical one\footnote{As we will explain in detail later, our main results for the asymptotic formulas for boundary structure constants will turn out to be \textit{unambiguous} with respect to the choice of such normalisations. Nevertheless, we will find it convenient at some point to express these results in a particular choice of normalisation.}.  

Mapping the UHP with boundary condition $a$ to the disk, the boundary condition on the circle defines a state in $\mathcal{H}_{\text{closed}}$ according to the usual radial quantization. This is the so-called \textit{boundary state} (or Cardy state) $\ket{B_{a}}$ which, via the mode expansion of (\ref{bdycond}), satisfies\footnote{In this work we will be interested in bosonic theories where the chiral algebra is just Virasoro. For theories with extended chiral algebras, e.g. Kac-Moody, superconformal, W-algebras etc., one can impose more general boundary conditions relating the holomorphic and anti-holomorphic parts of the corresponding
currents on the boundary (see e.g. \cite{Cardy:1989ir} for a more detailed discussion on extended chiral algebras in the BCFT set up).}:
\begin{equation}\label{cardyst}
\begin{aligned}
\left(L_{n}-\bar{L}_{-n}\right)\ket{B_{a}}=0\ , \ \ \ \ \ \ n\in\mathbb{Z}.
\end{aligned}
\end{equation}
The boundary state can be thought of as a coherent state in $\mathcal{H}_{\text{closed}}$ and is in general a non-normalisable state. Setting $n=0$ in (\ref{cardyst}) shows that $\ket{B_{a}}$ has zero spin (though \textit{not} an energy eigenstate), and hence it can be built in principle out of a basis of states belonging to the \textit{scalar} sector of $\mathcal{H}_{\text{closed}}$, that is, irreducible representations $\mathcal{V}_{\alpha}\otimes\bar{\mathcal{V}}_{\bar{\alpha}}$ with conformal dimensions $h=\bar{h}$. We will denote this sector of the closed Hilbert space in what follows as $\mathcal{H}^{sc.}_{\text{closed}}\subseteq\mathcal{H}_{\text{closed}}$. In particular, it will be important for us that the closed-sector (or simply, ``bulk'') identity operator with $h=\bar{h}=0$ by definition belongs into $\mathcal{H}^{sc.}_{\text{closed}}$ and comes with unit multiplicity. 
\par In rational CFTs, Ishibashi \cite{Ishibashi:1988kg} showed that one can built a basis of solutions of (\ref{cardyst}) in one to one correspondence (unique up to a constant) with a particular scalar primary state of the theory, with Liouville momentum $P_s$. In other words, there exists a state $| B,P_s \rangle \rangle$ -- called \textit{boundary Ishibashi state} -- satisfying
\begin{equation}
\begin{aligned}
\left(L_{n}-\bar{L}_{-n}\right)| B,P_s \rangle \rangle=0\ , \ \ \ \ \ \ n\in\mathbb{Z}.
\end{aligned}
\end{equation}
The general boundary state (\ref{cardyst}) is then a superposition of boundary Ishibashi states\footnote{One important comment is that boundary Ishibashi states are not true boundary states even though they satisfy the same condition as $\ket{B_{a}}$. A boundary state is further subject to important consistency conditions such as the open-closed duality on the cylinder\cite{Cardy:1989ir} or the more general bootstrap equations on the disk\cite{Cardy:1991tv,Lewellen:1991tb}, as we will extensively discuss later.}
\begin{equation}\label{expan}
\begin{aligned}
\ket{B_a}=\sum_{s\in\mathcal{H}^{sc.}_{\text{closed}}}\mathcal{B}_a^s \ | B,P_s \rangle \rangle
\end{aligned}
\end{equation}
where the (complex, in general) coefficients $\mathcal{B}_a^s$ are sometimes called \textit{reflection coefficients} or just \textit{disk one-point function coefficients}. In the case of rational CFTs there is a finite number of primary operators and one can explicitly determine these coefficients (and hence the boundary state itself) via the bootstrap conditions (see e.g. \cite{Cardy:1986gw,Cardy:1989ir,Behrend:1999bn}).
In Liouville theory one can do the same by using the nice analytic properties of the theory in conjunction with the bootstrap, which eventually gives rise to the celebrated FZZT or ZZ branes \cite{Fateev:2000ik,Teschner:2000md,Zamolodchikov:2001ah}. Nonetheless, in a general irrational compact 2d CFT analogous calculations are completely out of reach and so far we know almost nothing  about the operator content or the allowed set of boundary conditions of such theories.

\par It is worth emphasizing at this point that the coefficients $\mathcal{B}_a^s$ encode all the information about the boundary condition $a$, since the construction of Ishibashi states is solely based on the Virasoro algebra. In particular, they can be realized as infinite sums of products of holomorphic and anti-holomorphic states of the form: 
\begin{equation}\label{ishibst}
\begin{aligned}
| B,P_s \rangle \rangle&=\sum_{\vec{m}}\ket{P_s,\vec{m}}\otimes U\ket{\overline{P_s,\vec{m}}}
\\
&=\left(1+\frac{L_{-1}\bar{L}_{-1}}{2h_s}+\cdots\right)\ket{P_s}
\end{aligned}
\end{equation}
where $\vec{m}$ denotes the descendant state constructed by acting with $L_{-j}$ raising operator
$m_j$ times on the primary $\ket{P_s}$ (normalized as $\langle{P_s}'|P_s\rangle=\delta(P_s'-P_s)$), and $U$ is an anti-unitary operator \cite{Ishibashi:1988kg,Blumenhagen:2009zz}. They are also normalized as follows\footnote{An interesting way to regularize the norm of boundary Ishibashi states in the case of RCFTs was discussed in \cite{Behrend:1999bn}.}:
\begin{equation}\label{normishi}
\begin{aligned}
\langle\langle B, P_s' | e^{-a\left(L_0+\overline{L_0}-\frac{c}{12}\right)}| B,P_s \rangle \rangle&=\delta(P_s-P_s')\frac{e^{- 2aP_s^2}}{\eta(\frac{ia}{\pi})}\ , \ \ \ \ P\neq vac. \\
\langle\langle B, vac | e^{-a\left(L_0+\overline{L_0}-\frac{c}{12}\right)}|B, vac \rangle \rangle&=\frac{(1-e^{-2a})e^{\frac{aQ^2}{2}}}{\eta(\frac{ia}{\pi})},
\end{aligned}
\end{equation}
where $\eta(\tau)$ is the Dedekind eta function and $a$ some constant.

As we will see later, consistency conditions like the open-closed duality on the cylinder shows that the $\mathcal{B}_a^s$'s define physically distinct sets of boundaries only up to an overall rescaling $\mathcal{B}_a^s\rightarrow\lambda\mathcal{B}_a^s$ with $|\lambda|=1$. We will fix this ambiguity by explicitly choosing $\mathcal{B}^{\mathbb{1}}_a > 0$ for the identity operator in what follows\footnote{This choice also makes sense from the reflection positivity of the disk partition function which we will define shortly. Note also that for $s\neq\mathbb{1}$ we take $\mathcal{B}^{s}_a\in\mathbb{R}$, since we can always choose a Hermitian basis of (scalar) operators.}. 

\par For the rest of the paper, we're going to consider irrational ($c>1$) compact, unitary BCFTs on the UHP with a given discrete spectrum of scalar primaries in $\mathcal{H}^{sc.}_{\text{closed}}$, and we will assume that a decomposition of the form (\ref{expan}) exists and converges for the corresponding boundary state. Following similar logic as in Liouville theory \cite{Teschner:2001rv,Teschner:2000md}, we will find it convenient to re-write (\ref{expan}) in an equivalent form by introducing an even distribution $\mathcal{B}(P|a)$ as follows:
\begin{equation}\label{ishibdec}
\begin{aligned}
\ket{B_a}=\int\frac{dP}{2}\mathcal{B}(P|a)| B,P \rangle \rangle
\end{aligned}
\end{equation}
where\footnote{Distributions with support on imaginary values of $P$, i.e. states
with $h < \frac{c-1}{
24}$, require special care. As described nicely in \cite{Maxfield:2019hdt} there is a solid mathematical background for these kind of distributions which can be taken to live on an enlarged space or, equivalently, a more restricted space of test functions which at least includes the Gaussians (in the variable $P$). We refer to appendix A of \cite{Maxfield:2019hdt} for more details. }
\begin{equation}\label{adistr}
\begin{aligned}
\mathcal{B}(P|a):= \sum_{i\in\mathcal{H}^{sc.}_{\text{closed}}}\mathcal{B}_a^i\left[\delta(P-P_i)+\delta(P+P_i)\right]
\end{aligned}
\end{equation}
It will also be important for us later to assume a \textit{scalar gap} above the identity in $\mathcal{H}^{sc.}_{\text{closed}}$. This would mean that
\begin{equation}\label{}
\begin{aligned}
\ket{B_a}=g_a| B,vac \rangle \rangle+\sum_{s_{gap}}\mathcal{B}_a^{s}| B,P_s \rangle \rangle,
\end{aligned}
\end{equation}
where $s_{{gap}}$ can be either in the discrete or the continuous regime of conformal dimensions. Recently, the conformal dimension corresponding to $s_{{gap}}$ turned out to be a crucial parameter in applications of the bootstrap techniques on the cylinder \cite{Collier:2021ngi}. Furthermore, the term $g_a\equiv \mathcal{B}_a^{\mathbb{1}}$ is a distinguished quantity here and will play a central role in our discussion. As we will review in more detail in the next subsection, it defines the so called \textit{boundary entropy}.
\par In the open sector $\mathcal{H}^{a,b}_{\text{open}}$, a vacuum representation $\mathcal{V}^{a,b}_{vac.}$ with $h_{vac.}=0$ is \textit{not} guaranteed to exist in the spectrum in general. This depends highly on the boundary conditions $a,b$. We will denote the conformal dimension of the primary with the \textit{minimum} conformal dimension inside $\mathcal{H}_{\text{open}}^{a,b}$ as $h^{ab}_{min.}\geq0$. In the special case of identical boundary conditions, the identity can certainly propagate in the spectrum and hence we can have $h^{aa}_{min.}=0$. It will also be important for us to assume a \textit{boundary gap} in the spectrum above $h^{ab}_{min.}$, either in the discrete or the continuous regime of conformal dimensions. 

\subsection{Structure constants and basic correlation functions}\label{sec:structureconstandcorrfns}
In boundary theories in which the boundary conditions do not break conformal symmetry -- meaning that the stress-tensor satisfies (\ref{bdycond}) on the UHP -- the short distance expansions between bulk and boundary operators, and between boundary operators are completely fixed by conformal symmetry\footnote{Originally discussed in \cite{Diehl:1981jgg}.}. We define the bulk-to-boundary and the boundary structure constants $C^{(s)i}_{\alpha}, C^{(abc)k}_{ij}$ for \textit{primary operators} (either bulk or boundary) via the following expansions
\begin{equation}
\begin{aligned}
\text{Bulk-to-Boundary OPE:} \ \ \ \  \ \mathcal{O}_\alpha(z)&\sim\sum_{i\in\mathcal{H}^{s,s}_{\text{open}}}C^{(s)i}_{\alpha}(2\text{Im} z)^{h_i-h_\alpha-\overline{h_\alpha}}\Psi^{ss}_i(\text{Re} z)+\cdots \ , \ \ \text{Im} z >0\\
\text{Boundary OPE:} \ \ \ \  \ \Psi_i^{ab}(x)\Psi_j^{bc}(y)&\sim\sum_{k\in\mathcal{H}^{a,c}_{\text{open}}}C^{(abc)k}_{ij}(x-y)^{h_k-h_i-h_j}\Psi_k^{ac}(y)+\cdots \ , \ \ x>y,
\end{aligned}
\end{equation}
where $\cdots$ denote contributions from descendants which are completely fixed by conformal symmetry. 
These two types of structure constants provide the necessary CFT data that determine the following basic (and physical) correlation functions in any BCFT$_2$ \cite{Cardy:1989ir,Cardy:1991tv,Lewellen:1991tb}:
\begin{itemize}
\item \textbf{Bulk one-point function}\\
\\
The one-point functions on the UHP (or the disk) are only non-zero for bulk scalar operators $\mathcal{O}_{\alpha}$ with $h_\alpha=\overline{h}_\alpha$. Their expression reads:
\begin{equation}\label{disk1ptfn}
\begin{aligned}
\langle\mathcal{O}_{\alpha}(z)\rangle_s&=\frac{\mathcal{B}_s^\alpha}{|z-\bar{z}|^{2h_\alpha}},\\
\langle\mathbb{1}\rangle_s&= g_s
\end{aligned}
\end{equation}
where $\mathcal{B}_s^\alpha$ are the disk one-point function coefficients of the boundary state expansion (\ref{expan}).\\  The one-point function of the bulk identity operator captures the disk partition function, and defines a distinguished universal quantity that characterizes the boundary condition. We define 
\begin{equation}\label{bdyentropy}
\begin{aligned}
 \mathfrak{s}_i:=2\log{g_i}
 \end{aligned}
\end{equation}
as the \textit{boundary entropy} (or the \textit{g-function}) labelled by the boundary condition $i$. This quantity provides a measure of the boundary degrees of freedom, as it was originally explained in \cite{Affleck:1991tk} (see also \cite{Harvey:1999gq}). Therefore we can think of it as an analogue of the central charge for the boundary theory. In fact, away from the boundary conformal fixed point, the boundary entropy is proven to be monotonic under boundary RG flows \cite{Affleck:1992ng,Friedan:2003yc,Casini:2016fgb,Cuomo:2021rkm}, providing a boundary analog of Zamolodchikov’s $c$-function in the
case of bulk RG flows. Note that, given our choice for $\mathcal{B}_s^\alpha>0$,  the boundary entropy can in principle take any real value\footnote{Indeed, in the 2d Ising model for example one can calculate explicitly the boundary entropies corresponding to the three bulk operators $\mathbb{1}$, $\epsilon$, $\sigma$ using the elements of the S-matrix \cite{Cardy:1991tv}. The resulting expressions are either negative or zero. } as opposed to the central charge which has to satisfy $c>0$ for unitary theories. \\
\par Furthermore, this quantity deserves the name ``entropy'' since it captures the subleading piece in the high-temperature limit of the
thermal entropy of a 2d CFT on an interval\cite{Affleck:1991tk}, and it is crucially a constant as a function of the temperature for conformal boundaries. It moreover contributes a universal term in the ground state entanglement entropy of an interval, as was elucidated in \cite{Calabrese:2009qy}. In RCFTs defined by a diagonal modular invariant the value of $g$ takes a particularly simple expression in terms of the corresponding $S$-matrix of the theory
\begin{equation}
\begin{aligned}
g_i=\frac{S_{0i}}{\sqrt{S_{00}}}
\end{aligned}
\end{equation}
and hence one can systematically obtain the various boundary entropies in this case and study its properties. 
For generic irrational CFTs with $c>1$ it is still unknown whether an analogous simple formula exists, mainly because the landscape of conformal boundaries in this case is largely unclear. Recently, it was shown that conformal bootstrap techniques applied in the open-closed consistency condition on the annulus can provide interesting and non-trivial bounds on $g_i$ even for irrational theories \cite{Collier:2021ngi}.
\\

\item  \textbf{Boundary two-point function}\\
\\
The two-point function on the real line or the boundary circle of the disk is given by:
\begin{equation}
\begin{aligned}
\langle\Psi^{ab}_{i}(x)\Psi^{ba}_{j}(y)\rangle=\frac{\mathfrak{g}^{(ab)}_{ij}}{|x-y|^{2h_i}},
\end{aligned}
\end{equation}
where the (boundary) operator metric $\mathfrak{g}^{(ab)}_{ij}$ is defined as
\begin{equation}\label{metric}
\begin{aligned}
\mathfrak{g}^{(ab)}_{ij}:=C^{(aba)\mathbb{1}}_{ij}g_a\delta_{ij}.
\end{aligned}
\end{equation}
Note that, compared to the two point function in the bulk CFT case, the operator metric here looks quite non trivial because of the presence of the factor $g_a$ which can be different for different boundary conditions. In fact, the metric is symmetric under the exchange $a\leftrightarrow b$, and we can use it to raise or lower indices for the bulk-to-boundary or boundary structure constants. In particular we define the structure constants with lower indices via the following relations:
\begin{equation}
\begin{aligned}
C^{(s)}_{\alpha i}&=\sum_{i'}C_{\alpha}^{(s)i'}\mathfrak{g}^{(ss)}_{ii'}=C_{\alpha}^{(s)i}C^{(sss)\mathbb{1}}_{ii}g_s\\
C^{(abc)}_{ijk}&=\sum_{k'}C_{ij}^{(abc)k'}\mathfrak{g}^{(ac)}_{kk'}=C_{ij}^{(abc)k}C^{(aca)\mathbb{1}}_{kk}g_a.
\end{aligned}
\end{equation}
As it was explained in \cite{Lewellen:1991tb}, one could imagine setting $\mathfrak{g}^{(ab)}_{ij}=\delta_{ij}$ by canonically normalizing the boundary operators $\Psi^{ab}_{i}$. However, in the case of identical boundary conditions $a=b$ we could in principle consider the one-point function of the (boundary) identity operator where the boundary metric yields $\mathfrak{g}^{(aa)}_{ij}:=g_a\delta_{ij}$ (which coincides exactly with the disk partition function $\langle\mathbb{1}\rangle_a$ in (\ref{disk1ptfn})). Therefore, assuming that our theory has a non-trivial set of conformal boundary conditions, it is not at all natural to set simultaneously all of the corresponding disk partition functions to unity.
\par A natural normalization for boundary primary operators that we may occasionally adopt in the present work is the following:
 \begin{equation}\label{normalization}
\begin{aligned}
C^{(aba)\mathbb{1}}_{ii}&=\sqrt{\frac{g_b}{g_a}} \ \ \ \ \Rightarrow \ \ \ \mathfrak{g}^{(ab)}_{ij}=\sqrt{g_ag_b} \ \delta_{ij}.
\end{aligned}
\end{equation}
As we will see in detail later, our asymptotic formulas for the bulk-to-boundary and boundary OPE coefficients take some particularly neat and unified form in this normalization, which suggests that (\ref{normalization}) provides a natural choice in BCFT$_2$ in general. Be that as it may  -- and unless our particular choice to work with (\ref{normalization}) is explicitly stated -- for the most part of this work we will keep the dependence on the metric $\mathfrak{g}^{(ab)}_{ij}$ manifest and present results in a normalisation-independent fashion.\\
\item \textbf{Bulk-Boundary two-point function}\\
\\
The amplitude between a bulk and a boundary operator takes the form:
\begin{equation}
\begin{aligned}
\langle\mathcal{O}_{\alpha}(z)\Psi^{ss}_{i}(x)\rangle=\frac{C^{(s)}_{\alpha i}}{|z-\bar{z}|^{2h_\alpha-h_{i}}|z-x|^{2h_{i}}}.
\end{aligned}
\end{equation}
Note that in the special case where $\Psi^{ss}_{i}=\mathbb{1}$ we have the relation $C^{(s)}_{\alpha \mathbb{1}}=\mathcal{B}_s^\alpha=C^{(s)\mathbb{1}}_{\alpha}g_s$.\\
\item \textbf{Boundary three-point function}
\\
\\
Finally, the three point function on the line or the boundary circle reads:
\begin{equation}
\begin{aligned}
\langle\Psi^{ab}_{i}(x_3)\Psi^{bc}_{j}(x_2)\Psi^{ca}_{k}(x_1)\rangle=\frac{C^{(abc)}_{ijk}}{|x_{12}|^{h_{12}}|x_{23}|^{h_{23}}|x_{31}|^{h_{31}}},
\end{aligned}
\end{equation}
where $x_{ij}\equiv x_i-x_j$ and $h_{ij}\equiv h_i+h_j-h_k$.
\end{itemize}
\par The primary focus of the present work is on the behaviour of the bulk-to-boundary and boundary OPE coefficients in BCFT$_2$ for generic boundary conditions. In rational CFTs, the bootstrap conditions along with the fact that the theory possesses a finite number of primary operators have been proven sufficient to explicitly solve for those data. For example, this strategy was famously successful for the A and D series Minimal Models\cite{Runkel:1998he,Runkel:1999dz}. Furthermore, in Liouville theory the bulk-to-boundary structure constants were studied in \cite{Hosomichi:2001xc,Ponsot:2003ss} and the boundary structure constant in \cite{Ponsot:2001ng}, where the authors provided explicit analytic expressions. In this work we will be interested in irrational, unitary and compact BCFTs where almost nothing is known so far about these coefficients. By revisiting the crossing equations on Riemann surfaces with boundary we will be able to provide universal asymptotic formulas for $C^{(s)}_{\alpha i}$ and $C^{(abc)}_{ijk}$ in particular asymptotic regimes of the boundary operator conformal dimensions.

\section{Crossing equations on Riemann surfaces with boundary}\label{sec:crossingeqn}

Before deriving our universal formulas for the boundary OPE coefficients, in this section we will revisit the crossing equations and explain the basic legos for constructing a general CFT correlation function on a Riemann surface with open boundaries. At the end of this section, we will revisit the derivation of the asymptotic Cardy formula for the boundary spectrum by carefully studying the slightly more general open-closed duality of the cylinder one-point function.
\subsection{Sewing constraints, three elementary legos, and the doubling trick}\label{sec:legos}
\begin{figure}
\centering
 \includegraphics[width=.26\textwidth]{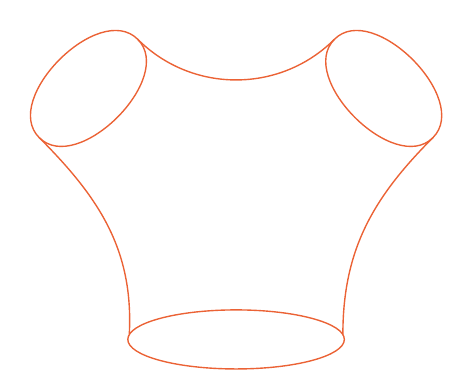}   \ \ \ \ \includegraphics[width=.15\textwidth]{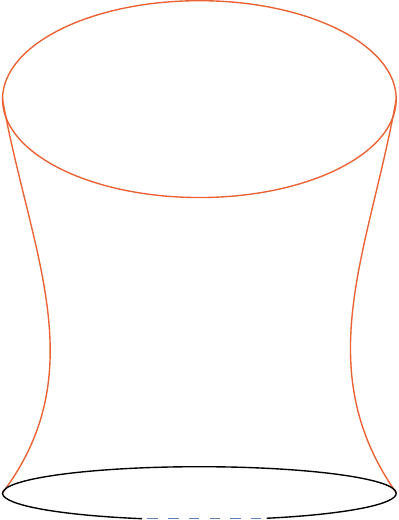}\ \ \ \ \ \ \includegraphics[width=.24\textwidth]{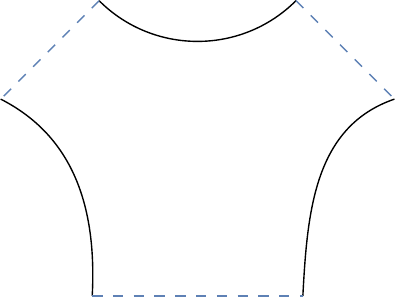}
	\caption{The three elementary ``legos'' out of which one can construct a BCFT$_2$ correlation function on a Riemann surface with open boundaries: (i) bulk OPE structure constant , (ii) bulk-to-boundary structure constant  and (iii) boundary structure constant. The solid orange lines represent points on the bulk surface, whereas solid black lines represent a conformal boundary. Local bulk operator insertions are depicted as orange circles, while local boundary operator insertions as dashed blue lines.}\label{fig:legos}
\end{figure}

We start by recalling briefly the definition of correlation functions of local operators in a 2d CFT on a \textit{compact} oriented surface $\Sigma_g$ of genus $g$ (equipped with some Riemannian metric), as it was pioneered in the early works \cite{Friedan:1986ua,Sonoda:1988fq,Moore:1988uz,Moore:1988qv,Moore:1989vd}. We may view the surface $\Sigma_g$ with $n$ local operators insertions as a surface with $n$ ``punctures''. We can now decompose the punctured surface into $2g-2+n$ \textit{pairs of pants}, with $n$ of the boundary circles shrunk to points. The $n$-point function on $\Sigma_g$ is then decomposed into the product of $2g-2+n$ three-point functions (see (i) in Fig.\ref{fig:legos}), appropriately Weyl transformed, summed over the basis of states inserted along the $3g-3+n$ circles. The consistency in defining correlation functions on $\Sigma_g$ requires that different pair-of-pants decompositions results in the same answer for the correlation function. These conditions are usually termed ``sewing constraints''. Furthermore, it can be shown\cite{Moore:1988qv,Bakalov:rp} that any two different pair-of-pants decompositions of the $n$-punctured surface $\Sigma_g$ can be related by a sequence of two types of simple \textit{crossing moves}, which consists of: (a) the crossing equation for the conformal block decomposition of sphere 4-point functions (or, in other words, the associativity of the OPE), and (b) the modular covariance of the torus one-point functions. As a result, the consistency of the $n$-point functions on $\Sigma_g$ follows from the crossing invariance of all sphere 4-point functions together with the modular covariance of all torus one-point functions.
\par The analogous construction of CFT correlation functions on non-compact Riemann surfaces with conformal boundaries was later analyzed in \cite{Lewellen:1991tb,Pradisi:1996yd}\footnote{We will review the basics of this construction in what follows without trying to be too rigorous at this stage since we are not going to need all the details of the construction. We refer the reader to the original papers for a more elaborate discussion.}. We want to study (Euclidean) correlation functions of $n_b$ local bulk operators on a genus $g$ Riemann surface which additionally has $B_i$ number of disconnected boundaries with conformal boundary conditions $s_i$ on each boundary. We call this surface $\Sigma^{\left\{s_i\right\}}_g$. On these boundaries we can also distribute a number of $n_{\partial}$ boundary operators. We denote this generic correlation function on such surface as 
\begin{equation}\label{notation}
\begin{aligned}
G_{g,n_b;B_i,n_{\partial}}=\langle\mathcal{O}_1(z_1)\cdots\mathcal{O}_{n_b}(z_{n_b});\Psi_1(x_1)\cdots\Psi_{n_{\partial}(x_{n_{\partial}})}\rangle_{\Sigma^{\left\{s_i\right\}}_g}.
\end{aligned}
\end{equation}
\par In parallel with the case of compact Riemann surfaces, we can imagine decomposing $G$ into simpler amplitudes via a series of ``cutting'' operations where we insert a complete set of states on each cutting. In the presence of conformal boundaries, however, one encounters two types of cuttings: we can either insert a complete set of \textit{bulk states} (i.e. primaries and their descendants under two copies of the maximal chiral algebra of the theory) when we cut along a closed loop of the surface, or we can insert a complete set of \textit{boundary states} (i.e. primaries and their descendants under a \textit{single} copy of the maximal chiral algebra) when we cut along a line joining two conformal boundaries. After an appropriate number of such cuttings the correlation function $G$ can be reduced to a collection of three types of building blocks which are depicted in Fig.\ref{fig:legos}: the correlation function is decomposed into the product of bulk three-point functions, bulk-to-boundary two-point functions and boundary three-point functions. In the string theory language -- where boundary operators correspond to open-string vertex operators -- the basic building blocks of amplitudes now include the open-string three-point function and the closed-string to open-string amplitude in addition to the closed-string three-point function. 
\par Different ways of cutting will produce different collections of the elementary legos of Fig.\ref{fig:legos} for the correlation function $G$. One then encounters additional non-trivial sewing constraints in the boundary case, which ensure the consistency of the CFT on such surfaces. As we recalled earlier in the case of compact Riemann surfaces one needs to impose two distinct sewing constraints -- namely sphere four-point crossing symmetry and modular covariance of the torus one-point functions -- to ensure crossing symmetry for all $n$-point functions on higher genus. It was further shown in \cite{Lewellen:1991tb,Pradisi:1996yd} that there are four additional necessary and sufficient basic sewing constraints that consistently define any arbitrary CFT correlation function on a surface with open boundaries. These elementary sewing constraints involve: (a) crossing symmetry of the boundary four-point function on the disk (or, in other words, the associativity of the boundary OPE), (b) crossing symmetry of the bulk-to-boundary three-point function with two boundary operators and a single bulk operator on the disk, (c) crossing symmetry of a different bulk-to-boundary three-point function with two bulk operators and a single boundary operator on the disk, and lastly (d) crossing symmetry of the boundary two-point function on the cylinder. We will encounter two of these sewing constraints in our analysis (namely cases (a) and (d)) as well as consistency conditions on more involved Riemann surfaces with open boundaries to study universal asymptotic results for the bulk-to-boundary structure constants and the boundary OPE coefficients. Our results therefore rely heavily on the pioneering construction of \cite{Lewellen:1991tb,Pradisi:1996yd}.
\par One important feature of every Riemann surface with open boundaries (or crosscaps) is that it admits a two-fold cover that is compact and orientable (see e.g. \cite{schiffer}). Under the lift to the covering surface, points in the bulk surface have two pre-images, while for boundary points the lift is unique. In boundary CFT one can see that concretely at the level of the symmetry: the Ward identities for $n$-point functions of operators in the bulk have the same form as those for \textit{chiral} conformal blocks in a bulk CFT with $2n$ insertions of chiral vertex operators carrying conformal dimensions $h_1, . . . , h_n , \bar{h}_1, . . . , \bar{h}_n$. This fact sometimes goes under the name ``doubling trick”, and it was first observed in \cite{Cardy:1989ir} (see also \cite{Recknagel:1997sb}). We will implement this trick throughout this work, in a way that we will make precise in later sections. In particular, as it was observed originally by Cardy \cite{Cardy:1989ir}, correlation functions in BCFT have an expansion into a \textit{linear} combination of the familiar \textit{holomorphic} conformal blocks which -- together with their anti-holomorphic counterparts -- are the building blocks of the usual bulk CFT correlation functions which correspond to the compact cover of the Riemann surface. One can then use known facts about conformal blocks, such as their duality relations, to infer useful information about the CFT data from the sewing constraints in BCFT. We will initiate an analytic study of this sort in the next subsection with a basic example, before moving on to study more involved sewing constraints later.

\subsection{Basic example: Asymptotics from the cylinder one-point function}\label{subsec:boundaryCardy}
\begin{figure}
\centering
 \includegraphics[width=.26\textwidth]{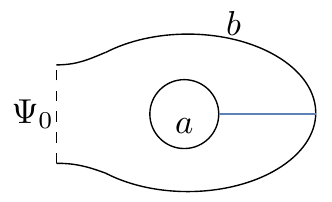} \   \raisebox{.053\textwidth}{\scalebox{1.5}{$=\int \frac{dP'}{2}  \modS_{P'P}[P_{0}]$}} \  \includegraphics[width=.26\textwidth]{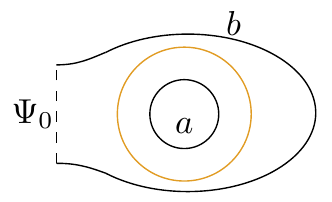}
	\caption{The open-closed duality of the cylinder one-point function with an external (boundary) operator labelled by $\Psi_0$. The blocks on the two channels are related with a \textit{holomorphic} copy of the modular kernel.}\label{fig:cyl1pt}
\end{figure}

One important example of the sewing constraints that we discussed in the previous subsection is the open-closed duality of the one-point function on the annulus or the cylinder. In particular we consider a single insertion of a boundary primary operator $\Psi_0^{bb}$ with conformal dimension $h_0$ (or Liouville momentum $P_0$) on the cylinder with boundary conditions $a,b$. Using the notation (\ref{notation}), we denote the correlation function as 
\begin{equation}
G_{0,0;2,1}(\tau)=\langle\Psi^{bb}_0\rangle_{\text{cyl}^{(ab)}}\ , \ \ \ \ \ \ \ \ \ \ \tau\equiv\frac{i\beta}{2\pi} \ , \ \beta\in\mathbb{R}.
\end{equation} 
From rotational invariance the amplitude depends only on the length of the cylinder $\beta$ and not on the location of the operator $\Psi_0^{bb}$ on the boundary $b$. We then get two equivalent expansions in conformal blocks, either in the open or in the closed sector as in Fig.\ref{fig:cyl1pt}:
\begin{equation}\label{cylexpansions}
\begin{aligned}
G^{\text{(open)}}_{0,0;2,1}(\tau)&=\sum_{\Psi_i\in \mathcal{H}^{a,b}_{\text{open}}}C^{(abb)i}_{i0} \ \mathcal{F}^{\text{cyl-1-pt}}\left(P_{0};P_{i}|\tau\right)\\
G^{\text{(closed)}}_{0,0;2,1}(-1/\tau)&=\sum_{\mathcal{O}_i\in\mathcal{H}^{sc.}_{\text{closed}}} C^{(a)}_{i\mathbb{1}}C^{(b)}_{i0}\ \mathcal{F}^{\text{cyl-1-pt}}\left(P_{0};P_i|-1/\tau\right).
\end{aligned}
\end{equation} 
The conformal data that enter in the open-sector expansion is a single factor of a boundary structure constant $C^{(abb)i}_{i0}=\sum_{i'}C^{(abb)}_{i0i'}\mathfrak{g}^{ii'}_{(ab)}=C^{(abb)}_{i0i}(C^{(aba)\mathbb{1}}_{ii}g_a)^{-1}$ indicating that a single boundary three-point function ``lego'' is needed to construct the amplitude in this channel. On the other hand, in the closed-sector expansion we have an internal \textit{scalar} bulk operator $\mathcal{O}_i$ which contributes a factor of a bulk-to-boundary structure constant $C^{(b)}_{i0}$ for the boundary $b$, and a disk one-point function coefficient $C^{(a)}_{i\mathbb{1}}=\mathcal{B}_a^i$ for the boundary $a$, appropriately contracted in the bulk operator indices\footnote{Note that upper or lower indices for bulk operators make no real difference because we have chosen a canonical normalization for their two-point function on the sphere.}.
Crucially, the two expansions are related via
\begin{equation}\label{cylrelation}
\begin{aligned}
G^{\text{(closed)}}_{0,0;2,1}(-1/\tau)=(-i\tau)^{h_{0}} \ G^{\text{(open)}}_{0,0;2,1}(\tau).
\end{aligned}
\end{equation} 
This is the crossing symmetry equation for the cylinder one-point function. This relation comes from the fact that the cylinder with one boundary puncture can arise from its twofold cover which we can take it to be a torus one-point function for a bulk primary operator with conformal dimensions $h_0=\overline{h_0}$. The cylinder one-point blocks $\mathcal{F}^{\text{cyl-1-pt}}$ are then naturally just the holomorphic half of the usual torus one-point blocks and all the information on the boundary conditions $a,b$ is incorporated into the CFT data that multiply the blocks, as in (\ref{cylexpansions}). Therefore, the blocks in the two channels of the cylinder one-point function are related via a usual modular $S$ transformation:
\begin{equation}\begin{aligned}\label{modularSTransform}
\tau^{h_0}\mathcal{F}^{\text{cyl-1-pt}}\left(P_{0};P|\tau\right) &= \int \frac{d P'}{2}\mathcal{F}^{\text{cyl-1-pt}}\left(P_{0};P'|-1/\tau\right)\modS_{P'P}[P_0]
\end{aligned}\end{equation}
The explicit expression for the modular S kernel for $c>1$ was obtained by Teschner in \cite{Teschner:2003at} (see also \cite{Nemkov:2015zha,Nemkov:2016ikx}). We reproduce the precise formula in Appendix \ref{torus1ptkernel}.
It is worth emphasizing that equation (\ref{cylrelation}) does \textit{not} imply in any sense that the amplitude $G^{ab}_{\text{cyl.}}[P_{0}](\tau)$ is a holomorphic modular form of weight $h_0$ as a function of $\tau$. This is because -- unlike in the case of the torus one-point functions -- equation (\ref{cylrelation}) relates two \textit{different} functions of $\tau$ as it is clear from the expansions (\ref{cylexpansions}).

\par We next introduce the even distributions for the CFT data in the two channels:
\begin{equation}
\begin{aligned}
\rho^{\text{(open)}}_{ab}(P;P_0)&:=\sum_{\Psi_i\in \mathcal{H}^{a,b}_{\text{open}}}C^{(abb)i}_{i0}[\delta(P-P_i)+\delta(P+P_i)]\\
\rho^{\text{(closed)}}_{ab}(P;P_0)&:=\sum_{\mathcal{O}_i\in\mathcal{H}^{sc.}_{\text{closed}}} C^{(a)}_{i\mathbb{1}}C^{(b)}_{i0}[\delta(P-P_i)+\delta(P+P_i)]
\end{aligned}
\end{equation} 
Using the transformation (\ref{modularSTransform}), we can write the crossing equation (\ref{cylrelation}) as a transform relating the two distributions
\begin{equation}\label{distrRelation}
\begin{aligned}
\rho^{\text{(open)}}_{ab}(P';P_0)=\int \frac{dP}{2}\modS_{P'P}[P_0] \  \rho^{\text{(closed)}}_{ab}(P;P_0).
\end{aligned}
\end{equation} 
Equation (\ref{distrRelation}) is the main result of this section: it encodes the open-closed duality of the cylinder one-point functions and relates the boundary CFT data $C^{(abb)i}_{i0}$ supported on the boundary spectrum, with $C^{(a)}_{i\mathbb{1}}C^{(b)}_{i0}$ supported on the scalar bulk spectrum for irrational theories with $c>1$. The novel feature of this equation is encoded exactly in the non-trivial form of the modular kernel $\modS_{P'P}[P_0]$\footnote{It is instructive to compare (\ref{distrRelation}) with the analogous expression in the bulk case where we consider the bootstrap condition of the torus one-point function\cite{Kraus:2016nwo,Collier:2019weq}. In that case, we have \textit{two copies} of the modular kernel (one for the holomorphic and one for the anti-holomorphic part) relating the spectral OPE density (c.f. equation (3.17) in \cite{Collier:2019weq}):
\begin{equation}\label{BulkdistrRelation}
	{\rho}[\op_0](P',\bar{P}') = \int \frac{dP}{2} \frac{d\bar{P}}{2} \modS_{P'P}[P_0]\modS_{\bar{P}'\bar{P}}[\bar{P}_0] \ \rho[\op_0](P,\bar{P}),
\end{equation}
where ${\rho}[\op_0](P,\bar{P}):=\sum_{i} C_{\mathcal{O}_i\mathcal{O}_0\mathcal{O}_i}[\delta(P-P_i)+\delta(P+P_i)]\times[\delta(\bar{P}-\bar{P}_i)+\delta(\bar{P}+\bar{P}_i)]$ is the primary OPE spectral density. Note that in (\ref{BulkdistrRelation}) the distribution $\rho[\op_0]$ is the same on both sides of the equation. This is in contrast with the open-closed duality in (\ref{distrRelation}) where the distributions are different on the two sides of the equation, having different supports and in general different amplitudes.}.
\par Before moving on to discussing implications of this relation we will consider first the special case where the external operator is $\Psi_0^{bb}=\mathbb{1}$, or $h_0\rightarrow0$ ($P_0\rightarrow i\frac{Q}{2}$). In that case we obtain the more familiar \textit{cylinder partition function} with boundary conditions $a,b$:
\begin{equation}
G_{0,0;2,0}(\tau)\equiv Z_{\text{cyl}^{(ab)}} \ , \ \ \ \ \ \ \ \ \ \ \tau\equiv\frac{i\beta}{2\pi} \ , \ \beta\in\mathbb{R}.
\end{equation} 
We then have the usual open-closed duality of the cylinder amplitude described by the following expansions
\begin{equation}\label{cylZexpansions}
\begin{aligned}
Z^{\text{(open)}}_{\text{cyl}^{(ab)}}(\tau)&=\sum_{\Psi_i\in \mathcal{H}^{a,b}_{\text{open}}}n^{ab}_i \ \chi_{i}(\tau)\\
&\equiv\int\frac{dP}{2}\rho^{\text{(open)}}_{ab}(P) \ \chi_P(\tau) \\
Z^{\text{(closed)}}_{\text{cyl}^{(ab)}}(-1/\tau)&=\sum_{\mathcal{O}_i\in\mathcal{H}^{sc.}_{\text{closed}}} \mathcal{B}_a^i\mathcal{B}_b^i \ \chi_{i}(-1/\tau)\\
&\equiv \int\frac{dP}{2}\rho^{\text{(closed)}}_{ab}(P) \ \chi_P(-1/\tau)
\end{aligned}
\end{equation} 
where $n^{ab}_i\in\mathbb{Z}_{>0}$ are the multiplicities of the boundary primary operators, and $\chi_P$ are the usual $c>1$ characters of the Virasoro algebra:
\begin{equation}\label{virchar}
\begin{aligned}
\chi_P(\tau)&=\frac{e^{2\pi i \tau P^2}}{\eta(\tau)} , \ \ \ \ \ i\neq\mathbb{1}\\
\chi_{\mathbb{1}}(\tau)&=\frac{(1-e^{2\pi i \tau})e^{-\pi i\tau Q^2/2}}{\eta(\tau)}\ , \ \ \ \ \ \ \tau=\frac{i\beta}{2\pi} \ , \ \beta\in\mathbb{R}.
\end{aligned}
\end{equation} 
Given the relation between the distributions in (\ref{distrRelation}), the open-closed duality in the case of the cylinder partition function becomes the following transform on the corresponding densities  
\begin{equation}\label{cylZrelation}
\begin{aligned}
\rho^{\text{(open)}}_{ab}(P')&=\int \frac{dP}{2}\modS_{P'P}[\mathbb{1}] \  \rho^{\text{(closed)}}_{ab}(P)\\
\modS_{P'P}[\mathbb{1}]&=2\sqrt{2}\cos{(4\pi PP')}
\end{aligned}
\end{equation} 
where the modular kernel asymptotes to the usual Fourier kernel in the variable $P$ in the limit $P_0\rightarrow i\frac{Q}{2}$, and relates the Virasoro characters (\ref{virchar}) in dual channels (see Appendix \ref{torus1ptkernel}). In particular, for the degenerate representation of the identity character we get the more subtle expression \cite{Jackson:2014nla,Maxfield:2019hdt,Collier:2019weq}:
\begin{equation}\label{char}
\begin{aligned}
\chi_{\mathbb{1}}(-1/\tau)&=\int_{-\infty}^{\infty} \frac{dP}{2}\modS_{P\mathbb{1}}[\mathbb{1}] \ \chi_{P}(\tau)
\end{aligned}
\end{equation}
where 
\begin{equation}\label{identchar}
\begin{aligned}
\modS_{P\mathbb{1}}[\mathbb{1}]=4\sqrt{2}\sinh{(2\pi b P)}\sinh{(2\pi b^{-1}P)}\equiv \rho_0(P).
\end{aligned}
\end{equation}
\par We can now reproduce the known Cardy formula for the boundary spectrum $\rho^{\text{(open)}}_{ab}(P')$ \cite{Cardy:1989ir,Affleck:1991tk,Hikida:2018khg} by working purely in terms of the distributions and their relation (\ref{cylZrelation}), following the logic of \cite{Maxfield:2019hdt,Collier:2019weq}; the density of states $\rho^{\text{(closed)}}_{ab}(P)$ is a sum of delta functions for each scalar primary operator dimension, so we may write (\ref{cylZrelation}) schematically as a sum over modular S-matrices with appropriate supports:
\begin{equation}\label{relexp}
\begin{aligned}
\rho^{\text{(open)}}_{ab}(P)&=g_ag_b \ \modS_{P\mathbb{1}}[\mathbb{1}]+\sum_{s_{gap}}\mathcal{B}_a^s\mathcal{B}_b^s \ \modS_{PP_s}[\mathbb{1}].
\end{aligned}
\end{equation}
The above equation should be taken with a grain of salt. The spectral density $\rho^{\text{(open)}}_{ab}(P)$ is a sum of delta functions whereas the RHS of (\ref{relexp}) is a sum of smooth functions. As it was explained thoroughly in \cite{Maxfield:2019hdt,Collier:2019weq}, the two sides make sense and converge only in the sense of distributions, i.e. when integrated against some appropriate space of test functions. We will return to this point shortly. We can now consider this sum in the limit of large $P$. Assuming a scalar gap $s_{gap}$ in the bulk scalar spectrum, in this limit the relative importance of the terms is determined by $P_i$. It is straightforward to show that
\begin{equation}\label{eq:Sratio}
\frac{\modS_{PP'}[\id]}{\modS_{P\id}[\id]} \sim \begin{cases}
 	e^{-4\pi\alpha'P} & \alpha' = \tfrac{Q}{2}+iP' \in (0,\tfrac{Q}{2}) \\
 	2\cos(4\pi P P')e^{-2\pi Q P} &  P' \in \RR
 \end{cases}
 \quad\text{as }P\to\infty
\end{equation}
Therefore, we find that the density of boundary states at large $P$ asymptotically approaches:
\begin{equation}\label{eq:BdyCardy}
	\rho^{\text{(open)}}_{ab}(P) \sim e^{\frac{1}{2}(\mathfrak{s}_a+\mathfrak{s}_b)} \rho_0(P) \text{ as }P\to \infty, \text{ where } \rho_0(P) = \modS_{P\id}[\id] \sim \sqrt{2} e^{2\pi QP}.
\end{equation}
This is the familiar Cardy formula for the asymptotic density of boundary states in BCFT$_2$\cite{Cardy:1989ir,Affleck:1991tk,Hikida:2018khg}, correct up to corrections exponential in $\sqrt{h}$ coming from the dimension of the lightest non-vacuum scalar primary state in the bulk. We also expressed the result in terms of the corresponding boundary entropies via (\ref{bdyentropy}).

\par  The most conservative statement is that (\ref{eq:BdyCardy}) applies in an integrated sense: the total number of states below a given (boundary) conformal dimension is asymptotic to the integral of the boundary Cardy formula. Another possibility is that (\ref{eq:BdyCardy}) would hold even when integrated over a small window around some large dimension. That result however would depend in general on the size of that window, something
that the asymptotic formula (\ref{eq:BdyCardy}) does not make explicit at all. A careful analysis of this sort (for the case of bulk asymptotic formulas) was performed recently in a series of nice papers\cite{Qiao:2017xif,Mukhametzhanov:2018zja,Mukhametzhanov:2019pzy,Pal:2019zzr,Pal:2019yhz,Ganguly:2019ksp,Mukhametzhanov:2020swe,Das:2020uax} implementing tools from the so-called Tauberian theory. It will be certainly interesting to apply analogous Tauberian theory
methods in the various boundary asymptotic formulas that we obtain here, though we will not
focus on that aspect in the present work.
\par Returning back to the general case of the cylinder one-point function, we would like to follow similar logic and obtain an asymptotic result for the boundary OPE coefficients via (\ref{distrRelation}). It is straightforward to show that the \textit{microcanonical average} of the diagonal heavy-heavy-light boundary structure constant $C^{(abb)}$ takes the form
\begin{equation}\label{HLHcyl1pt}
	\overline{C^{(abb)H}_{H0}} \sim  \left( C^{(a)\mathbb{1}}_{\chi}C^{(b)0}_{\chi}C^{(bbb)\mathbb{1}}_{00}\right) {\modS_{P_HP_\chi}[P_0]\over \rho_0(P_H)},~P_H\to\infty.
\end{equation}
This is the BCFT counterpart of the Kraus-Maloney analysis for the torus one-point function \cite{Kraus:2016nwo}. A few comments about this expression are in order. First, (\ref{HLHcyl1pt}) is an asymptotic formula which is finite in the central charge and the expression is unambiguous of the boundary operator normalisations, which in our notation is manifest from the fact that the heavy operator indices are appropriately contracted as upper and lower indices in the structure constant.
The operator $\mathcal{O}_\chi$ is the lightest non-trivial bulk scalar operator that couples to $\Psi_0$ on the boundary $b$, with fixed conformal dimension labelled by $P_{\chi}(=\pm\bar{P}_{\chi})$. To obtain the microcanonical average we divided with the boundary Cardy density (\ref{eq:BdyCardy}) for the heavy operator, namely $\rho_{ab}(P_H)=e^{\frac{1}{2}(\mathfrak{s}_a+\mathfrak{s}_b)}\rho_0(P_H)$, and we additionally used the relations $C^{(a)}_{\chi\mathbb{1}}=C^{(a)\mathbb{1}}_{\chi}g_a$, $C^{(b)}_{\chi0}=C^{(b)0}_{\chi}C^{(bbb)\mathbb{1}}_{00}g_b$. We emphasize that each of the three factors in the parenthesis of (\ref{HLHcyl1pt}) is \textit{order one} in the boundary entropy. Finally, note that the diagonal heavy-heavy-light boundary OPE coefficients are non-trivial only when the boundary conditions are of the form $(abb)$ (with either $a\neq b$ or $a=b$). Put it differently, the diagonal element for the structure constant $C^{(abc)H}_{H0}$ for generic boundary conditions $(abc)$ is by definition zero.
\par  The asymptotic formula (\ref{HLHcyl1pt}) holds provided that $\mathcal{O}_\chi$ is sufficiently light ($P_\chi$ lies in the discrete regime in the sense of \cite{Collier:2018exn}) and that there exists a gap above this lightest operator in the bulk scalar spectrum so that corrections due to the inversion of the contributions of other operators in the original channel are indeed suppressed, as it was shown explicitly in \cite{Collier:2019weq} (and we recall in Appendix \ref{app:torusOnePtAsymptotics}). The large $P$ asymptotics of this formula are straightforward to find by taking the large $P_H$ limit of the modular S kernel, namely
\begin{equation}\label{eq:modSOverRho0}
	{\modS_{P_HP_\chi}[P_0]\over\rho_0(P_H)}\approx e^{-4\pi(\frac{Q}{2}+iP_\chi) P_H}P_H^{h_0}.
\end{equation}

\par Ultimately, we would like to make a statement about the heavy-heavy-light limit of the boundary three-point function coefficient which is a physically relevant quantity. Equation (\ref{HLHcyl1pt}) captures the asymptotics of the structure constant, which is related with the boundary three-point function coefficient via an appropriate contraction with the operator metric. If we choose the normalization (\ref{normalization}) for the boundary operators, we find the following asymptotic formula: 
\begin{equation}\label{HHLnormal}
	\overline{C^{(abb)}_{H0H}} \sim  e^{\frac{1}{4}(\mathfrak{s}_a+\mathfrak{s}_b)} \left( C^{(a)\mathbb{1}}_{\chi}C^{(b)0}_{\chi}\right){\modS_{P_HP_\chi}[P_0]\over \rho_0(P_H)},~P_H\to\infty.
\end{equation}
This result captures the boundary three-point functions and is now of order $e^{\mathfrak{s}/2}$ in the boundary entropy, though this fact depended highly on the choice of our normalisation. It is only when we compare the \textit{relative} size of two or more OPE coefficients (under the same normalisation) that this expression might be of some interest. We will return to this point in section \ref{sec:bcftETH}.

\section{Bulk-to-boundary OPE asymptotics}\label{sec:Bbasymptotics}

After discussing asymptotics from the open-closed duality on the cylinder we will now move on to study crossing equations on slightly more involved Riemann surfaces with boundary that will lead us to universal asymptotics for the bulk-to-boundary structure constants $C^{(s)i}_{\alpha}$. For that purpose we will examine the crossing equations for: i) bulk two-point functions on the disk, and ii) the partition function on the torus with a hole. This study will lead us to universal asymptotic formulas for light bulk-heavy boundary and heavy bulk-heavy boundary structure constants respectively. We will see that these two distinct asymptotics originate essentially from a single formula. We will also make some comments at the end of the section for the -- somewhat distinct -- case of heavy bulk-light boundary asymptotics.
\subsection{Bulk two-point function on the disk: light-heavy}\label{sec:BbasymptoticsLH}
We start with the constraints coming from the two-point functions of bulk operators $\mathcal{O}_1 ,\mathcal{O}_2$ with conformal dimensions $(h_1,\bar{h}_1),(h_2,\bar{h}_2)$ on the UHP or the disk with boundary condition labelled by $a$. We denote the correlation function following the notation (\ref{notation}) as:
\begin{equation}
\begin{aligned}
G_{0,2;1,0}(\eta) &= \braket{\mathcal{O}_1 (z_1,\bar{z}_1)\mathcal{O}_2(z_2,\bar{z}_2)}_{\text{disk}^{(a)}}.
\end{aligned}
\end{equation}
The correlation function depends only on the conformal cross-ratio $\eta = \f{(z_1 -z_2)(\bar{z}_1 -\bar{z}_2)}{(z_1-\bar{z}_2)(\bar{z}_1-z_2)}$ and after grouping together several kinematic terms we can bring it to the general form:
\begin{equation}
\begin{aligned}
G_{0,2;1,0}(\eta) &= (2\text{Im}z_1)^{r - h_1 -\bar{h}_1}(2\text{Im}z_2)^{r - h_2 -\bar{h}_2} |z_1-z_2|^{2 (r-h_1-h_2)} |z_1-\bar{z}_2|^{2 (r-h_1-h_2)}  \notag \\
& \qquad (\bar{z}_2 -\bar{z}_1)^{h_1-\bar{h}_1 + h_2-\bar{h}_2} (z_2 - \bar{z}_1)^{h_1 - \bar{h}_1 - h_2 +\bar{h}_2 } Y(\eta),
\end{aligned}
\end{equation}
where $r = \f{1}{3}(h_1 + \bar{h}_1 + h_2 + \bar{h}_2)$ and $Y(\eta)$ is a function of the cross-ratio. We then get two equivalent expansions of this function which we call ``boundary OPE'' channel when $\eta\rightarrow0$, and ``bulk OPE'' channel when $\eta\rightarrow1$\cite{Cardy:1991tv,Lewellen:1991tb}:
\begin{equation}\label{bk2pt}
\begin{aligned}
Y^{\text{($\partial$OPE)}} (\eta) &=\sum_{\Psi_i\in \mathcal{H}^{a,a}_{\text{open}}} C_1^{(a)i}C^{(a)}_{2i} \ \mathcal{F}\sbmatrix{P_2 & \bar{P}_1 \\ P_1 & \bar{P}_2}(P_i|\eta)\\
&\equiv\int \frac{dP}{2} \rho^{\text{($\partial$OPE)}}_{12}(P) \ \mathcal{F}\sbmatrix{P_2 & \bar{P}_1 \\ P_1 & \bar{P}_2}(P|\eta)\\
Y^{\text{(bOPE)}} (1-\eta) &=\sum_{\mathcal{O}_i\in\mathcal{H}^{sc.}_{\text{closed}}} C_{12 i}C^{(a)}_{i\mathbb{1}} \  \mathcal{F}\sbmatrix{P_2 & P_1 \\ \bar{P}_1 & \bar{P}_2}(P_i|1-\eta) \\
&\equiv\int \frac{dP}{2}\rho^{\text{(bOPE)}}_{12}(P) \ \mathcal{F}\sbmatrix{P_2 & P_1 \\ \bar{P}_1 & \bar{P}_2}(P|1-\eta),
\end{aligned}
\end{equation}
and we introduced the even distributions for the CFT data:
\begin{equation}\label{distrbk2pt}
\begin{aligned}
\rho^{\text{($\partial$OPE)}}_{12}(P)&:=\sum_{\Psi_i\in \mathcal{H}^{a,a}_{\text{open}}} C_1^{(a)i}C^{(a)}_{2i}[\delta(P-P_i)+\delta(P+P_i)]\\
\rho^{\text{(bOPE)}}_{12}(P)&:=\sum_{\mathcal{O}_i\in\mathcal{H}^{sc.}_{\text{closed}}} C_{12 i}C^{(a)}_{i\mathbb{1}}[\delta(P-P_i)+\delta(P+P_i)].
\end{aligned}
\end{equation}
In the first line of (\ref{bk2pt}) the sum runs over boundary primary operators $\Psi^{aa}_i$ appearing in the bulk-to-boundary OPE of $\mathcal{O}_1$ and $\mathcal{O}_2$, and $C_1^{(a)i},C^{(a)}_{2i}$ are the corresponding structure constants (appropriately contracted in the boundary operator indices). The sum over $\mathcal{O}_i$ on the third line runs over bulk \textit{scalar} primaries appearing in the $\mathcal{O}_1\times\mathcal{O}_2$ OPE with $C_{12i}$ the corresponding bulk OPE coefficient, and $C^{(a)}_{i\mathbb{1}}=\mathcal{B}^i_a$ is the corresponding disk one-point function coefficient. Additionally, $\mathcal{F}\sbmatrix{P_2 & \overline{P_1} \\ P_1 & \overline{P_2}}(P|\eta)$ (similarly for $\mathcal{F}\sbmatrix{P_2 & P_1 \\ \overline{P_1} & \overline{P_2}}(P|1-\eta)$ ) is the \textit{holomorphic} S-channel (T-channel) sphere four-point Virasoro block with external dimensions $h_1,h_2,\bar{h}_1,\bar{h}_2$\footnote{The conformal blocks are normalized as $\mathcal{F}\sbmatrix{P_2 & P_1 \\ P_3 & P_4}(P|\eta)\sim\eta^{h_p-h_1-h_2}$ as $\eta\rightarrow0$. Here we follow the notation of \cite{Collier:2018exn} for the labelling of the external operators on the block.}.
\par The equivalence between the two expansions in (\ref{bk2pt}) leads to the sewing constraint: 
\begin{equation}\label{sewingbk2pt}
\begin{aligned}
Y^{\text{($\partial$OPE)}} (\eta)=Y^{\text{(bOPE)}} (1-\eta).
\end{aligned}
\end{equation}
\begin{figure}
\centering
 \includegraphics[width=.14\textwidth]{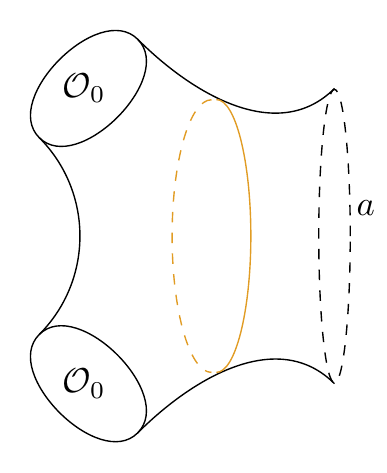} \   \raisebox{.073\textwidth}{\scalebox{1.5}{$=\int \frac{dP}{2}\fusion_{PP'}\sbmatrix{P_0 & \bar{P}_0 \\ P_0 & \bar{P}_0}$}} \includegraphics[width=.14\textwidth]{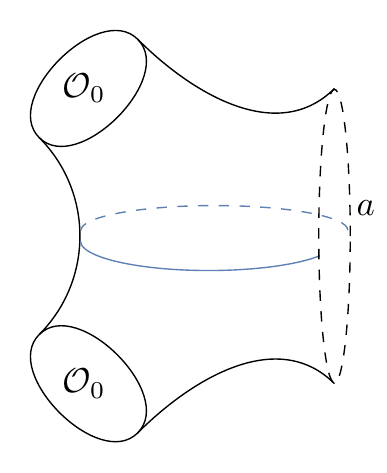}
\caption{The bulk OPE channel (left) and boundary OPE channel (right) of a bulk two-point function on the disk for identical bulk external operators. The corresponding conformal blocks are related with a \textit{holomorphic} copy of the fusion kernel.}\label{fig:bulk2ptcrossing}
\end{figure}In the original work \cite{Cardy:1991tv,Lewellen:1991tb},
 the authors studied (\ref{sewingbk2pt}) in the case of minimal models and, using the \textit{fusion matrices} that relate different channel conformal blocks, they were able to rewrite the sewing constraint as a constraint purely on the CFT data. We will now apply the same logic here, except we will implement the power of the \textit{fusion kernel} constructed by Ponsot and Teschner \cite{Ponsot:1999uf,Ponsot:2000mt,Teschner:2001rv} to make a statement about irrational CFTs with central charge $c>1$.
\par The defining  relation of the fusion kernel $\fusion_{P_sP_t}\sbmatrix{P_2 & P_1 \\ P_3 & P_4}$ is
\begin{equation}\label{eq:fusionTransformation}
	\mathcal{F}\sbmatrix{P_2 & P_3 \\ P_1 & P_4}(P_t|1-z) = \int_{C} \frac{d P_s}{2}\fusion_{P_sP_t}\sbmatrix{P_2 & P_1 \\ P_3 & P_4}\mathcal{F}\sbmatrix{P_2 & P_1 \\ P_3 & P_4}(P_s|z), \ \ \ \ \ \ \ \ z\in\mathbb{C}-{\{0,1\}}.
\end{equation}
The kernel expresses holomorphic Virasoro blocks on the T-channel as a linear combination of S-channel blocks, and $z\in\mathbb{C}-{\{0,1\}}$ is the usual sphere four-point cross-ratio. The kernel $\fusion_{P_sP_t}\sbmatrix{P_2 & P_1 \\ P_3 & P_4}$ is an explicit meromorphic function of $P_s,P_t$ and the support $C$ of the integral depends on the external operator unitary dimensions: if $Re\left(\alpha_1+\alpha_2\right)>\frac{Q}{2}$ the contour $C$ can be chosen to run along the whole real line $\mathbb{R}$, whereas if $\alpha_1+\alpha_2<\frac{Q}{2}$ ($\alpha_i$ necessarily real in the discrete regime) some poles of $\fusion_{P_sP_t}$ may cross the contour $C=\mathbb{R}$ and hence the integral acquires additional contributions from the residues of these poles. We review in detail the properties of the kernel in Appendix \ref{subsec:fusion}. 
\par For our purposes, we will consider the case of two identical external bulk operators $\mathcal{O}_0$ (see Fig.\ref{fig:bulk2ptcrossing}). In this case using the transformation (\ref{eq:fusionTransformation}) it is straightforward to re-write the sewing constraint (\ref{sewingbk2pt}) as a transform relating the two corresponding distributions (\ref{distrbk2pt}). One finds
\begin{equation}\label{distrrelbk2pt}
\begin{aligned}
\rho^{\text{($\partial$OPE)}}_{0}(P)=\int \frac{dP'}{2}\fusion_{PP'}\sbmatrix{P_0 & \bar{P}_0 \\ P_0 & \bar{P}_0}\rho^{\text{(bOPE)}}_{0}(P')
\end{aligned}
\end{equation}
We see that for identical external operators the distribution $\rho^{\text{($\partial$OPE)}}_{0}(P)=\sum_{\Psi_i\in \mathcal{H}^{a,a}_{\text{open}}} C_0^{(a)i}C^{(a)}_{0i}[\delta(P-P_i)+\delta(P+P_i)]$ captures the \textit{square} of the bulk-to-boundary structure constants. We can now repeat the logic of \cite{Maxfield:2019hdt,Collier:2019weq} to derive an asymptotic formula for this density; for a compact CFT the distribution $\rho^{\text{(bOPE)}}_{0}(P')$ is a sum of delta functions for each scalar primary operator dimension, so we may write (\ref{distrrelbk2pt}) schematically as a sum over fusion matrices with appropriate supports:
\begin{equation}
\begin{aligned}
\rho^{\text{($\partial$OPE)}}_{0}(P)=g_a \ \fusion_{P\mathbb{1}}\sbmatrix{P_0 & \bar{P}_0 \\ P_0 & \bar{P}_0}+\sum_{s_{gap}}C_{00 s}C^{(a)}_{s\mathbb{1}}\fusion_{PP_s}\sbmatrix{P_0 & \bar{P}_0 \\ P_0 & \bar{P}_0}.
\end{aligned}
\end{equation}
In the first term above we used the fact that $C_{00\mathbb{1}}=1$ and $C^{(a)}_{\mathbb{1}\mathbb{1}}=C^{(a)\mathbb{1}}_{\mathbb{1}}g_a=g_a$. Now we want to take the large $P$ limit of this expression. Assuming a scalar gap $s_{gap}$ in the bulk scalar spectrum, it was shown in \cite{Collier:2019weq} (and we review in Appendix \ref{sub:fusionasympt})
that\footnote{This result is accurate up to a factor independent of $P_s$ (see equation (\ref{eq:fusionSuppressionn})).}
\begin{equation}\label{eq:fusionSuppression}
\frac{\fusion_{P P_s}}{\fusion_{P\id}} \approx \begin{cases}
 	e^{-2\pi\alpha_s P} & \alpha_s = \tfrac{Q}{2}+iP_s \in (0,\tfrac{Q}{2}) \\
 	e^{-\pi Q P}\cos(2\pi P_s P) &  P_s \in \RR
 \end{cases}
 \quad\text{as }P\to\infty
\end{equation}
\\
Therefore, we find that the distribution of the boundary data at large $P$ asymptotically approaches:
\begin{equation}\label{eq:LHbktobd}
\rho^{\text{($\partial$OPE)}}_{0}(P)\sim g_a \ \fusion_{P\mathbb{1}}\sbmatrix{P_0 & \bar{P}_0 \\ P_0 & \bar{P}_0}=e^{\frac{\mathfrak{s}_a}{2}} \ \fusion_{P\mathbb{1}}\sbmatrix{P_0 & \bar{P}_0 \\ P_0 & \bar{P}_0}, ~~~~ P\rightarrow\infty
\end{equation}
Just as for the boundary Cardy formula explained in subsection \ref{subsec:boundaryCardy}, (\ref{eq:LHbktobd}) should be interpreted as a microcanonical statement about the asymptotic spectral density integrated over a window of energies. We can translate the result to a microcanonical \textit{average} of bulk-to-boundary OPE coefficients squared, by dividing with the boundary Cardy formula (\ref{eq:BdyCardy}) for the dimension of the boundary operator $\Psi_i^{aa}$ giving the asymptotic density of boundary states $\rho^{\text{(open)}}_{aa}(P) \sim e^{\mathfrak{s}_a} \rho_0(P)$ in the relevant limit. Furthermore, as elucidated in \cite{Collier:2018exn,Collier:2019weq}, the identity fusion kernel $\fusion_{P\mathbb{1}}$ takes a particularly simple form which we review in Appendix \ref{subsec:fusion}. We can write
\begin{equation}\label{eq:idFusion}
	\fusion_{P\id}\sbmatrix{P_2 & P_1 \\ P_2 & P_1} = \rho_0(P) C_0(P_1,P_2,P),
\end{equation}
where $\rho_0(P)$ is the density of states appearing as the modular S-transform of the vacuum \eqref{identchar}. The factor $C_0$ was dubbed the ``universal OPE density'' in \cite{Collier:2019weq} and is a symmetric function under the exchange of all three of its arguments. It has a simple explicit expression in terms of the special function $\Gamma_b$:
\begin{equation}\label{eq:C0}
	C_0(P_1,P_2,P_3) := \frac{1}{\sqrt{2}}{\Gamma_b(2Q)\over \Gamma_b(Q)^3}\frac{\prod_{\pm\pm\pm}\Gamma_b\left(\tfrac{Q}{2}\pm iP_1\pm iP_2 \pm iP_3\right)}{\prod_{k=1}^3\Gamma_b(Q+2iP_k)\Gamma_b(Q-2iP_k)}.
\end{equation}
The $\prod$ in the numerator denotes the product of the eight combinations related by the reflections $P_k\to -P_k$. The function $\Gamma_b$ is a `double' gamma function, which is meromorphic, with no zeros, and with poles at argument $-mb-nb^{-1}$ for nonnegative integers $m,n$ (similarly to the usual gamma function, which has poles at nonpositive integers). 

The microcanonical average of the bulk-to-boundary coefficients squared is therefore captured by $C_0$ accompanied by a universal factor exponential in the boundary entropy:
\begin{equation}\label{LHbktobdyResult}
	\overline{\left|C^{(a)i}_{0 }\right|^2} \sim e^{-\mathfrak{s}_a/2}  \ C_0\left(P_{0},\bar{P}_0,P_i\right),~~~~~P_i \to \infty.
\end{equation}
This result is valid at finite central charge and finite boundary entropy for any fixed bulk operator $\op_0$, averaging over operators $\Psi^{aa}_i$ in a large (boundary) dimension limit. The asymptotic form of $C_0$ in the limit where one of its arguments is large was first computed in \cite{Collier:2018exn} and in our set up reads
\begin{equation}\label{eq:C0HLL}
	C_0(P_0,\bar{P}_0,P_i) \sim 2^{-4P_i^2}e^{-\pi Q P_i} P_i^{4(h_0+\bar{h}_0)-{3Q^2+1\over 2}}{2^{Q^2-2\over 6}\Gamma_0(b)^6\Gamma_b(2Q)\over\Gamma_b(Q)^3\Gamma_b(Q+2iP_0)\Gamma_b(Q-2iP_0)\Gamma_b(Q+2i\bar{P}_0)\Gamma_b(Q-2i\bar{P}_0)},
\end{equation}
where $\Gamma_0(b)$ is a special function that appears in the large-argument asymptotics of $\Gamma_b$ (see equation (\ref{g0})). 

We emphasize here that by ``square'' of the structure constants we mean $\left|C^{(a)i}_{0 }\right|^2\equiv C^{(a)i}_{0 }C^{(a)}_{0 i}=\sum_{i'}\mathfrak{g}_{(aa)}^{i'i}C^{(a)}_{0i'}C^{(a)}_{0i}$. Hence, the expression (\ref{LHbktobdyResult}) holds for any choice of the operator metric $\mathfrak{g}$ defined in (\ref{metric}).
If we explicitly choose the normalization (\ref{normalization}), we can write an asymptotic formula for the bulk-to-boundary two-point functions (i.e. the structure constants with all indices lowered) which in this case gives:
\begin{equation}\label{LHbktobdyResultnorm}
\overline{\left|C^{(a)}_{0i}\right|^2} \sim C_0\left(P_{0},\bar{P}_0,P_i\right),~~~~~P_i \to \infty.
\end{equation}Therefore in this special normalisation the dependence on the boundary entropy completely disappears and  the asymptotic formula for the bulk-to-boundary two-point functions only depends on the bulk central charge and the dimensions of the bulk operator $\op_0$ through the universal OPE coefficient $C_0$. Somewhat remarkably, we will discover next that under the normalisation (\ref{normalization}) all the different heavy limits of the bulk-to-boundary two-point functions as well as all the different heavy limits of the boundary three-point functions will be governed by a single factor of $C_0$, exactly as in (\ref{LHbktobdyResultnorm}).

\subsection{Torus with a hole: heavy-heavy}\label{sec:BbasymptoticsHH}

We will next consider the partition function on a torus with an open conformal boundary labelled by $a$. This surface can be thought of as arising from the previously considered bulk two-point function on the disk where we ``glue'' together (i.e. identify and sum over a complete set of) the bulk operators. We will establish the sewing constraint on this surface which will then allow us to derive an asymptotic formula for bulk-to-boundary structure constants in the heavy-heavy limit.

We denote the partition function on the torus with a hole as $G_{1,0;1,0}$, following (\ref{notation}). This correlation function depends in general on three real moduli parameters which we will call collectively $\{\beta_i\}$, $i=1,2,3$. As we will see, the exact dependence on this moduli will not be relevant for our discussion on the asymptotic formulas since we will only care about the crossing moves that relate different ``cuttings'' of the surface.
\par The partition function admits the following two equivalent decompositions:
\ba\label{torusholedecomp}
 G^{(\text{OC-loop})}_{1,0;1,0}(\beta_i)&= \sum_{\op_0\in\mathcal{H}_{\text{closed}}}\sum_{\Psi_i\in \mathcal{H}^{a,a}_{\text{open}}} C_{0}^{(a)i}C^{(a)}_{0i} \ \mathcal{F}^{(\text{OC-loop})}(P_{0},\bar{P}_{0};P_{i}|\beta_i)  \notag \\
&\equiv \int \f{dP_{0}}{2} \f{d\bar{P}_{0}}{2}\f{d P}{2} \rho^{(\text{OC-loop})}(P_{0},\bar{P}_{0};P) \ \mathcal{F}^{(\text{OC-loop})}(P_{0},\bar{P}_{0};P|\beta_i)   \notag \\
G^{(\text{tadpole})}_{1,0;1,0}(\tilde{\beta}_i)&=\sum_{\op'_0\in\mathcal{H}_{\text{closed}}}\sum_{\mathcal{O}'_i\in\mathcal{H}^{sc.}_{\text{closed}}} C_{0'0' i'}C^{(a)}_{i'\mathbb{1}} \ \mathcal{F}^{(\text{tadpole})}(P'_{0},\bar{P}'_{0},P'_i|\tilde{\beta}_i)  \notag \\
&\equiv  \int \f{dP'_{0}}{2} \f{d\bar{P}'_{0}}{2}\f{d P'}{2} \rho^{(\text{tadpole})}(P'_{0},\bar{P}'_{0},P')  \mathcal{F}^{(\text{tadpole})}(P'_{0},\bar{P}'_{0},P'|\tilde{\beta}_i)
\ea
\begin{figure}
		\qquad\qquad\, \includegraphics[width=.2\textwidth]{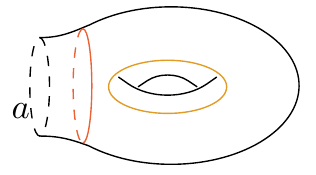}  \raisebox{.034\textwidth}{\scalebox{1.5}{$=\int \frac{dP_1}{2}\frac{d\bar{P}_1}{2}\modS_{P_1P_1'}[P'_2]\modS_{\bar{P}_1 \bar{P}_1'}[P'_2]$}} \includegraphics[width=.2\textwidth]{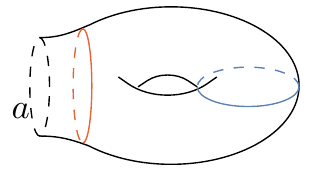} 
		\flushright\raisebox{.034\textwidth}{\scalebox{1.5}{$=\int \frac{dP_1}{2}\frac{d\bar{P}_1}{2}\frac{dP_2}{2}\modS_{P_1P_1'}[P'_2]\modS_{\bar{P}_1 \bar{P}_1'}[P'_2] \fusion_{P_2P_2'}\sbmatrix{P_1&\bar{P}_1\\P_1&\bar{P}_1}$}}
		\includegraphics[width=.2\textwidth]{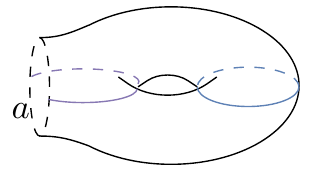}
	\caption{The decomposition of the`` tadpole'' channel conformal block in terms of ``OC-loop'' channel conformal blocks for the genus-one surface with a conformal boundary labelled by $a$: first we interchange cycles on the torus with an S-transformation, and then we change to the dual channel of the resulting disk two-point function (of identical bulk operators, suitably summed over) with a holomorphic fusion transformation.\label{fig:TorusHole}}
\end{figure}In the first line, we called ``OC-loop'' (open-closed loop) channel the decomposition where we construct the surface out of two bulk-to-boundary legos as in the bottom right picture of Fig.\ref{fig:TorusHole}. The expansion involves a summation on both boundary (labelled by the Liouville momentum $P_i$) and bulk primary operators (labelled by the Liouville momenta $(P_0,\bar{P}_0)$) and we denoted the corresponding conformal blocks as $\mathcal{F}^{(\text{OC-loop})}$. In analogous sense, in the so-called ``tadpole'' channel we construct the surface out of one bulk OPE lego and one bulk-to-boundary lego as in the top left picture of Fig.\ref{fig:TorusHole}. The expansion now involves a summation over purely bulk spectra where one of the two summations should necessarily be on the scalar sector $\mathcal{H}^{sc.}_{\text{closed}}$ (because of the factor $C^{(a)}_{i'\mathbb{1}}=\mathcal{B}_a^{i'}$). We again denote formally the corresponding conformal blocks in this channel as $\mathcal{F}^{(\text{tadpole})}$ which depend in principle on different moduli parametrized as $\{\tilde{\beta}_i\}$. The second and fourth lines in (\ref{torusholedecomp}) define the ``OC-loop'' and ``tadpole'' spectral densities $\rho^{(\text{OC-loop})},\rho^{(\text{tadpole})}$ for the BCFT data.

Proceeding with the same logic as before, we will consider the crossing kernel that decomposes the  tadpole channel conformal block in terms of OC-loop channel conformal blocks. This sewing procedure is illustrated in Figure \ref{fig:TorusHole}, from which we see that one first gets a factor of the torus one-point kernel (along with its anti-holomorphic counterpart) since we interchange between equivalent cycles on the torus\footnote{Note that the holomorphic part $\modS_{P_1P_1'}[P'_2]$ and the anti-holomorphic part $\modS_{\bar{P}_1 \bar{P}_1'}[P'_2]$ of the torus kernel contribution in the total kernel have the same ''external'' momentum $P'_2$. This is because the operator corresponding to $P'_2$ is necessarily scalar since it fuses to the boundary.}, and secondly we get a holomorphic copy of the sphere four-point kernel since at this stage we essentially decompose a disk two-point function for identical bulk external operators as we did in subsection \ref{sec:BbasymptoticsLH}. The total kernel that relates the corresponding spectral densities reads
\begin{equation}\label{HHdistrrel}
\begin{aligned}
\mathbb{K}^{\text{(open torus)}}_{P_{1},\bar{P}_{1},P_2;P'_1,\bar{P}'_1,P'_2}:=&\modS_{\bar{P}_1 \bar{P}_1'}[P'_2]\modS_{P_1P_1'}[P'_2] \fusion_{P_2P_2'}\sbmatrix{P_1&\bar{P}_1\\ P_1&\bar{P}_1}\\
\rho^{(\text{OC-loop})}(P_{1},\bar{P}_{1};P_2)=\int \frac{dP'_1}{2}\frac{d\bar{P}'_1}{2}&\frac{dP'_2}{2} \ \mathbb{K}^{\text{(open torus)}}_{P_{1},\bar{P}_{1},P_2;P'_1,\bar{P}'_1,P'_2} \ \rho^{(\text{tadpole})}(P'_1,\bar{P}'_1,P'_2)
\end{aligned}
\end{equation}
In an appropriate kinematic limit the BCFT data on the tadpole channel will be dominated by the bulk identity operator propagating in both internal circles on the top left picture of Figure \ref{fig:TorusHole}. Assuming a gap in the dimension above the identity operator in the bulk, in the limit $P_1,\bar{P}_1,P_2\rightarrow\infty$ the spectral density $\rho^{(\text{OC-loop})}(P_{1},\bar{P}_{1},P_2)$ will be approximated by
 \begin{equation}\label{HHasymptdensity}
\begin{aligned}
&\rho^{(\text{OC-loop})}(P_{1},\bar{P}_{1};P_2)\sim g_a \ \mathbb{K}^{\text{(open torus)}}_{P_{1},\bar{P}_{1},P_2;\mathbb{1},\mathbb{1},\mathbb{1}} , ~~~~~~~~ P_1,\bar{P}_1,P_2\rightarrow\infty\\
\mathbb{K}^{\text{(open torus)}}_{P_{1},\bar{P}_{1},P_2;\mathbb{1},\mathbb{1},\mathbb{1}}&=\rho_0(\bar{P}_1)\rho_0(P_1)\fusion_{P_2\mathbb{1}}\sbmatrix{P_1&\bar{P}_1\\ P_1&\bar{P}_1}=\rho_0(\bar{P}_1)\rho_0(P_1)\rho_0(P_2)C_0(P_1,\bar{P}_1,P_2).
\end{aligned}
\end{equation}  
Corrections to this identity contribution due to the exchange of non-vacuum bulk primaries in the tadpole channel are actually exponentially suppressed when we only take $P_1,P_2$ to be large with $\bar{P}_1$ fixed (or equivalently if we take $\bar{P}_1,P_2\rightarrow\infty$ with $P_1$ fixed). If we define $\mathcal{K}_{P_1P_2;P'_1P'_2}[\bar{P}_1]:=\modS_{P_1P'_1}[P'_2]\fusion_{P_2P'_2}\sbmatrix{P_1&\bar{P}_1\\ P_1&\bar{P}_1}$, then the kernel in (\ref{HHdistrrel}) can be written as $\mathbb{K}^{\text{(open torus)}}=\modS_{\bar{P}_1 \bar{P}_1'}[P'_2]\times \mathcal{K}_{P_1P_2;P'_1P'_2}[\bar{P}_1]$. In \cite{Collier:2019weq} the following result was shown, which we also review in Appendix \ref{sub:fusionasympt}:
\begin{equation}\label{eq:torusTwoPointId}
	\frac{\mathcal{K}_{P_1P_2;P'_1P'_2}[\bar{P}_1]}{\mathcal{K}_{P_1P_2;\id\id}[\bar{P}_1]} \approx e^{-2\pi\alpha_1'P_1}
\end{equation}
in the limit $P_1,P_2\to\infty$, with either the ratio or difference of $P_1$ and $P_2$ held fixed. Therefore this result immediately shows that the asymptotic formula (\ref{HHasymptdensity}) is also valid in a `large spin' regime for the bulk operator where e.g. we fix $\bar{P}_1$ and we take $P_1\rightarrow\infty$. In this limit the relative suppression (\ref{eq:torusTwoPointId}) of non-vacuum contributions is controlled by $P_1$ only, so we require the additional assumption of a \textit{twist gap} in this case\footnote{The twist of a (primary) state is defined as $t=2\text{min}(h,\bar{h})$.}.

Therefore, the kernel $\mathbb{K}_{P_{1},\bar{P}_{1},P_2;\mathbb{1},\mathbb{1},\mathbb{1}}$ encodes an asymptotic formula for bulk-to-boundary structure constants in the somewhat unusual regime of heavy bulk-heavy boundary, averaged over both heavy bulk and heavy boundary dimensions. Dividing by the corresponding bulk Cardy formula $\rho_0(\bar{P}_1)\rho_0(P_1)$ for the heavy bulk operator, and the boundary Cardy formula $e^{\mathfrak{s}_a} \rho_0(P_2)$ for the boundary operator we arrive at the microcanonical averaged asymptotic result:
 \begin{equation}
\begin{aligned}
\overline{\left|C^{(a)2}_{1}\right|^2} \sim e^{-\mathfrak{s}_a/2}  \ C_0\left(P_{1},\bar{P}_1,P_2\right),~~~~~P_1,\bar{P_1},P_2 \to \infty.
\end{aligned}
\end{equation} 
The square of the structure constants is defined again as $\left(C^{(a)2}_{1 }\right)^2\equiv C^{(a)2}_{1 }C^{(a)}_{12}=\sum_{2'}\mathfrak{g}_{(aa)}^{2'2}C^{(a)}_{12'}C^{(a)}_{12}$. We emphasize that in the presence of a nonzero twist gap in the bulk spectrum the above asymptotic formula also holds in the large spin regime where only $P_1,P_2$ or $\bar{P}_1,P_2$ are taken to be heavy. Once again, the asymptotics of $C_0$ universally governs the asymptotics of bulk-to-boundary structure constants, this time in the heavy bulk-heavy boundary regime. We find the same asymptotic formula (with the same boundary entropy factor) as in the case of light bulk-heavy boundary limit (\ref{LHbktobdyResult}). As before, we can make a statement about the bulk-to-boundary two-point functions by lowering the indices in the corresponding structure constants. In our special normalization (\ref{normalization}) we find again that the boundary entropy factor drops out and we get the simple formula:
\begin{equation}
\overline{\left|C^{(a)}_{12}\right|^2} \sim C_0\left(P_{1},\bar{P}_1,P_2\right),~~~~~P_1,\bar{P_1},P_2 \to \infty.
\end{equation}
which is again the same as the expression (\ref{LHbktobdyResultnorm}) in the light bulk-heavy boundary limit.

There are various ways to study the limit of large arguments of $C_0$ that is relevant in the present case. These limits were discussed extensively in \cite{Collier:2019weq} and we will not repeat them here. We refer the reader to that paper for the corresponding expressions.

\subsection{Cylinder two-point function: heavy-light}\label{sec:BbasymptoticsHL}
\begin{figure}
	\centering
		 \includegraphics[width=.31\textwidth]{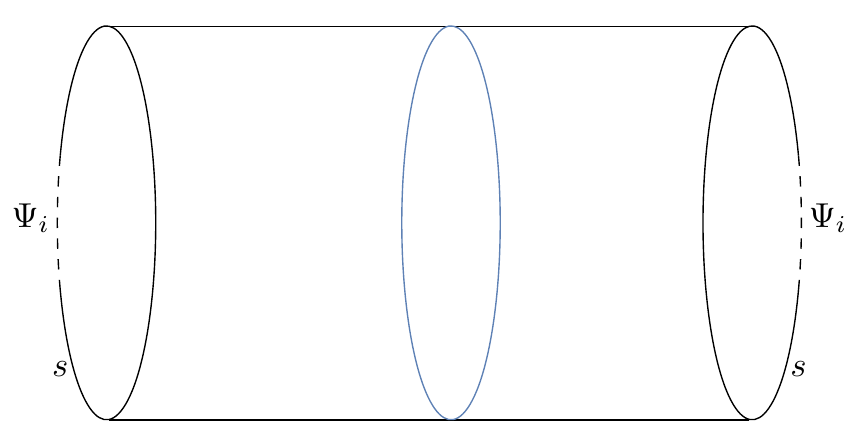} \qquad\qquad\,\qquad  \includegraphics[width=.31\textwidth]{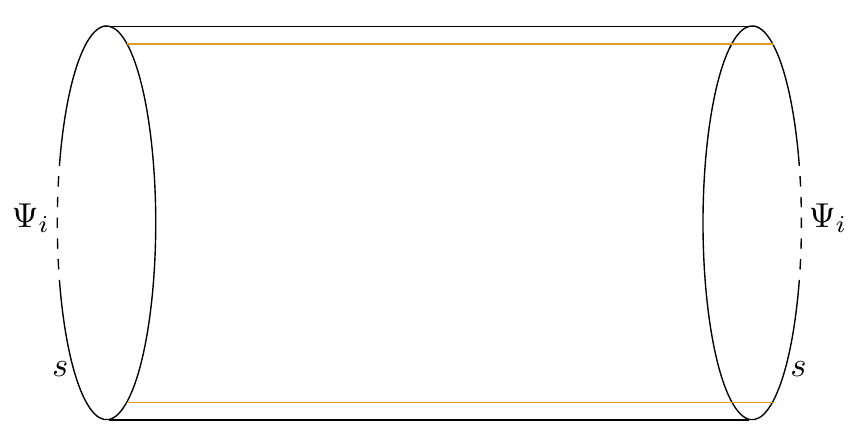} 
	\caption{The ``boundary necklace'' channel (left) and the ``boundary bagel'' channel (right) of the cylinder two-point function with identical boundary operators and identical boundary conditions on each boundary.\label{fig:Cly2ptCase1}}
\end{figure}

We will conclude the discussion on the asymptotics of bulk-to-boundary structure constants by examining the remaining case of the heavy bulk-light boundary limit. For that purpose it is easy to see that the relevant Riemann surface which will give us the desired asymptotics is the cylinder two-point function with two identical external (boundary) primary operators and identical boundary conditions on each boundary (see Figure \ref{fig:Cly2ptCase1}). We denote the amplitude as
\begin{equation}
G_{0,0;2,2}(\beta,\theta)=\langle\Psi_i^{ss}\Psi_i^{ss}\rangle_{\text{cyl}^{(ss)}}, \ \ \ \ \ \ \ \ \ \ \beta\in\mathbb{R}, \ \theta\in[0,\pi].
\end{equation}
The moduli of the surface consists of two real parameters which is the length of the cylinder $\beta$ and the relative angle $\theta$ of the insertions of the two operators on the boundary circles. As we mentioned in section \ref{sec:crossingeqn}, according to \cite{Lewellen:1991tb,Pradisi:1996yd} the sewing constraint on this amplitude is one of the four basic sewing constraints that are necessary and sufficient to ensure consistency of BCFT correlation functions in any Riemann surface with open boundaries. Nonetheless, this correlation function is not so well studied in the literature\footnote{See Appendix B of \cite{Balthazar:2018qdv} for a discussion on the crossing relations of the cylinder two-point function in the case of Liouville theory.} and, as we will see shortly, the construction of a crossing kernel that implements the sewing constraint of the correlation function is not going to be straightforward as in the previous cases.  

We will consider two equivalent constructions of the cylinder two-point function as follows:
\ba\label{cyl2ptCase1decompos}
G^{(\partial \text{necklace})}_{0,0;2,2}(\beta,\theta) &= \sum_{\op_\alpha\in\mathcal{H}_{\text{closed}} } C^{(s)}_{\alpha i}C^{(s)}_{\alpha i} \mathcal{F}^{(\partial \text{necklace})}(P_{\alpha},\bar{P}_{\alpha};P_i|\beta,\theta)  \notag \\
&\equiv \int \f{dP}{2} \f{d\bar{P}}{2} \rho^{(\partial \text{necklace})}(P,\bar{P};P_i)\mathcal{F}^{(\partial \text{necklace})}(P,\bar{P};P_i|\beta,\theta)  \notag \\
G^{(\partial \text{bagel})}_{0,0;2,2}(\tilde{\beta},\tilde{\theta})&= \sum_{\Psi_j,\Psi_k\in\mathcal{H}^{s,s}_{\text{open}}} C^{(sss)}_{i jk}C^{(sss)kj}_{i}\mathcal{F}^{(\partial\text{bagel})}(P_j,P_k;P_i|\tilde{\beta},\tilde{\theta})  \notag \\
&\equiv  \int \f{dP_1}{2} \f{dP_2}{2} \rho^{(\partial\text{bagel})}(P_1,P_2;P_i) \mathcal{F}^{(\partial\text{bagel})}(P_1,P_2;P_i|\tilde{\beta},\tilde{\theta}) 
\ea
In the so-called ``boundary necklace'' channel we construct the surface out of two bulk-to-boundary legos summed over the bulk primary operators, as in the left of Figure \ref{fig:Cly2ptCase1}. We denoted the corresponding conformal blocks as $\mathcal{F}^{(\partial \text{necklace})}$ and, as before, it will not be important for us to know their precise form but only their crossing transformations. In the ``boundary bagel'' channel we are sewing the surface out of two boundary OPE coefficients, as in the right of Figure \ref{fig:Cly2ptCase1}, and the corresponding conformal blocks are dubbed $\mathcal{F}^{(\partial\text{bagel})}$. The spectral densities of the BCFT data $\rho^{(\partial\text{necklace})},\rho^{(\partial\text{bagel})}$ are defined in the second and forth lines of (\ref{cyl2ptCase1decompos}).

We would like to find a crossing kernel that expresses conformal blocks in the boundary bagel channel as a linear combination of conformal blocks in the boundary necklace channel for the particular case of identical external boundary operators $\Psi_i$ (labelled by the Liouville momentum $P_i$) and identical boundary conditions (labelled here by $s$) at the two ends of the cylinder. Such kernel will then allow us to relate the spectral densities $\rho^{(\partial\text{necklace})},\rho^{(\partial\text{bagel})}$ exactly as we did in the previous sections. However we do not manage to find an obvious crossing kernel that relates the blocks by implementing the crossing moves that we have understood so far. In Appendix \ref{sec:CykKerDouble} we argue that one can actually implement the ``doubling trick'' and lift to the compact covering surface (which is a torus with two punctures) which then allows us to guess the form of the crossing kernel from the conformal blocks of the compact case\footnote{The imaginative names ``necklace'' and ``bagel'' that we dubbed the two distinct channels are somewhat motivated from the compact covering case as we will see in Appendix \ref{sec:CykKerDouble}. }. We propose that the crossing kernel in this case takes the following slightly complicated form:
\begin{equation}\label{HLdistrrel}
\begin{aligned}
 &\mathbb{K}^{\text{(cyl--2pt)}}_{P\bar{P};P'\bar{P}'}[P_i] := \int \f{d\tilde{P}}{2} 
 \fusion^{-1}_{\tilde{P} \bar{P}'}\sbmatrix{P'&P_i\\ P'&P_i}
 \mathbb{S}_{P P'}[\tilde{P}]\fusion_{\bar{P} \tilde{P}}\sbmatrix{P&P_i\\ P&P_i}, \\
\rho&^{(\partial \text{necklace})}(P,\bar{P};P_i) =\int \frac{dP'}{2}\frac{d\bar{P}'}{2} \ \mathbb{K}^{\text{(cyl--2pt)}}_{P\bar{P};P'\bar{P}'}[P_i] \ \rho^{(\partial\text{bagel})}(P',\bar{P}';P_i).
\end{aligned}
\end{equation}
As we will see in Appendix \ref{sec:CykKerDouble} this kernel passes some nontrivial consistency checks which makes us believe that it gives the correct formula (for the particular case of identical external boundary operators and identical boundary conditions), even though we did not manage to prove it rigorously. Interestingly, the kernel involves an integral which is composed by both the fusion and modular kernel as well as the inverse of the fusion kernel. In the boundary bagel channel the leading contribution in the BCFT data does \textit{not} come from $\Psi_j=\Psi_k=\mathbb{1}$ this time, since the boundary OPE coefficient $C^{(sss)}_{i\mathbb{1}\mathbb{1}}$ is zero (for non-trivial external operator $\Psi_i$). It is reasonable to expect that the leading nontrivial contribution is when either one of the internal operators is itself equal to $\Psi_i$ and the other internal operator is the identity. We take $\Psi_{j}=\mathbb{1}$ and $\Psi_k=\Psi_i$. In this case the contribution of the kernel gives
\begin{equation}
\begin{aligned}
\mathbb{K}^{\text{(cyl--2pt)}}_{P\bar{P};\mathbb{1}P_i}[P_i]&=\int \f{d\tilde{P}}{2} 
 \fusion^{-1}_{\tilde{P} P_i}\sbmatrix{\mathbb{1}&P_i\\ \mathbb{1}&P_i}
 \mathbb{S}_{P \mathbb{1}}[\tilde{P}]\fusion_{\bar{P} \tilde{P}}\sbmatrix{P&P_i\\ P&P_i}\\
 &= \mathbb{S}_{P \mathbb{1}}[\mathbb{1}]\fusion_{\bar{P} \mathbb{1}}\sbmatrix{P&P_i\\ P&P_i}\\
 &=\rho_0(P)\rho_0(\bar{P})C_0(P,\bar{P},P_i)
\end{aligned}
\end{equation}
where we used the fact that $ \fusion^{-1}_{\tilde{P} P_i}\sbmatrix{\mathbb{1}&P_i\\ \mathbb{1}&P_i}$ should by definition localize the integral over $\tilde{P}$ on the identity contribution\footnote{Note this is a highly nontrivial statement to prove starting from the analytic expression of the fusion kernel (\ref{eq:explicitFusion}) after taking the limit where two of the external operators are the identity operator. Nevertheless we expect such statement to be true from the very definition of the two-point functions on the sphere. Rigorously what one should show for $\fusion_{P_sP_t}\sbmatrix{ P_2 & P_1 \\ P_3 & P_4 }$ is that in the continuous limit where two of the external operator dimensions go to the identity, say $P_2=P_3\rightarrow i\frac{Q}{2}$ with $P_1=P_4\equiv P$ fixed, then the limit $P_t\rightarrow P$ should yield
\begin{equation}
\begin{aligned}
\lim_{P_t\rightarrow P}\fusion_{P_sP_t}\sbmatrix{  \frac{iQ}{2} & P \\ \frac{iQ}{2}  & P }=\delta\left(P_s-i\frac{b+b^{-1}}{2}\right)-\delta\left(P_s-i\frac{b^{-1}-b}{2}\right)+(P_s\leftrightarrow-P_s)\equiv \rho_{\mathbb{1}}(P_s). 
\end{aligned}
\end{equation}
The expression on the RHS captures properly the fact that the identity module has an additional negative contribution that subtracts the null state at level one.
 }. As in the previous sections, we can write (\ref{HLdistrrel}) schematically with the rest of the contributions in the bagel channel as
 \begin{equation}\label{sumcyl2pt}
\begin{aligned}
\rho^{(\partial \text{necklace})}(P,\bar{P};P_i) =\left( \mathfrak{g}^{(ss)}_{ii}g^{-1}_s\right)\mathbb{K}^{\text{(cyl--2pt)}}_{P\bar{P};\mathbb{1}P_i}[P_i]+\sum_{j,k}\left(C^{(sss)}_{i jk}C^{(sss)kj}_{i}\right)\mathbb{K}^{\text{(cyl--2pt)}}_{P\bar{P};P_jP_k}[P_i]
\end{aligned}
\end{equation}
Notice that there is a nontrivial prefactor multiplying the kernel in the first term, since
 \begin{equation}
 C^{(sss)}_{i \mathbb{1}i}C^{(sss)i\mathbb{1}}_{i}=\underbrace{C^{(sss)i}_{i\mathbb{1}}}_{=1}\mathfrak{g}^{(ss)}_{ii}\underbrace{C^{(sss)i}_{i\mathbb{1}}}_{=1}\left(\mathfrak{g}^{(ss)}_{\mathbb{1}\mathbb{1}}\right)^{-1}=\mathfrak{g}^{(ss)}_{ii}g^{-1}_s. 
\end{equation}
We now want to study the sum (\ref{sumcyl2pt}) in the large $P$ and/or $\bar{P}$ limit when we keep $P_i$ fixed. This will give us the heavy bulk-light boundary limit of the structure constants squared in the spectral density $\rho^{(\partial \text{necklace})}$ which is what we are after. For that purpose we need first to assume the existence of a gap in the boundary spectrum, and second, to ensure that the the first contribution in (\ref{sumcyl2pt}) is dominant in this limit compared to the contributions coming from other boundary operators. Due to the complicated form of the crossing kernel (\ref{HLdistrrel}) however we were not able to establish such a suppression in the limit $P,\bar{P}\rightarrow\infty$. If we assume for a moment that this is indeed the case, the boundary necklace spectral density would be approximated by  

 \begin{equation}\label{asymptcyl2ptt}
\begin{aligned}
\rho^{(\partial \text{necklace})}(P,\bar{P};P_i)&\sim\left( \mathfrak{g}^{(ss)}_{ii}g^{-1}_s\right)\mathbb{K}^{\text{(cyl--2pt)}}_{P\bar{P};\mathbb{1}P_i}[P_i]\\
&=\left( \mathfrak{g}^{(ss)}_{ii}g^{-1}_s\right)\rho_0(P)\rho_0(\bar{P})C_0(P,\bar{P},P_i)
,~~~~~P,\bar{P}\to \infty.
\end{aligned}
\end{equation}
We can actually eliminate the factor of the operator metric $\mathfrak{g}^{(ss)}_{ii}$ in (\ref{asymptcyl2ptt}) by rewriting the definition of the spectral density on the necklace channel as
 \begin{equation}\label{}
\begin{aligned}
\rho^{(\partial \text{necklace})}(P,\bar{P};P_i)&:=\sum_{\op_\alpha\in\mathcal{H}_{\text{closed}} } C^{(s)}_{\alpha i}C^{(s)}_{\alpha i}\left[\delta(P-P_\alpha)+\delta(P+P_\alpha)\right]\\
=\mathfrak{g}&^{(ss)}_{ii}\sum_{\op_\alpha\in\mathcal{H}_{\text{closed}} } C^{(s)}_{\alpha i}C^{(s)i}_{\alpha }\left[\delta(P-P_\alpha)+\delta(P+P_\alpha)\right].
\end{aligned}
\end{equation}
Therefore, if we divide by the bulk Cardy formula $\rho_0(P)\rho_0(\bar{P})$ for the heavy bulk operator propagating in the boundary necklace channel, we arrive at a universal asymptotic result for the microcanonical average of the square of bulk-to-boundary structure constants $\left(C^{(s)i}_{\alpha }\right)^2\equiv C^{(s)i}_{\alpha }C^{(s)}_{\alpha i}$ in the heavy bulk-light boundary limit:
 \begin{equation}\label{HLfinal}
\begin{aligned}
\overline{\left|C^{(s)i}_{\alpha}\right|^2} \sim e^{-\mathfrak{s}_s/2}  \ C_0\left(P_{\alpha},\bar{P}_\alpha,P_i\right),~~~~~P_\alpha,\bar{P}_{\alpha}\to \infty.
\end{aligned}
\end{equation}Remarkably, the result is described again by the same universal factor of $C_0$ and the same exponential factor in the boundary entropy with the other two distinct asymptotic regimes of light bulk-heavy boundary and heavy bulk-heavy boundary that we considered in the previous sections. We should however bare in mind that this last result is on less rigorous footing compared to the previous sections, since we did not manage to establish properly any suppression from contributions of other nontrivial boundary operators in this asymptotic limit, mainly due to the complicated formula of the crossing kernel. It is definitely a worthy goal to study further this construction and establish the validity of the result (\ref{HLfinal}) in a more rigorous way.

\section{Boundary OPE asymptotics}\label{sec:basymptotics}

We will now discuss crossing equations of three particular correlation functions on Riemann surfaces with boundaries, namely: i) the the four-point function of boundary operators on the disk, ii) the two-point function on the cylinder with boundary operators inserted on a single boundary, and iii) the partition function on the sphere with three holes. This study will lead us to universal asymptotic formulae for the boundary structure constants $C_{ij}^{(abc)k}$ for generic boundary conditions $a,b,c$ when one or more of the boundary operators are taken to be heavy compared to the central charge or the corresponding boundary entropies. We will again find a single universal formula governing the three distinct heavy regimes.
\subsection{Boundary four-point function on the disk: heavy-light-light}\label{sec:basymptoticsHLL}
We first examine the case of four-point functions on the disk. The most general configuration involves the array of four boundary operators with mutually adjusted boundary labels and arbitrary scaling dimensions $h_i$, $i=1,\cdots4$. Boundary Virasoro symmetry fixes the form of the four-point function up to a function of the boundary cross-ratio\cite{Lewellen:1991tb}:
\begin{equation}\label{bdy4ptgeneral}
\begin{aligned}
G_{0,0;1,4}(\eta):=\langle \Psi_1^{ab}(x_1) \Psi_2^{bc}(x_2) \Psi_3^{cd}(x_3) \Psi_4^{da}(x_4)\rangle=\prod_{i<j}\left(x_i-x_j\right)^{r-h_i-h_j} G(\eta),
\end{aligned}
\end{equation}
with $x_1<x_2<x_3<x_4$, $r\equiv\frac{1}{3}\sum_{i=1}^4h_i $ and the cross-ratio is $\eta=\frac{(x_1-x_2)(x_3-x_4)}{(x_1-x_3)(x_2-x_4)}$. 

The function $G(\eta)$ can be computed in two equivalent ways: one can either take the boundary OPE limit between 1,2 and 3,4 with $\eta\rightarrow0$, or the boundary OPE limit between 4,1 and 2,3 with $\eta\rightarrow1$. We then get the following expansions
\begin{equation}\label{G1}
\begin{aligned}
G(\eta)&=\sum_{\Psi_i\in \mathcal{H}^{a,c}_{\text{open}}} C^{(abc)i}_{12}C^{(cda)}_{34i}  \ \mathcal{F}\sbmatrix{P_2 & P_1 \\ P_3 & P_4}(P_i|\eta)\\
&\equiv\int \frac{dP}{2} \ \rho^{(a,c)}_{12,34}(P) \ \mathcal{F}\sbmatrix{P_2 & P_1 \\ P_3 & P_4}(P|\eta)\\
G(1-\eta)&=\sum_{\Psi_i\in \mathcal{H}^{d,b}_{\text{open}}} C^{(dab)i}_{41}C^{(bcd)}_{23i}  \ \mathcal{F}\sbmatrix{P_2 & P_3 \\ P_1 & P_4}(P_i|1-\eta)\\
&\equiv\int \frac{dP}{2} \ \tilde{\rho}^{(d,b)}_{41,23}(P) \ \mathcal{F}\sbmatrix{P_2 & P_3 \\ P_1 & P_4}(P|1-\eta),
\end{aligned}
\end{equation}
where $\mathcal{F}\sbmatrix{P_2 & P_1 \\ P_3 & P_4}(P|z)$, $\mathcal{F}\sbmatrix{P_2 & P_3 \\ P_1 & P_4}(P|1-z)$ are the \textit{holomorphic} $S$-channel and $T$-channel sphere four-point Virasoro blocks, and we defined the distributions:
\begin{equation}\label{r11}
\begin{aligned}
\rho^{(a,c)}_{12,34}(P)&:=\sum_{\Psi_k\in \mathcal{H}^{a,c}_{\text{open}}}C^{(abc)k}_{12}C^{(cda)}_{34k}\left[\delta(P-P_k)+\delta(P+P_k)\right]\\
\tilde{\rho}^{(d,b)}_{41,23}(P)&:=\sum_{\Psi_k\in \mathcal{H}^{d,b}_{\text{open}}}C^{(dab)k}_{41}C^{(bcd)}_{23k}\left[\delta(P-P_k)+\delta(P+P_k)\right].
\end{aligned}
\end{equation}
We refer to these two expansions of $G(\eta)$ as the \textit{boundary $S$} and $T$ channels respectively.

\begin{figure}
\centering

 \includegraphics[width=.14\textwidth]{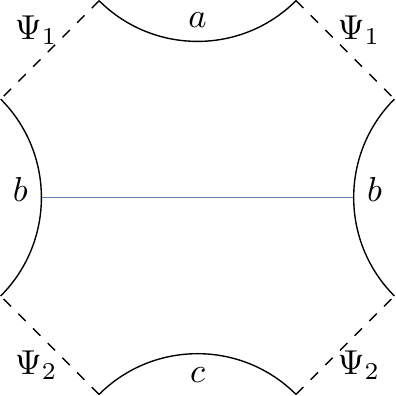} \   \raisebox{.055\textwidth}{\scalebox{1.5}{$=\int \frac{dP}{2}\fusion_{PP'}\sbmatrix{P_2 & P_1 \\ P_2 & P_1}$}} \includegraphics[width=.14\textwidth]{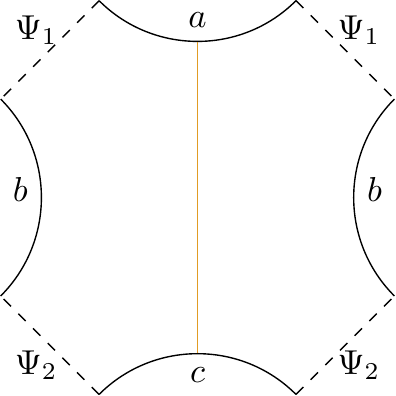}

	\caption{The boundary T-channel (left) and boundary S-channel (right) decomposition of a boundary four-point function in the case where two (out of the total of four) boundary conditions are identical and pairwise-identical boundary operators. The two channels are related with a \textit{holomorphic} copy of the fusion kernel.}\label{fig:bbdy4ptcrossing}
\end{figure}

The associativity of the boundary OPE implies that the two decompositions in (\ref{G1}) are equal. Using the fusion transformation of the sphere four-point blocks (\ref{eq:fusionTransformation}), we can express the crossing symmetry constraint as a transform relating the spectral densities (\ref{r11}): 
\begin{equation}\label{bdycrosssymm}
\begin{aligned}
\rho^{(a,c)}_{12,34}(P)=\int \frac{dP'}{2}\tilde{\rho}^{(d,b)}_{41,23}(P') \ \fusion_{PP'}\sbmatrix{P_2 & P_1 \\ P_3 & P_4}
\end{aligned}
\end{equation}
This is a highly non-trivial constraint on the boundary spectrum and OPE coefficients for a consistent BCFT$_2$. This relation has been discussed before in the literature mostly in the cases of Minimal Models or RCFTs where $\fusion_{PP'}$ is a known finite dimensional matrix relating a finite number of conformal blocks. The new ingredient here is the fact that we can extend this to the case of generic irrational BCFTs, where $\fusion_{PP'}$ is the Ponsot-Teschner crossing kernel \cite{Ponsot:1999uf,Ponsot:2000mt}. 
\par It will be intructive to compare the above expression of boundary crossing symmetry with the usual crossing symmetry of bulk four-point functions on the sphere. First, we note that (\ref{bdycrosssymm}) relates spectra on the Hilbert space $\mathcal{H}^{d,b}_{\text{open}}$ with spectra on $\mathcal{H}^{a,c}_{\text{open}}$. These two Hilbert spaces contain different states and hence equation (\ref{bdycrosssymm}) implies that the fusion matrix provides a non-trivial link between them. In our notation, this is reflected on the fact that the arguments $P,P'$ of the fusion kernel are Liouville momenta with support on $\mathcal{H}^{a,c}_{\text{open}}$ and $\mathcal{H}^{d,b}_{\text{open}}$ respectively. Therefore, equation (\ref{bdycrosssymm}) contains some highly non-trivial information about the structure of boundary conditions in an irrational CFT given the fact that the Ponsot-Teschner crossing kernel has an explicit analytic structure in $P,P'$. This point definitely deserves some further exploration. In the bulk case, the analogous relation between bulk OPE in the S and T channels (see e.g. \cite{Collier:2018exn,Collier:2019weq}) relates again spectra with different supports (the S-channel OPE density is in general different from the T-channel one) but now these are states on the closed Hilbert space on a circle. Another comment worth emphasizing here is that in (\ref{bdycrosssymm}) a \textit{single} copy of the crossing kernel (i.e. its holomorphic half) relates \textit{quadratic} expressions in the (boundary) OPE coefficients on the two channels, whereas in the case of sphere four-point crossing we have the product of holomorphic and  anti-holomorphic parts of the fusion kernel that relates quadratic expressions in the (bulk) OPE coefficients.
\par We will now be interested in the fusion transform of the boundary vacuum in (\ref{bdycrosssymm}). This can only appear in the case where two of the boundary conditions, either $a,c$ or $d,b$, are identical and in addition the operators are identical in pairs, namely $\Psi_1=\Psi_2, \Psi_3=\Psi_4$ or $\Psi_1=\Psi_4, \Psi_2=\Psi_3$ for the cases $a=d$ or $d=b$ respectively. We will consider the case $d=b$ in what follows, see Fig.\ref{fig:bbdy4ptcrossing}. 

\par Crucially, with this particular configuration we get a squared OPE density in the boundary S-channel as follows:\begin{equation}\label{sdesnity}
 \begin{aligned}
\rho&^{(a,c)}_{12,21}(P):=\sum_{\Psi_k\in \mathcal{H}^{a,c}_{\text{open}}}C^{(abc)k}_{12}C^{(cba)}_{21k}\left[\delta(P-P_k)+\delta(P+P_k)\right]\\
&=\mathfrak{g}^{(ab)}_{11}\mathfrak{g}^{(bc)}_{22}\sum_{\Psi_k\in \mathcal{H}^{a,c}_{\text{open}}}C^{(abc)k}_{12}C^{(cba)21}_{k}\left[\delta(P-P_k)+\delta(P+P_k)\right]
\end{aligned}
\end{equation}
where in the second line we pulled out some appropriate factors of the boundary operator metric in order to have fully contracted structure constants inside the sum. As we will see later this will cancel some analogous factors in the T-channel density.\\
\\
In the T-channel the OPE data of the vacuum simplifies to
\begin{equation}\label{}
 \begin{aligned}
\tilde{\rho}^{(b,b)}_{11,22}(\mathbb{1})=C^{(bab)\mathbb{1}}_{11}C^{(bcb)}_{22\mathbb{1}}=g^{-1}_b\mathfrak{g}^{(ab)}_{11}\mathfrak{g}^{(bc)}_{22}
\end{aligned}
\end{equation}
Repeating our arguments from earlier, and assuming a gap on the spectrum in the boundary T-channel, we obtain an asymptotic result for the boundary S-channel spectral density via (\ref{bdycrosssymm}) that reads
 \begin{equation}\label{hll}
\begin{aligned}
\rho^{(a,c)}_{12,21}(P)\sim \left(g^{-1}_b\mathfrak{g}^{(ab)}_{11}\mathfrak{g}^{(bc)}_{22}\right)  \fusion_{P\mathbb{1}}\sbmatrix{P_2 & P_1 \\ P_2 & P_1} \ , \ \ \ P\rightarrow\infty.
\end{aligned}
\end{equation}
The non-vacuum kernels with boundary T-channel dimension $h_t>0$ will be exponentially suppressed in the limit $P\rightarrow\infty$ due to the key result we discussed in (\ref{eq:fusionSuppression}), and we repeat here for convenience:
\begin{equation}\label{eq:fusionSuppression}
\frac{\fusion_{P P_t}}{\fusion_{P\id}} \approx \begin{cases}
 	e^{-2\pi\alpha_t P} & \alpha_t = \tfrac{Q}{2}+iP_t \in (0,\tfrac{Q}{2}) \\
 	e^{-\pi Q P}\cos(2\pi P_t P) &  P_t \in \RR
 \end{cases}
 \quad\text{as }P\to\infty
\end{equation}
\par We can now translate our asymptotic expression (\ref{hll}) to a microcanonical average of OPE coefficients by dividing with the asymptotic Cardy formula (\ref{eq:BdyCardy}) for states in the $\mathcal{H}^{a,c}_{\text{open}}$ Hilbert space, namely  $e^{\frac{1}{2}\left(\mathfrak{s}_a+\mathfrak{s}_c\right)} \ \rho_0(P)$. Writing the identity fusion kernel in the form \eqref{eq:idFusion} of the universal density $\rho_0(P)$ times $C_0(P_1,P_2,P)$, and cancelling the common factors in front of (\ref{sdesnity}) and (\ref{hll}), we find that the microcanonical average of the OPE coefficients is given by:
\begin{equation}\label{averegefinalhll}
	\overline{\left|C_{12}^{(abc)P}\right|^2} \sim e^{-\frac{1}{2}\left(s_a+s_b+s_c\right)}C_0(P_1,P_2,P), \ \ \ ~P \to \infty.
\end{equation}
We emphasize that by the ``square'' we really mean $\left|C_{12}^{(abc)P}\right|^2\equiv C^{(abc)P}_{12}C^{(cba)21}_{P}$, i.e. the fully contracted quantity in the operator indices. This result is valid for any two fixed operators $\Psi_1,\Psi_2$, averaging over operators $\Psi_P$ in the large boundary dimension limit. Interestingly, the result is nice and symmetric under the exchange of the three boundary entropies as well as in the Liouville momenta $P_i$, $i=1,2,3$ (due to the symmetry of $C_0$).

If we explicitly choose the normalization (\ref{normalization}) for the boundary operators, we can write an asymptotic formula for the boundary three-point functions squared (i.e. the structure constants with all indices \textit{lowered}) which in this case gives:
\begin{equation}
\overline{\left|C_{12P}^{(abc)}\right|^2} \sim C_0(P_1,P_2,P), \ \ \ ~P \to \infty.
\end{equation}
where $\left|C_{12P}^{(abc)}\right|^2\equiv C^{(abc)}_{12P}C^{(cba)}_{P21}$. We see that the dependence on the boundary entropy completely drops out in this normalization, and remarkably it seems to be equal asymptotically with the analogous formulas for the bulk-to-boundary two-point functions (i.e. the bulk-to-boundary structure constants with, again, all indices lowered) that we obtained in the previous sections.

\subsection{Cylinder two-point function: heavy-heavy-light}\label{sec:basymptoticsHHL}
We next turn to the case of the cylinder two-point function where we now place two boundary operators on a single boundary of the cylinder. The general amplitude takes the form
\begin{equation}\label{def2pt}
\begin{aligned}
G_{0,0;2,2}(q):=\langle\Psi_1^{ab}\Psi_2^{ba}\rangle_{\text{cyl}^{(ac)}}
\end{aligned}
\end{equation}
where $q$ is a collective label for the moduli of the surface which we won't need to make explicit for our purposes\footnote{As we saw in section \ref{sec:BbasymptoticsHL} the moduli $q$ consists of two real parameters: the length $\beta$ of the cylinder, and the relative angle $\theta$ between the insertions of the two external operators. }. We can think of this correlation function as starting from the cylinder with boundary conditions $a,c$ at its two ends, and insert two boundary changing operators $\Psi^{ab}_1,\Psi^{ba}_2$ on the boundary $a$. The boundary condition $b$ is in general different from $a$. If $b=a$, we could choose $\Psi_{1,2}=\mathbb{1}$ and then the amplitude reduces to the usual cylinder partition function with boundary conditions $a,c$.
\par The amplitude (\ref{def2pt}) has in general a nontrivial expansion in terms of conformal blocks. Nonetheless we will only be interested in the crossing moves that relate different dissections of the amplitude. In particular we will next focus on the case where $\Psi_1=\Psi_2\equiv \Psi_0$.

\begin{figure}
		\qquad\qquad\, \includegraphics[width=.2\textwidth]{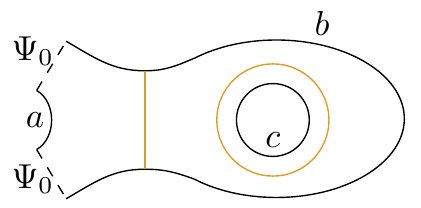}  \raisebox{.034\textwidth}{\scalebox{1.5}{$=\int \frac{dP_1}{2}\modS_{P_1P_1'}[P'_2]$}} \includegraphics[width=.2\textwidth]{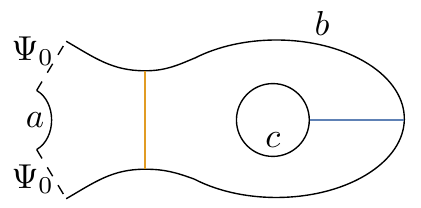} 
		\flushright\raisebox{.034\textwidth}{\scalebox{1.5}{$=\int \frac{dP_1}{2}\frac{dP_2}{2}\modS_{P_1P_1'}[P'_2] \fusion_{P_2P_2'}\sbmatrix{P_1&P_0\\P_1&P_0}$}}
		\includegraphics[width=.2\textwidth]{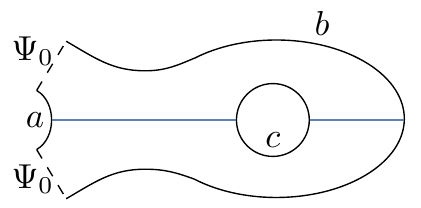}
	\caption{The sequence of moves expressing the boundary OPE channel of the cylinder two-point block in terms of the boundary necklace channel block: a \textit{holomorphic} modular kernel, followed by a \textit{holomorphic} fusion kernel. \label{fig:Cyl2pt}}
\end{figure}

\par There are two obvious and qualitatively distinct ways we can decompose such a correlation function into conformal blocks, see Fig.\ref{fig:Cyl2pt}. We call ``boundary OPE'' channel the decomposition where we first take the boundary OPE between the two identical external operators $\Psi_0$ and insert a complete set of states in $\mathcal{H}^{b,b}_{\text{open}}$. We then study a one-point function on the cylinder (i.e. a single insertion of a boundary operator on one side of the cylinder) and quantize along a circle in the closed Hilbert space (top left of Fig.\ref{fig:Cyl2pt}). On the other hand, we call ``boundary necklace'' channel the decomposition where we just quantize along $\mathcal{H}^{a,c}_{\text{open}}$ and $\mathcal{H}^{c,b}_{\text{open}}$, namely along the blue lines going from one boundary of the cylinder to the other, as in bottom right picture of Fig.\ref{fig:Cyl2pt}.
\par We therefore have the two equivalent expansions:
\begin{equation}
\begin{aligned}
G^{(\text{$\partial$N})}_{0,0;2,2}(q)&=\sum_{\Psi_1\in \mathcal{H}^{b,c}_{\text{open}}}\sum_{\Psi_2\in \mathcal{H}^{a,c}_{\text{open}}}C^{(abc)}_{012}C^{(cba)21}_{0}\mathcal{F}^{(\text{$\partial$N})}[P_0]\left(P_1,P_2|q\right)\\
&\equiv\int \frac{dP_1}{2}\frac{dP_2}{2}\ \rho_{(\text{$\partial$N})}(P_1,P_2)\ \mathcal{F}^{(\text{$\partial$N})}[P_0]\left(P_1,P_2|q\right)\\
G^{(\text{$\partial$OPE})}_{0,0;2,2}(\tilde{q})&=\sum_{\mathcal{O}'_1\in \mathcal{H}^{sc.}_{\text{closed}}}\sum_{\Psi_2'\in \mathcal{H}^{b,b}_{\text{open}} }C^{(bab)2'}_{00} \ C^{(b)}_{1'2'}C^{(c)}_{1'\mathbb{1}} \ \mathcal{F}^{(\text{$\partial$OPE})}[P_0]\left(P'_1,P'_2|\tilde{q}\right)\\
&\equiv\int \frac{dP'_1}{2}\frac{dP'_2}{2} \ \rho_{(\text{$\partial$OPE})}(P'_1,P'_2)\ \mathcal{F}^{(\text{$\partial$OPE})}[P_0]\left(P'_1,P'_2|\tilde{q}\right).
\end{aligned}
\end{equation}
The crossing kernel that relates the conformal blocks -- and hence the spectral densities -- in the two aforementioned decompositions is easy to figure out and is depicted in Fig.\ref{fig:Cyl2pt}: we first use a crossing move for the cylinder one-point function that we described in section \ref{subsec:boundaryCardy} and involves a holomorphic copy of the modular S, and then we use a crossing move for the resulting disk four-point function (with two of its ``external'' operators identified) which involves a holomorphic fusion move. The total crossing kernel that relates the corresponding spectral densities reads
\begin{equation}\label{HHLdistrrel}
\begin{aligned}
\mathbb{K}^{\text{(cyl$^{'}$--2pt)}}_{P_{1}, P_2;P'_1,P'_2}[P_0]:=&\modS_{P_1P_1'}[P'_2] \fusion_{P_2P_2'}\sbmatrix{P_1&P_0\\P_1&P_0}\\
\rho_{(\text{$\partial$N})}(P_1,P_2)=\int \frac{dP'_1}{2}&\frac{dP'_2}{2} \ \mathbb{K}^{\text{(cyl$^{'}$--2pt)}}_{P_{1}, P_2;P'_1,P'_2}[P_0] \ \rho_{(\text{$\partial$OPE})}(P'_1,P'_2)
\end{aligned}
\end{equation}
The boundary necklace channel density describes a positive definite distribution in the boundary OPE data, which we can write as
\begin{equation}\label{bdyneckrew}
\begin{aligned}
\rho_{(\text{$\partial$N})}(P_1&,P_2):=\sum_{\Psi_i\in \mathcal{H}^{b,c}_{\text{open}}}\sum_{\Psi_j\in \mathcal{H}^{a,c}_{\text{open}}}C^{(abc)}_{0ij}C^{(cba)ji}_{0}\left[\delta(P_1-P_i)+\delta(P_1+P_i)\right]\left[\delta(P_2-P_j)+\delta(P_2+P_j)\right]\\
&=\mathfrak{g}^{(ab)}_{00}\sum_{\Psi_i\in \mathcal{H}^{b,c}_{\text{open}}}\sum_{\Psi_j\in \mathcal{H}^{a,c}_{\text{open}}}C^{(abc)0}_{ij}C^{(cba)ji}_{0}\left[\delta(P_1-P_i)+\delta(P_1+P_i)\right]\left[\delta(P_2-P_j)+\delta(P_2+P_j)\right].
\end{aligned}
\end{equation}
On the other hand, the vacuum contribution in the boundary OPE channel yields
\begin{equation}
\begin{aligned}
\rho_{(\text{$\partial$OPE})}(\mathbb{1},\mathbb{1})=C^{(bab)\mathbb{1}}_{00} \ C^{(b)}_{\mathbb{1}\mathbb{1}}C^{(c)}_{\mathbb{1}\mathbb{1}}=\mathfrak{g}^{(ab)}_{00}g_c.
\end{aligned}
\end{equation}
Assuming a gap in the boundary spectrum above the identity as well as a gap in the bulk scalar spectrum above the (bulk) identity, we can use the result (\ref{eq:torusTwoPointId}) from earlier to show that the vacuum contribution in the boundary OPE channel in (\ref{HHLdistrrel}) will be dominant in the limit $P_1,P_2\rightarrow\infty$ while we keep $P_0$ fixed. Therefore, we obtain the following asymptotic result for the boundary necklace spectral density:
\begin{equation}\label{hhlasymptrel}
\begin{aligned}
\rho_{(\text{$\partial$N})}(P_1,P_2)&\sim \left(\mathfrak{g}^{(ab)}_{00}g_c\right) \modS_{P_1\mathbb{1}}[\mathbb{1}] \fusion_{P_2\mathbb{1}}\sbmatrix{P_1&P_0\\P_1&P_0}\\
&=\left(\mathfrak{g}^{(ab)}_{00}g_c\right)\rho_0(P_1)\rho_0(P_2)C_0(P_1,P_2,P_0), \ \ \ \ \ \ \ \ \ P_1,P_2\rightarrow\infty.
\end{aligned}
\end{equation}
Stripping off the density of states of the heavy operators after dividing with the corresponding boundary Cardy formulas, namely $e^{\frac{1}{2}(\mathfrak{s}_b+\mathfrak{s}_c)}\rho_0({P_1})$ and $e^{\frac{1}{2}(\mathfrak{s}_a+\mathfrak{s}_c)}\rho_0({P_2})$, and cancelling the common factor in front of (\ref{bdyneckrew}) and (\ref{hhlasymptrel}), we find that the heavy-heavy-light limit of the averaged boundary OPE coefficients is given by
\begin{equation}
\begin{aligned}
\overline{\left|C_{01}^{(abc)2}\right|^2} \sim e^{-\frac{1}{2}\left(s_a+s_b+s_c\right)}C_0(P_1,P_2,P_0), \ \ \ ~P_1,P_2 \to \infty.
\end{aligned}
\end{equation}
where, again, by the square we mean $\left|C_{01}^{(abc)2}\right|^2\equiv C^{(abc)0}_{12}C^{(cba)21}_{0} $. This is the same symmetric formula in both the boundary entropies and the Liouville momenta that we obtained also in the previous case of the heavy-light-light limit. Choosing the special normalisation (\ref{normalization}) we can obtain a result for the boundary three-point function coefficients which again takes the form
\begin{equation}
\begin{aligned}
\overline{\left|C_{012}^{(abc)}\right|^2} \sim C_0(P_1,P_2,P_0), \ \ \ ~P_1,P_2 \to \infty.
\end{aligned}
\end{equation}

\subsection{Sphere with three holes: heavy-heavy-heavy}\label{sec:basymptoticsHHH}

\begin{figure}
	\centering
	\includegraphics[width=.25\textwidth]{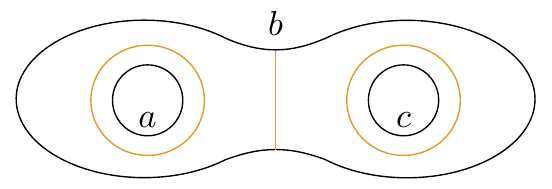} \raisebox{.03\textwidth}{\scalebox{1.5}{$=\int \frac{dP_1}{2}\frac{dP_2}{2}\modS_{P_1P_1'}[P'_3]\modS_{P_2P_2'}[P'_3]$}}\includegraphics[width=.25\textwidth]{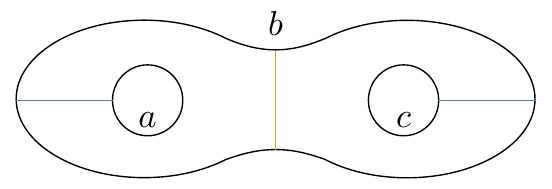} \\
		\flushright\raisebox{.03\textwidth}{\scalebox{1.5}{$=\int \frac{dP_1}{2}\frac{dP_2}{2}\frac{dP_3}{2}\modS_{P_1P_1'}[P'_3]\modS_{P_2P_2'}[P'_3] \fusion_{P_3P_3'}\sbmatrix{P_1&P_2\\P_1&P_2}$}}\includegraphics[width=.25\textwidth]{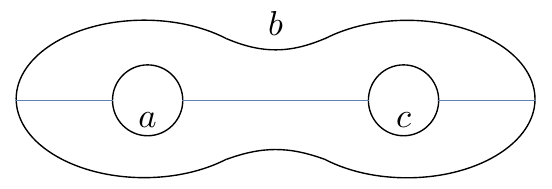}
	\caption{The sequence of moves expressing a boundary dumbbell channel block (top left) in terms of boundary sunset channel blocks (bottom right) for the decomposition of the partition function of the sphere with three holes.\label{fig:sphere3holes}}
\end{figure}

Our final case of study involves the partition function on the sphere with three holes, where we have three distinct boundary conditions $a,b,c$ on each hole. We denote the partition function by $G_{0,0;3,0}(\beta_i)$, where $\beta_i$ denotes collectively the moduli\footnote{The moduli of this surface is in general described by three real parameters $\beta_i\in\mathbb{R}$, $i=1,2,3$.}. It will be convenient for our purposes to picture the sphere with three holes as in Figure \ref{fig:sphere3holes}, that is, as a surface with an outer boundary (labelled by $b$ in our case) and two inner boundaries labelled by $a$ and $c$. As we will see, in this way of thinking of the surface, the various crossing moves that relate different dissections of the amplitude will become easier to visualize.
\par We want to study the relation of the conformal block decomposition between what we call the ``boundary dumbbell'' channel and the ``boundary sunset'' channel, as depicted in Fig.\ref{fig:sphere3holes}. We get the following two equivalent expansions:
\begin{equation}
\begin{aligned}
G^{(\text{$\partial$sunset})}_{0,0;3,0}(\beta_i)&=\sum_{\Psi_1\in \mathcal{H}^{a,b}_{\text{open}}}\sum_{\Psi_2\in \mathcal{H}^{b,c}_{\text{open}}}\sum_{\Psi_3\in \mathcal{H}^{a,c}_{\text{open}}}\left(C^{(abc)}_{123}C^{(cba)321}\right)\mathcal{F}^{(\text{$\partial$sunset})}\left(P_1,P_2,P_3|\beta_i\right)\\
&\equiv\int \frac{dP_1}{2}\frac{dP_2}{2}\frac{dP_3}{2}\ \rho_{(\text{$\partial$sunset})}(P_1,P_2,P_3)\ \mathcal{F}^{(\text{$\partial$sunset})}\left(P_1,P_2,P_3|\beta_i\right)\\
G^{(\text{$\partial$dumbbell})}_{0,0;3,0}(\tilde{\beta}_i)&=\sum_{\mathcal{O}'_1\in\mathcal{H}^{sc.}_{\text{closed}}}\sum_{\mathcal{O}'_2\in \mathcal{H}^{sc.}_{\text{closed}}}\sum_{\Psi_3'\in \mathcal{H}^{b,b}_{\text{open}} }\left(C^{(a)}_{1'\mathbb{1}}C^{(b)3'}_{1'}\right)\left(C^{(b)}_{2'3'}C^{(c)}_{2'\mathbb{1}}\right) \ \mathcal{F}^{(\text{$\partial$dumbbell})}\left(P'_1,P'_2,P'_3|\tilde{\beta}_i\right)\\
&\equiv\int \frac{dP'_1}{2}\frac{dP'_2}{2}\frac{dP'_3}{2} \ \rho_{(\text{$\partial$dumbbell})}(P'_1,P'_2,P'_3)\ \mathcal{F}^{(\text{$\partial$dumbbell})}\left(P'_1,P'_2,P'_3|\tilde{\beta}_i\right).
\end{aligned}
\end{equation}
The crossing kernel that relates the relevant conformal blocks is depicted in Fig.\ref{fig:sphere3holes}: it consists of two modular S moves for the two cylinder one-point functions on the left and the right of the surface, accompanied by a fusion move for the resulting disk four-point function (with its ``external'' operators identified in pairs). We therefore get
\begin{equation}\label{HHHdistrrel}
\begin{aligned}
\mathbb{K}^{\text{(sphere--3holes)}}_{P_{1}, P_2,P_3;P'_1,P'_2,P'_3}:=&\modS_{P_1P_1'}[P'_3]\modS_{P_2P_2'}[P'_3] \fusion_{P_3P_3'}\sbmatrix{P_1&P_2\\P_1&P_2}\\
\rho_{(\text{$\partial$sunset})}(P_1,P_2,P_3)=\int \frac{dP'_1}{2}&\frac{dP'_2}{2}\frac{dP'_3}{2} \ \mathbb{K}^{\text{(sphere--3holes)}}_{P_{1}, P_2,P_3;P'_1,P'_2,P'_3} \ \rho_{(\text{$\partial$dumbbell})}(P'_1,P'_2,P'_3).
\end{aligned}
\end{equation}
The propagation of the identity in all the lines of the boundary dumbbell channel gives a contribution of the form:
\begin{equation}\label{}
\rho_{(\text{$\partial$dumbbell})}(\mathbb{1},\mathbb{1},\mathbb{1})=\left(C^{(a)}_{\mathbb{1}\mathbb{1}}C^{(b)\mathbb{1}}_{\mathbb{1}}\right)\left(C^{(b)}_{\mathbb{1}\mathbb{1}}C^{(c)}_{\mathbb{1}\mathbb{1}}\right)=g_ag_bg_c
\end{equation}
Corrections to this contribution due to the exchange of non-vacuum primaries in the boundary dumbbell channel are exponentially suppressed by the following ratio which we can calculate explicitly:
\begin{equation}
	{\mathbb{K}^{\text{(sphere--3holes)}}_{P_1 P_2 P_3;P'_1 P'_2 P'_3}\over\mathbb{K}^{\text{(sphere--3holes)}}_{P_1 P_2 P_3;\id\id\id}} \approx e^{-2\pi(\alpha_1'P_1+\alpha_3'P_3)}
\end{equation}
in the limit where the ratios or differences between the $P_i$ are held fixed (see Appendix \ref{sec:asymptcrosskernnls} for more details). We therefore conclude that the boundary sunset channel density is well-approximated by the exchange of the vacuum in the boundary dumbbell channel when the internal boundary operators become heavy:
\begin{equation}\label{HHH}
\begin{aligned}
\rho_{(\text{$\partial$sunset})}(P_1,P_2,P_3)&\sim \left(g_ag_bg_c\right)\modS_{P_1\mathbb{1}}[\mathbb{1}]\modS_{P_2\mathbb{1}}[\mathbb{1}] \fusion_{P_3\mathbb{1}}\sbmatrix{P_1&P_2\\P_1&P_2}\\
&=\left(g_ag_bg_c\right)\rho_0(P_1)\rho_0(P_2)\rho_0(P_3)C_0(P_1,P_2,P_3), \ \ \ \ \ P_1,P_2,P_3\rightarrow\infty.
\end{aligned}
\end{equation}
To obtain the microcanonical average of the heavy-heavy-heavy boundary structure constants we divide with the asymptotic Cardy density of the boundary states $P_1,P_2,P_3$, namely with $e^{\frac{1}{2}(\mathfrak{s}_a+\mathfrak{s}_b)}\rho_0({P_1})$, $e^{\frac{1}{2}(\mathfrak{s}_b+\mathfrak{s}_c)}\rho_0({P_2})$ and $e^{\frac{1}{2}(\mathfrak{s}_a+\mathfrak{s}_c)}\rho_0({P_3})$ respectively, to obtain:
\begin{equation}
\begin{aligned}
\overline{\left|C_{12}^{(abc)3}\right|^2} \sim e^{-\frac{1}{2}\left(\mathfrak{s}_a+\mathfrak{s}_b+\mathfrak{s}_c\right)}C_0(P_1,P_2,P_0), \ \ \ ~P_1,P_2,P_3 \to \infty.
\end{aligned}
\end{equation}
Once again, we obtain the same universal formula for the fully contracted boundary structure constants as we showed in the heavy-light-light and heavy-heavy-light limits. Furthermore, in the normalization (\ref{normalization}) the asymptotic result for the square of the boundary three-point functions -- i.e. the structure constants with all indices lowered -- once again gets the universal form 
\begin{equation}
\begin{aligned}
\overline{\left|C_{123}^{(abc)}\right|^2} \sim C_0(P_1,P_2,P_0), \ \ \ ~P_1,P_2,P_3 \to \infty.
\end{aligned}
\end{equation}
It is quite a remarkable fact that there is \textit{a} normalisation of boundary operators, namely (\ref{normalization}), where all the distinct heavy limits of \textit{both} types of boundary correlation functions -- i.e. bulk-to-boundary two-point functions and boundary three-point functions -- are captured by the same universal formula which is $C_0$. 

\section{Eigenstate Thermalization Hypothesis in BCFT$_2$}\label{sec:bcftETH}

The Eigenstate Thermalization Hypothesis (ETH) \cite{Srednicki:1994ne,deutsch1991quantum} states that in an isolated quantum system with a sufficiently chaotic Hamiltonian the matrix elements of a ``simple'' operator $\mathcal{O}$ should obey
\begin{equation}\label{ETH}
\begin{aligned}
\langle i|\mathcal{O}|j\rangle\approx f^{\mathcal{O}}(E_i)\delta_{ij}+g^{\mathcal{O}}(E_i,E_j)R_{ij}
\end{aligned}
\end{equation}
for states $i$ and $j$ of fixed energy density in a large volume thermodynamic limit. The functions $f^{\mathcal{O}},g^{\mathcal{O}}$ are smooth functions of the energy related to the microcanonical one and two-point functions, and $R_{ij}$ is a pseudo-random variable of zero mean and unit variance. In particular, if the one- and two-point functions are of order one, then $f^{\mathcal{O}}$ is of order one and $g^{\mathcal{O}}$ of order $e^{-S/2}$ where $S$ is the microcanonical entropy. In a scale-invariant theory, the large volume thermodynamic limit is equivalent to a large energy limit at fixed volume, which is the heavy limit we have been studying in the present work. When $\mathcal{O}$ is a local operator in the CFT, ETH is a nontrivial statement about the statistics of OPE coefficients for the corresponding \textit{primary operators} in the theory (see \cite{Asplund:2015eha,deBoer:2016bov,Lashkari:2016vgj,He:2017vyf,Lashkari:2017hwq,Faulkner:2017hll,Guo:2018pvi,Maloney:2018hdg,Maloney:2018yrz,Dymarsky:2018lhf,Dymarsky:2018iwx,Anous:2019yku,Dymarsky:2019etq,Datta:2019jeo,Besken:2019bsu} for discussions of ETH in the CFT context).
\par In this section, we would like to initiate an analogous investigation in the context of BCFT. We will see that our asymptotic results (at least in the case of one dimensional boundary) indicate a natural extension of the structure of the ETH ansatz when we consider the matrix elements of primary operators in the presence of conformal boundaries. Furthermore, the fact that we are obtaining a single asymptotic formula, not only for distinct asymptotically heavy regimes, but also for \textit{both types} of boundary structure constants (under the particular normalization (\ref{normalization})):
\begin{equation}\label{totalasympt}
\begin{aligned}
\overline{\left|C_{i\bar{i}j}^{(abc)}\right|^2}\sim \overline{\left|C^{(s)}_{ij}\right|^2} \sim C_0(P_i,\bar{P}_i,P_j)
\end{aligned}
\end{equation}
seems to suggest that, at least in two dimensions, the ETH ansatz should be generalized to all distinct heavy regimes for both bulk-to-boundary and boundary OPE coefficients. A similar argument for the bulk case was presented in \cite{Collier:2019weq}, where again one finds a single asymptotic formula for the bulk OPE coefficients in three distinct heavy regimes.

An extension of the ETH ansatz to the case of boundary OPE coefficients (for primary boundary operators) is quite natural to expect since on the boundary we have a distinct Hilbert space $\mathcal{H}^{a,b}_{\text{open}}$ with nontrivial three point functions of operators which live solely there. However, an ETH-like proposal is far less intuitive in the case of bulk-to-boundary OPE coefficients. We will return to this point towards the end of this section. For now we will discuss the case of boundary OPE coefficients and what our asymptotic results suggest for the structure of the ETH ansatz in this case. 

One should start by appropriately adjusting the ansatz (\ref{ETH}) in the boundary setup. We consider the eigenstates of a fixed Hamiltonian, say the energy eigenstates in the open sector Hilbert space $\mathcal{H}^{a,b}_{\text{open}}$ between boundaries $a,b$, and we want to make an ansatz about the matrix elements in that eigenbasis of a ''simple'' operator $\Psi_0$. Therefore we want to study the matrix element:
\begin{equation}\label{ethbdy}
\begin{aligned}
{}_{ab}\langle \Psi_i|\Psi_0|\Psi_j\rangle_{ba}\approx  f_{ab}^{0}(h_i)\delta_{ij}+g_{ab}^{0}(h_i,h_j)R_{ij}
\end{aligned}
\end{equation}
in the large energy limit $h_{i},h_j\rightarrow\infty$. In BCFT we can take the states $\ket{\Psi_j}_{ba}$ to be primary states and similarly $\Psi_0=\Psi_0^{bb}$ to be a primary operator. Then, equation (\ref{ethbdy}) becomes a statement about the boundary three-point function coefficients
$
{}_{ab}\langle \Psi_i|\Psi_0|\Psi_j\rangle_{ba}\equiv C^{(abb)}_{i0j}
$.
In this case our asymptotic formulas for the diagonal heavy-heavy-light boundary OPE in Section \ref{subsec:boundaryCardy} (captured by the cylinder one-point functions), and off-diagonal heavy-heavy-light boundary OPE in Section \ref{sec:basymptoticsHHL} (captured by the cylinder two-point functions) determine the functions $f_{ab}^{0}$ and $g_{ab}^{0}$ as:
\begin{equation}
\begin{aligned}
\overline{C^{(abb)}_{i0i}}=f^{0}_{ab}(h_i), \qquad \overline{\left|C^{(abb)}_{i0j}\right|^2}=(g_{ab}^{0}(h_i,h_j))^2
\end{aligned}
\end{equation}
Therefore our formulas provide a precise formulation of ETH for BCFTs in two dimensions with finite central charge $c$ and finite boundary entropies\footnote{It is important to emphasize that our asymptotic formulas predict the form of the smooth functions $f^0_{ab}$ and $g^0_{ab}$, but say nothing about the statistics of the remainder term $R_{ij}$.  The statement that $R_{ij}$ has zero mean and unit variance, severely constraining the fluctuations of matrix elements, is an important component of ETH but unfortunately one that is invisible using crossing symmetry techniques, as we did in the present work.}. This means that the order of the off-diagonal component $g_{ab}^{0}$ in the ``boundary'' ETH ansatz should naturally depend on the relevant boundary entropies which contribute additional factors to the microcanonical entropy. Indeed, this is captured by our asymptotic formulas! Working in the normalisation (\ref{normalization}), in Sections \ref{subsec:boundaryCardy} and \ref{sec:basymptoticsHHL} we obtained the following results:
\begin{equation}\label{hierarchy}
\begin{aligned}
	\overline{C^{(abb)}_{H0H}} &\sim  e^{\frac{1}{4}(\mathfrak{s}_a+\mathfrak{s}_b)} \left( C^{(a)\mathbb{1}}_{\chi}C^{(b)0}_{\chi}\right){\modS_{P_HP_\chi}[P_0]\over \rho_0(P_H)},~P_H\to\infty\\
	\overline{\left|C^{(abb)}_{H0H'}\right|^2}&\sim C_0(P_H,P_{H'},P_0),~~~~~~~ P_H,P_{H'}\to\infty.
	\end{aligned}
\end{equation}
where $C^{(a)\mathbb{1}}_{\chi}C^{(b)0}_{\chi}$ are order one in the boundary entropy as we explained in \ref{subsec:boundaryCardy}. Therefore we see that the hierarchy between the diagonal and the off-diagonal term is controlled by the boundary entropy in a specific way: the diagonal term is either enhanced or suppressed (depending on the sign of the boundary entropy) compared to the off-diagonal term with an extra factor of $e^{\frac{1}{4}(\mathfrak{s}_a+\mathfrak{s}_b)}$. This result actually makes a lot of sense if we remind ourselves that the boundary entropy determines the subleading piece in the high-temperature limit of the thermal entropy of a 2d CFT on an interval of length $L$ with inverse temperature $\beta$. In general, with different conformal boundary conditions at the two ends of the interval, the entropy is determined by the cylinder partition function $Z_{\text{cyl}^{(ab)}}$ and takes the form
\begin{equation}
\begin{aligned}
S(\beta)&\equiv (1-\beta\partial_{\beta})\log{\left(Z_{\text{cyl}^{(ab)}}(\beta)\right)}\\
&=\frac{\pi c L}{3\beta}+\frac{1}{2}\left(\mathfrak{s}_{a}+\mathfrak{s}_b\right)+\cdots
\end{aligned}
\end{equation}
As we discussed earlier, the ETH ansatz suggests that the smooth function $g^0$ is of order $e^{-S/2}$ (if $f^0$ is order one). Thus we see exactly this $S/2$ appearing in (\ref{hierarchy}), properly captured by our asymptotic results.

We will close this section with some comments on a potential ETH-like proposal for bulk-to-boundary OPE coefficients. Our asymptotic formulas suggested that the bulk-to-boundary two-point functions are captured by similar asymptotics as the boundary three-point function coefficients (\ref{totalasympt}). This leads us to suspect that there might some BCFT setup where we can imagine the boundary spectrum to act as a thermal bath with some effective temperature for the bulk spectrum or vice versa. In fact, the case where an operator in the bulk CFT acts as a bath for a ``simple'' boundary operator is more intuitive, since from the doubling trick, we can think of the bulk-to-boundary two-point function as a bulk three-point function on the sphere after an appropriate quotient (see discussion in Section \ref{sec:legos}). Therefore, in this case we can expect an ETH-like proposal to hold for the bulk-to-boundary structure constants since it originates from the usual ETH ansatz of the bulk OPE coefficients. On the other hand, the case where a boundary operator can effectively describe a thermal background for a simple bulk operator is far less instinctive to expect, but it certainly opens an exciting new possibility to explore, as suggested by our formulas. In recent lower dimensional setups of evaporating black holes, 
it has been proposed  (see e.g.\cite{Almheiri:2019yqk,Rozali:2019day,Sully:2020pza,Reeves:2021sab}) that BCFTs can serve as natural models for such a construction where essentially the boundary degrees of freedom in the CFT on their own describe an equilibrium
black hole, but the coupling to the bulk CFT degrees of freedom allows this black hole to
evaporate. It will be extremely interesting to understand the relevance of our asymptotic formulas for the bulk-to-boundary  OPE coefficients in these constructions in the future.

\section{Semiclassical limits and AdS$_3$/BCFT$_2$}\label{sec:semiclassical}

In former sections we derived asymptotic formulas for the structure constants of boundary primary operators.
Our formulas are universal and apply in any irrational two dimensional BCFT with $c>1$.
On the other hand, it is interesting to study BCFTs at large central charge to gain insights about a potential holographic gravity dual.
As a bottom up model, the semiclassical Einstein gravity on the bulk with an end-of-the-world (ETW) brane is well studied \cite{Takayanagi:2011zk,Fujita:2011fp}, where one considers an action localized on the ETW brane.
A simple case is to choose the brane tension as a boundary action.
In that model,  
the boundary entropy is given by the brane tension $T$ and the AdS radius $R$:
\be
\log g = \f{c}{6} \text{arctanh}(TR), \label{eq:AdSBCFTbdyentropy}
\ee 
We can also consider top down models in 6d type 4b supergravity \cite{Chiodaroli:2012vc,Chiodaroli:2011fn} where we can embed the solutions into string theory (see also \cite{Martinec:2022ofs}). 
In those models, it is possible to study the boundary entropy through the holographic entanglement entropy \cite{Ryu:2006bv,Ryu:2006ef}.
Another interesting example of BCFTs with large central charge is the symmetric orbifold CFTs where recently some boundary states were explicitly constructed \cite{Belin:2021nck}.

Our findings showed that all universal formulas involving boundary operators are captured at finite central charge by the factor $C_0(P_1,P_2,P_3)$ (c.f. equation (\ref{eq:C0})) up to powers of the boundary entropy.
The large central charge asympotitcs of $C_0$ were studied extensively in \cite{Collier:2019weq}.
In this section we will apply those limits to our BCFT asymptotic formulas and consider their interpretation and implications for theories with gravity duals. We will not attempt to understand precisely when these formulas apply, in terms of constraints on the theory and regime of validity of operator dimensions (in the spirit of \cite{Hartman:2014oaa}). In fact, in the AdS/BCFT case this picture is still quite unclear; as it was discussed recently, even the simple ETW model\cite{Takayanagi:2011zk} requires a special fine-tuned spectrum in the BCFT side \cite{Reeves:2021sab} which shows that the quest for a holographic BCFT is still a bit uncertain\footnote{See also \cite{Kastikainen:2021ybu} for some recent calculations in the bulk.}.

Before moving on to the large $c$ analysis, we state again our main results on the square of bulk-to-boundary two-point functions and boundary three-point functions -- i.e. the boundary structure constants with all indices \textit{lowered} -- in the specific normalization (\ref{normalization}) with $\mathfrak{g}_{ij}^{(ab)} = \s{g_ag_b}\delta_{ij}$.
Our asymptotic formulas for generic boundary conditions take the simple unified form:
\begin{equation}\label{formsumm}
\begin{aligned}
\overline{\left|C^{(s)}_{\alpha i}\right|^2}&\sim C_0(P_{\alpha},\bar{P}_{\alpha},P_i)\  \notag \\ 
\overline{\left|C^{(abc)}_{ijk}\right|^2}&\sim C_0(P_i,P_j,P_k)\ .
	\end{aligned}
\end{equation}
As we explained in the relevant sections, for the bulk-to-boundary two-point functions the asymptotic formula (\ref{formsumm}) is valid in three distinct heavy regimes of conformal dimensions: light bulk-heavy boundary, heavy bulk-light boundary and heavy bulk-heavy boundary. Similarly for the three-point functions the same formula is valid in the heavy-light-light, heavy-heavy-light, and heavy-heavy-heavy regimes. Under the particular normalization (\ref{normalization}), the asymptotic formulas do \textit{not} involve any factors of the boundary entropy and, quite interestingly, both types of structure constants are captured by the same formula, namely by a factor of $C_0$.
Therefore we can discuss the large $c$ limits of these results without caring about the order of the boundary entropy as a function of $c$.
Note however that in the case where the boundary two point-functions are \textit{not} canonically normalized -- and in particular when the normalization depends on the boundary entropy, as in (\ref{normalization}) --
the two- and three-point functions $\overline{\left|C^{(s)}_{\alpha i}\right|^2}$, $\overline{\left|C^{(abc)}_{ijk}\right|^2}$ can be large or small compared to them in the large $c$ limit.
We will elaborate more on that point in section \ref{sec:BEntropyLargeC}.

\subsection{Boundary OPEs in the large $c$ limit}\label{sec:ThreeLargeC}
\subsubsection{Spectral density of black holes \label{subsec:ThreeHHL}}
First we consider a large $c$ limit of $C_0$ which probes the physics of black holes.
The setup we are imagining is a CFT on an interval with boundary conditions $a$ and $b$. Holographic duals of these setups were considered e.g. in \cite{Geng:2021iyq}, which can be seen as a wedge holography with a regularization \cite{Akal:2020wfl} (see also \cite{Miao:2020oey,Miao:2021ual}).
We take two operators $\Psi^{ab}$, $\Psi^{ac}$ with heavy dimensions $h_1,h_2$ scaling with $c$, to correspond to black hole states, but with $h_1-h_2$ fixed as $c\to \infty$.
The dimension of the third boundary operator $\Psi^{bc}$, which probes the bulk geometry, has $h$ fixed in that limit.
In terms of the momentum variables we can write 
\be
P_1 = b^{-1} p + b \delta, \qquad P_2 =b^{-1}p - b\delta, \qquad P_3= i\Big(\f{Q}{2} -bh \Big),
\ee
and take $b\to 0$ while $p,\delta, h$ are fixed.
Then, we can interpret $C_0$ as governing the matrix elements
${}_{ab}\langle BH_1|\Psi^{bc}|BH_2\rangle_{ac}$
of the probe operator $\Psi^{bc}$ with dimension $h$. Here $\ket{BH_1}_{ab}$ represents a black hole microstate that equilibrates with the common bath in the context of communicating black holes\cite{Geng:2021iyq}.

In this limit, the fusion kernel takes the form \cite{Collier:2018exn,Collier:2019weq}:
\be\label{hhllargec}
\rho_0(b^{-1}p)C_0(P_1,P_2,P_3) \sim \f{(2p)^{2h}}{2\pi b} \f{\Gamma(h + 2i\delta)\Gamma(h - 2i\delta)}{\Gamma(2h)}.
\ee
Note that in our case a \textit{single} copy of $C_0$ gives the asymptotic formula for the \textit{squared} boundary three-point functions, rather than two copies of $C_0$ which is the case for the bulk three-point functions \cite{Collier:2019weq}. The behaviour (\ref{hhllargec}) is actually similar to the spectral density of a CFT$_1$ and correspondingly the matter on a black hole in AdS$_2$, which is natural from the symmetry perspective since both the BCFT$_2$ and CFT$_1$ have the same global conformal symmetry $SL(2,\mathbb{R})$.
 
\subsubsection{Near extremal limit and the Schwarzian theory \label{subsec:ThreeHHLext}}
Next we consider a different limit where $h - \f{c-1}{24}$ is of order $c^{-1}$ while the third operator remains a light probe.
This limit was considered in \cite{Collier:2019weq,Ghosh:2019rcj} to study the near extremal limit of dual BTZ black holes. 
In this limit we can write 
\be
P_1 = bk_1, \qquad P_2 = bk_2 ,\qquad P_3  = i \Big(\f{Q}{2} - bh \Big)
\ee
and we take $k_1,k_2 ,h$ fixed in the $b\to 0$ limit.
The density of states and the universal OPE coefficient $C_0$ are then given by
\ba
\rho^{\text{(open)}}_{ab}(bk) &\sim 8 \s{2} e^{\f{1}{2}(\mathfrak{s}_a+\mathfrak{s}_b)} \pi b^2 k \sinh (2\pi k) , \\
C_0(bk_1,bk_2 , i (\f{Q}{2}-bh)) &\sim \f{b^{4h}}{\s{2} (2\pi b)^3} \f{\prod_{\pm \pm} \Gamma(h \pm ik_1 \pm ik_2)}{\Gamma(2h)}. 
\ea
These expressions turn out to be the same with the analogous expressions in Schwarzian theory, which captures the dynamics of both explicitly and spontaneously broken 1d conformal symmetry \cite{Maldacena:2016hyu, Maldacena:2016upp, Stanford:2017thb, Mertens:2017mtv}.
In the gravity side, this corresponds to the appearance of the nearly AdS$_2$ geometry which is governed by the Jackiw-Teitelboim (JT) gravity \cite{Maldacena:2016upp,Mertens:2017mtv,Ghosh:2019rcj}.
Notably, a reduction of  AdS$_3$ to JT gravity was studied recently in \cite{Verheijden:2021yrb}, where the spatial direction is an $S^1$. In a similar fashion, we could imagine constructing the same type of solutions where the spatial direction is an interval \cite{Geng:2021iyq} with boundary conditions $a$ and $b$ in the holographic BCFT models with ETW branes.
We would expect that an analogous reduction from AdS$_3$ to JT gravity will take place, and the near extremal limit of our formula will exactly capture this reduction.

\subsubsection{Conical defect limit \label{subsec:ThreeHHH}}
Finally we consider a regime where all the operators have dimensions of order $c$.
In this limit, we take 
\be
\alpha_i = b^{-1} \eta_i, \qquad b\to 0, \qquad \text{fixed   } \eta_i, \qquad \text{for} \ i = 1 ,2,3
\ee
and $\eta_i < 1$.
Then, one can show \cite{Collier:2019weq} that $C_0$ asymptotes to: 
\be
\log C_0 \sim b^{-2} ( - 1/2 S_{\text{grav}}(\eta_1,\eta_2,\eta_3) + i \theta(\eta_1,\eta_2,\eta_3))
\ee
where $S_{\text{grav}}$ and $\theta(\eta_1,\eta_2,\eta_3)$ are given by
\ba
-\f{1}{2}S_{\text{grav}} &:= (F(2\eta_1) - F(\eta_2 + \eta_3 - \eta_1) + (1-2\eta_2)\log (1-2\eta_2) + (2 \text{ permutations})) \notag \\
& \ \ \ \ + F(0) - F(\eta_1 + \eta_2 + \eta_3) - 2(1-\eta_1 - \eta_2 - \eta_3) \log (1-\eta_1-\eta_2-\eta_3) \\
\theta & := \pi( \eta _1 + \eta_2 + \eta_3 -1), 
\ea
and $F(z) = I(z) + I(1-z)$ for $I(z) = \int _\f{1}{2}^zdy \log \Gamma(y)$.
The term  $b^{-2}S_{\text{grav}}$ which looks like an on shell action actually appears as the gravitational action for a conical defect network in 3d gravity \cite{Chang:2016ftb}. 
For the case of bulk three point functions, the analogous semiclassical limit precisely agrees with the conical defect action \cite{Collier:2019weq}.
On the other hand, in our boundary three-point function cases the asymptotic formula contains only a single copy of $C_0$ and apparently seems to capture the square root of that conical defect action.
It is definitely an interesting open problem to study the corresponding bulk calculations in the AdS/BCFT setup (e.g. in the AdS/BCFT model with simple brane tension actions) and try to match those with our asymptotic formulas.

\subsection{Bulk-to-boundary OPEs in the large $c$ limit}\label{sec:BbLargeC}
So far we discussed the asymptotics of boundary three-point functions in the large $c$ limit.
Since the asymptotics of bulk-to-boundary correlation functions are also captured by $C_0$, one can in principle study the same limits as before.
For example, we can study the universal dynamics of the correlation function 
\be
{}_s\langle0|\op_{\alpha}\Psi_i^{ss}|0\rangle_s
\ee
where $\ket{0}_s$ is the ground state on a finite interval with boundary conditions labeled by $s$ on both sides.
When we take the length of the interval to be infinite, we are studying the correlation function on a half-line with boundary condition $s$.
This setup is close to the recent toy models of black holes coupled to a bath, where the total system is at zero temperature \cite{Almheiri:2019yqk}.
Mapping to the cylinder, we can bring the system to a thermofield double state.
This is a toy model where black holes equilibrate with a finite temperature bath.
We can also consider the moving mirror models of \cite{Davies:1976hi,Akal:2020twv,Akal:2021foz}, which describe a time-dependent boundary trajectory and hence serving as toy models of evaporating black holes \cite{Almheiri:2019psf,Almheiri:2019hni}.
In these contexts, a boundary operator in BCFT corresponds to an operator in the dual of the black hole, and a bulk operator is an operator in the bath system.
Our universal formula for the bulk-to-boundary two-point functions captures the asymptotics when either or both of these operators are heavy. It will be extremely fruitful to study more the consequences of our (explicit) asymptotic formulas in the context of black hole evaporation and information paradox as encoded in those simple models.

\subsection{The order of the boundary entropy}\label{sec:BEntropyLargeC}

In this final section we will make some comments on how the order of the boundary entropy controls the size of the correlation functions in AdS/BCFT (and solely in BCFT in general) given our universal asymptotic results for the boundary structure constants in various heavy regimes.
In the bulk CFT case, we usually canonically normalize the operators so that the operator metric is purely diagonal, i.e. $\braket{\op_i(0)\op_j(1)} = \delta_{ij}$. 
As we discussed in section \ref{sec:structureconstandcorrfns}, when we have boundary operators, it is not quite natural to consider a purely diagonal operator metric for the boundary two-point functions. We will instead choose to work in the normalization (\ref{normalization}).
To gain some intuition on the relation between the order of the boundary entropy and the boundary OPE coefficients, we will study the ratio of our asymptotic results \eqref{formsumm} with the corresponding two-point functions. One finds
\begin{align}
\label{eq:ratios}
&~~~~~~~~~~~
\f{\overline{\left|C^{(s)}_{\alpha i}\right|^2}}{\braket{\Psi_{i}^{aa} \Psi_{i}^{aa}}}  \sim g_s^{-1} C_0(P_\alpha,\bar{P}_{\alpha},P_i),  \notag   \ \ \  \f{\overline{\left|C^{(abc)}_{ijk}\right|^2}}{\braket{\Psi_{i}^{ab} \Psi_{i}^{ba}}\braket{\Psi_{j}^{bc} \Psi_{j}^{cb}}\braket{\Psi_{k}^{ca} \Psi_{k}^{ac}} } \sim g_a^{-1}g_b^{-1}g_c^{-1}  C_0(P_i,P_j,P_k) .\\ 
\end{align}
In the case of symmetric orbifold CFTs \cite{Belin:2021nck} with rational seed theories, a typical boundary entropy is of order $c$, in other words $g \sim e^{c}$.
Therefore in this case we get an overall $e^{-c}$ suppression in \eqref{eq:ratios}.
For seed theories with an infinite number of  boundary states, typically $g$ scales as $e^{ c \log c}$ and hence the bulk-to-boundary two-point functions and  boundary three-point functions are further suppressed by an additional factor of $\log c$.
For atypical boundary states, it turns out that we get $g \sim e^{-c\log c}$ and hence in that case the boundary entropy acts as an enhancement factor.
In the context of AdS/BCFT, positive boundary entropies correspond to \textit{creation} of additional spacetime \cite{Almheiri:2018ijj,Numasawa:2018grg} since their action increases the angle between the AdS boundary and the ETW brane.
Therefore, in this case we expect that bulk-to-boundary correlation functions and boundary three-point functions are relatively suppressed.
On the other hand, when the ETW brane ``eats up'' the spacetime due to the presence of negative boundary entropy, the ratios (\ref{eq:ratios}) are expected to be enhanced.
More concretely, for the AdS/BCFT model with a brane tension on the ETW brane  the boundary entropy is given by (\ref{eq:AdSBCFTbdyentropy}).
Therefore in that case the boundary entropy is always order $c$ -- assuming that the tension is order one in AdS units -- and its sign depends on the sign of the tension.
In a top down model in type 4b supergravity approximation, the calculation of the holographic entanglement entropy gives a diverging  boundary entropy \cite{Chiodaroli:2011fn}.
This fact suggests that the boundary three-point functions and the bulk-to-boundary correlation functions are almost negligible in that case.

It is definitely an interesting open problem to compute boundary three-point functions and bulk-to-boundary correlation functions in either bottom up models of AdS/BCFT or top down models in a supergravity approximation. We hope that the computations done in this work will motivate research towards this fascinating direction.

\section*{Acknowledgements}
We would like to thank Scott Collier, Yuya Kusuki, Alex Maloney, Henry Maxfield, Eric Perlmutter, Sylvain Ribault, Tadashi Takayanagi, Satoshi Yamaguchi for interesting discussions and correspondence on the draft. We would like to thank especially Alex Maloney for initial collaboration, and Scott Collier for illuminating discussions up to the late stages of the project. TN is supported by the JST CREST Grant (No.JPMJCR19T3). IT is supported by the ERC starting grant 679278 Emergent-BH.

\begin{appendix}

\section{Explicit forms of elementary crossing kernels}\label{app:explicitForms}

In this appendix we briefly review the explicit forms of the elementary crossing kernels in the irrational case ($c>1$).

\subsection{Fusion four-point kernel}\label{subsec:fusion}
We will start by reviewing the explicit form of the fusion kernel, which implements the fusion transformation relating sphere four-point Virasoro conformal blocks in different OPE channels. The fusion kernel was derived by Ponsot and Teschner \cite{Ponsot:1999uf,Ponsot:2000mt}. The expression involves the special functions $\Gamma_b(x)$, which is a meromorphic function with no zeros that one may think of as a generalization of the ordinary gamma function, but with simple poles at $x = -(mb+nb^{-1})$ for $m,n\in\mathbb{Z}_{\ge 0}$, and 
\begin{equation}
	S_b(x) = {\Gamma_b(x)\over \Gamma_b(Q-x)}.
\end{equation}
Many properties of these special functions, including large argument and small $b$ asymptotics, were summarized in Appendix A of \cite{Collier:2018exn}. The explicit expression for the kernel involves a contour integral and is given by
\begin{equation}\label{eq:explicitFusion}
	\fusion_{P_sP_t}\sbmatrix{ P_2 & P_1 \\ P_3 & P_4 } = P_b(P_i;P_s,P_t)P_b(P_i;-P_s,-P_t)\int_{\mathcal{C}'}{ds\over i}\prod_{k=1}^4{S_b(s+U_k)\over S_b(s+V_k)},
\end{equation}
where the prefactor $P_b$ is given by
\begin{equation}
\begin{aligned}\label{eq:fusionPrefactor}
	&P_b(P_i;P_s,P_t)\\
	=& {\Gamma_b({Q\over 2}+i(P_s+P_3-P_4))\Gamma_b({Q\over 2}+i(P_s-P_3-P_4))\Gamma_b({Q\over 2}+i(P_s+P_2-P_1))\Gamma_b({Q\over 2}+i(P_s+P_1+P_2))\over \Gamma_b({Q\over 2}+i(P_t+P_1-P_4))\Gamma_b({Q\over 2}+i(P_t-P_1-P_4))\Gamma_b({Q\over 2}+i(P_t+P_2-P_3))\Gamma_b({Q\over 2}+i(P_t+P_2+P_3))}{\Gamma_b(Q+2iP_t)\over \Gamma_b(2iP_s)}
\end{aligned}
\end{equation}
and the arguments of the special functions in the integrand are
\begin{equation}
	\begin{split}
		U_1&=i(P_1-P_4)\\
		U_2&=-i(P_1+P_4) \\
		U_3&= i(P_2+P_3)\\
		U_4&=i(P_2-P_3)
	\end{split}
	\qquad
	\begin{split}
		V_1 &= Q/2+i(-P_s+P_2-P_4)\\
		V_2 &= Q/2+i(P_s+P_2-P_4) \\
		V_3 &= Q/2+iP_t \\
		V_4 &= Q/2-iP_t
	\end{split}
\end{equation}
The contour $\mathcal{C}'$ runs from $-i\infty$ to $i\infty$, traversing between the towers of poles running to the left at $s = -U_i-mb-nb^{-1}$ and to the right at $s = Q-V_j +mb+nb^{-1}$ in the complex $s$ plane, for $m,n\in\mathbb{Z}_{\ge0}$.

Viewed as a function of the internal weight $P_s$, the kernel (\ref{eq:explicitFusion}) has eight semi-infinite lines of poles extending to both the top and bottom of the complex plane
\begin{equation}
\begin{aligned}\label{eq:fusionPoles}
	\fusion_{P_sP_t}\sbmatrix{ P_2 & P_1 \\ P_3 & P_4 }\text{: }&\text{simple poles at }P_s = \pm i\left({Q\over 2}+iP_0+mb+nb^{-1}\right),\text{ for }m,n\in\mathbb{Z}_{\ge 0},\\
	&\text{where }P_0 = P_1+P_2,\, P_3+P_4 \text{ (and six permutations under reflection $P_i\to-P_i$)}.
\end{aligned}
\end{equation}
In the case particularly relevant for this paper of pairwise identical operators $P_4 = P_1,~P_3 =P_2$, these singularities are enhanced to double poles, although there is an exception when the T-channel internal weight $P_t$ is degenerate ($P_t = \pm {i\over 2} ((m+1)b+(n+1)b^{-1}),~m,n\in\mathbb{Z}_{\ge 0}$), in which case the poles remain simple when the external operators have weights consistent with the fusion rules.

In the special case of pairwise identical operators with T-channel exchange of the identity, the contour integral can be computed very explicitly and the fusion kernel takes the following simple form, which makes the analytic structure manifest
\begin{equation}\begin{aligned}\label{eq:vacFusion}
	\fusion_{P_s\id}\sbmatrix{ P_2 & P_1 \\ P_2 & P_1 } &= {\Gamma_b(2Q)\over \Gamma_b(Q)^3}{\Gamma_b({Q\over 2}+i(P_1+P_2-P_s))\times(\text{7 permutations under reflection $P\to -P$})\over \Gamma_b(2iP_s)\Gamma_b(-2iP_s)\Gamma_b(Q+2iP_1)\Gamma_b(Q-2iP_1)\Gamma_b(Q+2iP_2)\Gamma_b(Q-2iP_2)}\\
	&= \rho_0(P_s)C_0(P_1,P_2,P_s),
\end{aligned}\end{equation}
with 
\begin{equation}\begin{aligned}\label{eq:rho0C0}
\rho_0(P)&\equiv 4\sqrt{2}\sinh{(2\pi b P)}\sinh{(2\pi b^{-1} P)}\\
C_0(P_i,P_j,P_k) &\equiv
\frac{1}{\sqrt{2}}{\Gamma_b(2Q) \over \Gamma_b(Q)^3} {\prod_{\pm\pm\pm}\Gamma_b\left({Q\over 2} \pm i P_i \pm i P_j\pm i P_k\right)  \over \prod_{a\in\{i,j,k\}} 
\Gamma_b(Q +2iP_a) \Gamma_b(Q -2iP_a)}~.
\end{aligned}\end{equation}

\subsection{Torus one-point kernel}\label{torus1ptkernel}
The crossing kernel that implements the modular S transformation on torus one-point Virasoro blocks was worked out by Teschner \cite{Teschner:2003at}. Similarly to the fusion kernel, its explicit form involves a contour integral and is given by
\begin{equation}\begin{aligned}\label{eq:explicitModularS}
	\modS_{PP'}[P_0] =& {\rho_0(P)\over S_b({Q\over 2}+iP_0)}{\Gamma_b(Q+2iP')\Gamma_b(Q-2iP')\Gamma_b({Q\over 2}+i(2P-P_0))\Gamma_b({Q\over 2}-i(2P+P_0))\over\Gamma_b(Q+2iP)\Gamma_b(Q-2iP)\Gamma_b({Q\over 2}+i(2P'-P_0))\Gamma_b({Q\over 2}-i(2P'+P_0))}\\
	&\int_C{d\xi\over i}e^{-4\pi P'\xi}{S_b(\xi+{Q\over 4}+i(P+\half P_0))S_b(\xi+{Q\over 4}-i(P-\half P_0))\over S_b(\xi+{3Q\over 4}+i(P-\half P_0))S_b(\xi+{3Q\over 4}-i(P+\half P_0))}\\
	\equiv& Q_b(P,P',P_0)\int_C{d\xi\over i}e^{-4\pi P'\xi}T_b(\xi,P,P_0).
\end{aligned}\end{equation}
This integral representation only converges when 
\begin{equation}\label{eq:alphaPrimeCondition}
	 {1\over 2}{\rm Re}(\alpha_0)<{\rm Re}(\alpha')<{\rm Re}\left(Q-{1\over 2}\alpha_0\right).
\end{equation}
Outside of this range, the kernel is defined via analytic continuation, using the fact that it satisfies shift relations\cite{Nemkov:2015zha,Collier:2019weq}.

The integral contributes the following series of poles in the $P$ plane, one extending to the top and the other extending to the bottom
\begin{equation}\begin{aligned}
	\text{integral: poles at }P = \pm{i\over 2}\left({Q\over 2}+iP_0+mb+nb^{-1}\right),~m,n\in\mathbb{Z}_{\ge 0}.
\end{aligned}\end{equation}
Together with the prefactor, the full kernel has the following polar structure in the $P$ plane
\begin{equation}\begin{aligned}
	\modS_{PP'}[P_0]:\text{ poles at }P = {i\over 2}\left({Q\over 2}-iP_0+mb+nb^{-1}\right),~m,n\in\mathbb{Z}_{\ge 0}, \text{ and all possible reflections (in $P,P_0$)}.
\end{aligned}\end{equation}
One can think of these poles as arising in the case that the external operator is a (Virasoro) double-twist of the internal operator\cite{Collier:2018exn}.

Similarly to the case of the fusion kernel, the modular S kernel can be straightforwardly evaluated in the case that the external operator is the identity, $P_0 = i{Q\over 2}$. In this case, the prefactor vanishes and so we only need to extract the singularities of the contour integral. By carefully studying this limit, one finds
\begin{equation}
	\modS_{PP'}[\id] = 2\sqrt{2}\cos(4\pi P P'),
\end{equation}
precisely reproducing the non-degenerate modular S matrix for the Virasoro characters (\ref{virchar}),(\ref{cylZrelation}). To study the limit in which the internal operator in the original channel is also the identity one should make use of the shift relations (see discussion in Appendix  of \cite{Collier:2019weq}) in which case one can reproduce the result:
\begin{equation}
\begin{aligned}
\modS_{P\mathbb{1}}[\id] =  4\sqrt{2}\sinh(2\pi b P)\sinh(2\pi b^{-1}P),
\end{aligned}
\end{equation}
which defines the universal spectral density $\rho_0$ in (\ref{eq:rho0C0}).

\section{Asymptotics of crossing kernels}\label{sec:asymptcrosskernnls}

In this section we will review some results obtained in \cite{Collier:2018exn, Collier:2019weq} for the asymptotic form of the elementary crossing kernels when some of the weights are taken to be heavy. These results are important for establishing both the form of our asymptotic formulas and their validity, via the suppression of corrections due to the propagation of non-vacuum primaries.

\subsection{Fusion kernel}\label{sub:fusionasympt}

In \cite{Collier:2018exn}, the asymptotic form of the fusion kernel when the S-channel internal weight $P_s$ was taken to be heavy with fixed external weights was extensively studied. The main result of that analysis was the following asymptotic form of the vacuum fusion kernel (\ref{eq:vacFusion}) with pairwise identical operators, which follows directly from the asymptotics of the special function $\Gamma_b$ that were established in that paper
\begin{equation}\begin{aligned}\label{eq:vacFusionAsymptotics}
	\fusion_{P_s\id}\sbmatrix{ P_2 & P_1 \\ P_2 & P_1 } \sim& 2^{-4P_s^2}e^{\pi QP_s}P_s^{4(h_1+h_2)-{3Q^2+1\over 2}}\\
	&\times{2^{Q^2+1\over 6}\Gamma_0(b)^6\Gamma_b(2Q)\over \Gamma_b(Q)^3\Gamma_b(Q+2iP_1)\Gamma_b(Q-2iP_1)\Gamma_b(Q+2iP_2)\Gamma_b(Q-2iP_2)},\,P_s\to\infty
\end{aligned}\end{equation}	
where
\begin{equation}\label{g0}
	\log\Gamma_0(b) = -\int_0^\infty{dt\over t}\left({e^{-Qt/2}\over(1-e^{-bt})(1-e^{-b^{-1}t})}-t^{-2}-{Q^2-2\over 24}e^{-t}\right)
\end{equation}
appears in the large-argument asymptotics of $\Gamma_b(x)$.

By carefully studying the asymptotics of the contour integral in the definition of the fusion kernel, in \cite{Collier:2018exn} it was also established that the fusion kernel with non-zero T-channel weight is exponentially suppressed at large $P_s$ compared to the vacuum kernel 
\begin{equation}\begin{aligned}\label{eq:fusionSuppressionn}
	{\fusion_{P_sP_t}\sbmatrix{ P_2 & P_1 \\ P_2 & P_1 }\over \fusion_{P_s\id}\sbmatrix{ P_2 & P_1 \\ P_2 & P_1 }}\sim& e^{-2\pi\alpha_t P_s}\left({\Gamma_b(Q+2iP_1)\Gamma_b(Q-2iP_1)\over \Gamma_b({Q\over 2}+i(2P_1-P_t))\Gamma_b({Q\over 2}-i(2P_1+P_t))}\times(P_1\to P_2)\right)\\
	&\times {\Gamma_b(Q-2iP_t)\Gamma_b(-2iP_t)\Gamma_b(Q)^3\over\Gamma_b(2Q)\Gamma_b({Q\over 2}-iP_t)^4},\, P_s\to\infty.
\end{aligned}\end{equation}
Thus we learn that corrections to either the light bulk-heavy boundary bulk-to-boundary structure constants (\ref{LHbktobdyResult}) or the boundary heavy-light-light OPE asymptotic formula (\ref{averegefinalhll}) due to the exchange of non-vacuum primaries in the corresponding dual channels are exponentially suppressed.

\par In \cite{Collier:2019weq}, it was further shown that the propagation of non-vacuum primaries is suppressed compared to that of the vacuum when one or both of the external operators $P_1,P_3$ are taken to be heavy along with the S-channel internal weight $P_2$:
\begin{equation}
	{\fusion_{P_2P_2'}\sbmatrix{ P_1 & P_3 \\ P_1 & P_3 }\over \fusion_{P_2\id}\sbmatrix{P_1 & P_3 \\ P_1 & P_3 }}
\end{equation}
In particular, for the heavy-heavy-light case, when $\alpha_1,\alpha_2 = {Q\over 2}+iP,~P\to\infty$, with $\alpha_3\equiv\alpha_0$ and $\alpha_2'$ fixed (with $0<\alpha_2'<{Q\over 2}$),
 it was shown explicitly that
\begin{equation}
	{\fusion_{P_2P_2'}\sbmatrix{P_0 & P_1 \\ P_0 & P_1 }\over \fusion_{P_2\id}\sbmatrix{P_0 & P_1 \\ P_0 & P_1 }} \sim \text{(order-one)}P^{-h_2'} \ .
\end{equation}
The analysis is similar for corrections to the heavy-heavy-heavy boundary OPE asymptotics due to propagation of non-vacuum primaries in the boundary dumbbell channel, which in the case where $\alpha_1,\alpha_2,\alpha_3 = {Q\over 2}+i P$ are taken to be heavy, one obtains
\begin{equation}
	{\fusion_{P_2P_2'}\sbmatrix{P_1 & P_3 \\ P_1 & P_3}\over\fusion_{P_2\id}\sbmatrix{P_1 & P_3 \\ P_1 & P_3}} \sim \text{(order-one)}P^{-2h_2'} \ .
\end{equation}

\subsection{Torus kernel}\label{app:torusOnePtAsymptotics}

In order to establish the validity of the heavy bulk-heavy boundary bulk-to-boundary structure constants as well as the heavy-heavy-light and heavy-heavy-heavy boundary OPE asymptotics, we also need to study the asymptotics of the torus one-point kernel (\ref{torus1ptkernel}) in the limit that the internal weight in one of the channels becomes heavy, namely the limit $P\to\infty$. 

It was established in \cite{Collier:2019weq} that in the regime (\ref{eq:alphaPrimeCondition}) the kernel obeys the following asymptotics
\begin{equation}\label{kern1ptasympt}
\begin{aligned}
	\modS_{PP'}[P_0]\approx& \left({{Q\over 2}-iP_0\over \sqrt{2}\pi(-2iP')({Q\over 2}+i(2P'-P_0))}{\Gamma_b(Q+2iP')\Gamma_b(Q-2iP')\over S_b({Q\over 2}+iP_0)\Gamma_b({Q\over 2}+i(2P'-P_0))\Gamma_b({Q\over 2}-i(2P'+P_0))}\right)\\
	&\times e^{-4\pi iPP'}(2P)^{h_0}
\end{aligned}\end{equation}
To compute the kernel when $\alpha'$ is outside of the regime (\ref{eq:alphaPrimeCondition}), we can make use of the shift relations for the kernel, where one finds that the leading asymptotic behaviour at large $P$ is basically the same as in (\ref{kern1ptasympt}) (see Appendix B of \cite{Collier:2019weq} for more details).

\section{Details on the crossing kernel for cylinder two-point functions from the doubling trick \label{sec:CykKerDouble}}

In this appendix we will implement the ``doubling trick'' discussed in Section \ref{sec:legos} to justify the proposal for the crossing kernel in Section \ref{sec:BbasymptoticsHL} for the cylinder two-point functions in the particular case of identical external boundary operators and identical boundary conditions at the two ends of the cylinder.

\begin{figure}
\centering
\includegraphics[width=3.3cm]{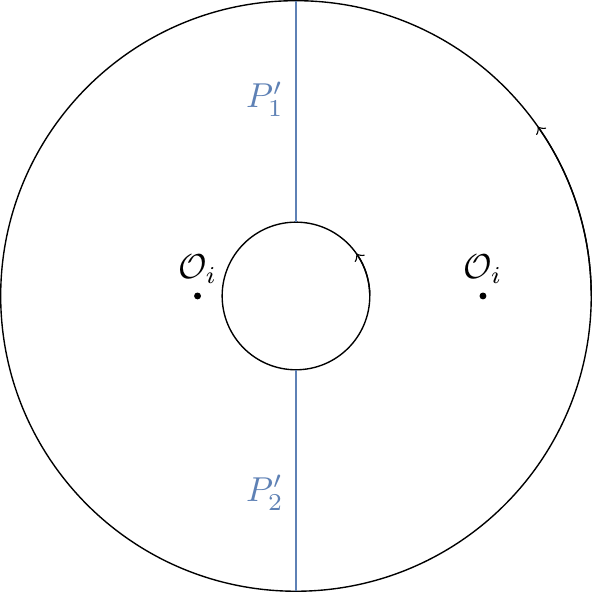}  \qquad \qquad
\includegraphics[width=3.3cm]{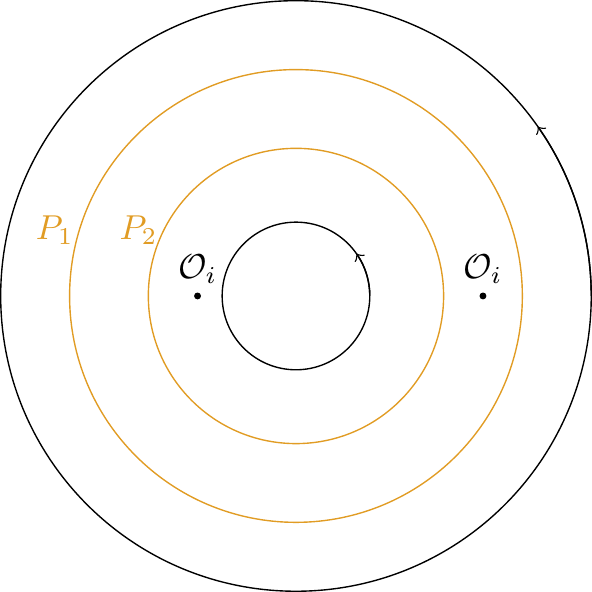}
\caption{Bird's-eye view of a torus two-point function with identical external operators in two different (bulk) decompositions: the ``necklace'' channel (left) and the ``bagel'' channel (right). }  
\label{fig:torus2ptChannels}
\end{figure}

The two-fold compact, oriented cover of a cylinder two-point function is the torus with two punctures.
As illustrated in figure \ref{fig:torus2ptChannels}, we will try to understand the torus two-point functions of identical bulk external operators in two different channels in order to gain intuition for the case of the cylinder two-point functions.
The first channel we will consider is the so-called ``necklace'' channel, which was studied extensively in \cite{Collier:2019weq}.
The second one is the "bagel'' channel, in which we span the Hilbert space in the orthogonal direction comparing to the necklace channel.

\begin{figure}
\centering
\includegraphics[width=3.7cm]{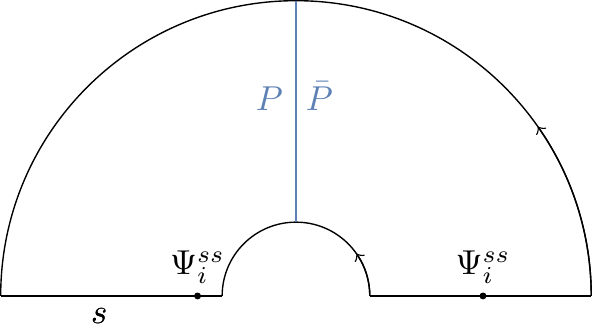}  \qquad \qquad
\includegraphics[width=3.7cm]{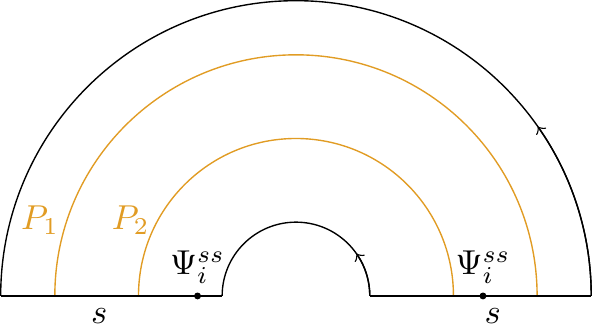}
\caption{Different illustration of the ``boundary necklace'' and ``boundary bagel'' decompositions of the cylinder two-point function with operators on each boundary (see Figure \ref{fig:Cly2ptCase1} in Section \ref{sec:BbasymptoticsHL}). The two-fold compact cover of these ``cuttings'' are the channels depicted in Figure \ref{fig:torus2ptChannels} for the torus two-point function with identical external operators.}   
\label{2ptcyldiffillustr}
\end{figure}

In the cylinder two-point function case, the channels we considered back in Section \ref{sec:BbasymptoticsHL} were intentionally dubbed ``boundary necklace'' and ``boundary bagel'' channel, and we illustrate them again in a different way in Figure \ref{2ptcyldiffillustr}. We notice that the two-fold compact universal covers of the ``boundary necklace'' and ``boundary bagel'' cuttings of the surface are exactly the corresponding necklace and bagel channels of the torus two-point function in Figure \ref{fig:torus2ptChannels}. Therefore, via the doubling trick, the decomposition between the necklace and bagel channels in the torus two-point function case will give us an important clue for the crossing kernel in the cylinder two-point function case.

\begin{figure}
	~~~~~~~~~~~~~~~~~~\includegraphics[width=.15\textwidth]{torus2bagel.pdf} \ \ \raisebox{.07\textwidth}{\scalebox{1.5}{$=\int \frac{dP_2''}{2} \fusion_{P_2P_2''}\sbmatrix{P_1&P_i\\ P_1&P_i} $}} \ \ \ \includegraphics[width=.15\textwidth]{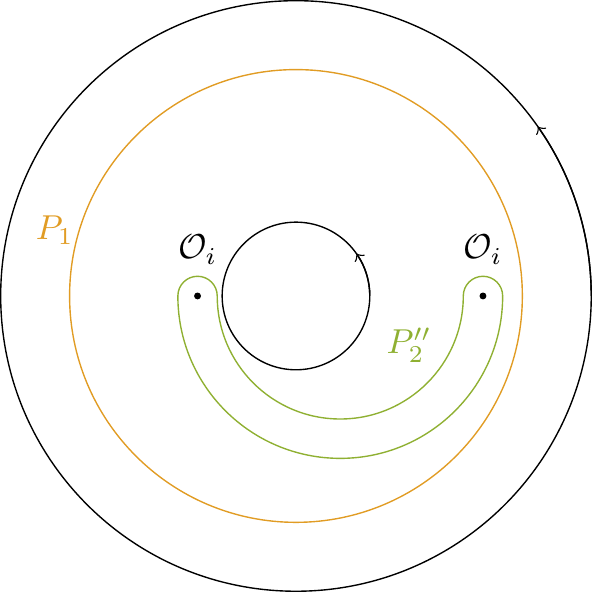} \
	\flushright\raisebox{.07\textwidth}{\scalebox{1.5}{$=\int \frac{dP_1'}{2}\frac{dP_2''}{2}\modS_{P_1P_1'}[P''_2] \fusion_{P_2P_2''}\sbmatrix{P_1&P_i\\ P_1&P_i}$}}\ \ \ \includegraphics[width=.15\textwidth]{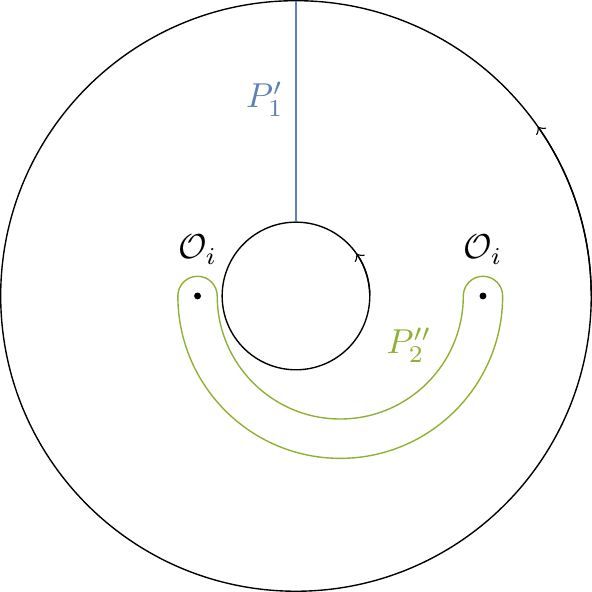} ~~~~~~\\
		\flushright\raisebox{.07\textwidth}{\scalebox{1.5}{$=\int \frac{dP_2'}{2}\frac{dP_1'}{2}\frac{dP_2''}{2} \fusion_{P_2''P_2'}^{-1}\sbmatrix{P_1'&P_i\\ P_1'&P_i} \modS_{P_1P_1'}[P''_2] \fusion_{P_2P_2''}\sbmatrix{P_1&P_i\\ P_1&P_i}$}} \ \  \includegraphics[width=.15\textwidth]{torus2necklace.pdf}
	\caption{The sequence of moves expressing a ``bagel'' channel conformal block (top left) in terms of ``necklace'' channel conformal blocks (bottom right) for a torus two-point function with identical bulk external operators. The corresponding crossing kernel is given in (\ref{kernelcyl2ptt}). \label{fig:Torus2ptDecomp2}}
\end{figure}

The decomposition of bagel channel conformal blocks for the torus two-point function in terms of necklace channel conformal blocks is shown in Figure \ref{fig:Torus2ptDecomp2}. It is straightforward to see that the crossing kernel consists of a convoluted set of moves which includes a fusion move, followed by a modular S move, and finally by an inverse fusion move. The corresponding spectral densities of the bulk OPE data in the two channels are related as
\be\label{densitiesbagelneck}
\rho^{(P_i,\bar{P}_i)}_{\text{necklace}}(P_1,\bar{P}_1;P_2,\bar{P}_2) = \int \f{dP_1'}{2} \f{d\bar{P}_1'}{2}\f{dP_2'}{2}\f{d\bar{P}_2'}{2} \ \mathbb{K}_{P_1P_2;P_1'P_2'}[P_i]\mathbb{K}_{\bar{P}_1\bar{P}_2;\bar{P}_1'\bar{P}_2'}[\bar{P}_i] \ \rho^{(P_i,\bar{P}_i)}_{\text{bagel}}(P_1',\bar{P}_1';P_2',\bar{P}_2'), 
\ee
where the crossing kernel is given by
\be\label{kernelcyl2ptt}
\mathbb{K}_{P_1P_2;P_1'P_2'}[P_i] := \int \frac{dP_2''}{2} \fusion_{P_2''P_2'}^{-1}\sbmatrix{P_1'&P_i\\ P_1'&P_i} \modS_{P_1P_1'}[P''_2] \fusion_{P_2P_2''}\sbmatrix{P_1&P_i\\ P_1&P_i}.
\ee
In (\ref{densitiesbagelneck}) we encounter two copies of the crossing kernel relating the bulk spectral densities (or the conformal blocks), one for each holomorphic part of the spectrum. When we take the appropriate quotient of the torus two-point function to obtain the cylinder two-point function we expect that only the corresponding \textit{holomorphic} conformal blocks will span the cylinder two-point functions, following \cite{Cardy:1989ir}. Therefore, the crossing kernel that relates ``boundary necklace'' and ``boundary bagel'' conformal blocks in the case of cylinder two-point functions with identical external operators on each boundary will be given by (\ref{kernelcyl2ptt}).

As a consistency check, notice that in the case where $\mathcal{O}_i=\mathbb{1}$ one should be able to recover the familiar modular invariance of the torus partition function in the bulk, i.e. the kernel (\ref{kernelcyl2ptt}) should asymptote to the modular kernel $\mathbb{S}$. This is indeed the case since, by the definition of the two point functions on the sphere, first the factor $\fusion_{P_2P_2''}\sbmatrix{P_1&\mathbb{1}\\ P_1&\mathbb{1}}$ localizes $P^{''}_2$ on the identity contribution and sets $P_2=P_1$, which then similarly sets $\fusion_{\mathbb{1}P_2'}^{-1}\sbmatrix{P_1'&\mathbb{1}\\ P_1'&\mathbb{1}}=\delta(P'_1-P'_2)$ and at the end leaves only a single factor of $\mathbb{S}_{P_1P'_1}[\mathbb{1}]$, which is what we wanted. In a similar fashion, it is straightforward also to verify that taking (\ref{kernelcyl2ptt}) to be the crossing kernel between the boundary necklace and boundary bagel channels of the cylinder two-point function reproduces in the relevant limits the kernels for the cylinder one-point function or the cylinder partition function we encountered in Section \ref{subsec:boundaryCardy}.

\section{Relation between the bulk-to-boundary structure constant in Liouville theory and the modular kernel \label{sec:ZZmodularK}}

In this last appendix we prove an exact relation between the bulk-to-boundary structure constant in Liouville theory with Neumann boundary conditions\cite{Hosomichi:2001xc} and the irrational version of the modular kernel that relates torus one-point blocks\cite{Teschner:2003at}. This result is new as far as we can tell. Our motivation originates from a similar relation between the bulk-to-boundary structure constants and the modular matrix of the torus one-point functions for the A-series Minimal Models \cite{Runkel:1998he}.
\par The bulk-to-boundary structure constant of Liouville theory on the disk with Neumann boundary conditions labelled by $s$ (which encodes the boundary cosmological constant), a bulk operator labelled by $a=\frac{Q}{2}+iP_{a}$, and a boundary operator labelled by $\beta=\frac{Q}{2}+iP_{\beta}$ was first obtained in \cite{Hosomichi:2001xc}\footnote{The case of Dirichlet boundary condition in Liouville theory was later studied by B. Ponsot in \cite{Ponsot:2003ss}.}. The expression reads:
\begin{equation}\label{hosom}
\begin{aligned}
A(P_a;P_\beta|s)&=\int_{-i\infty}^{i\infty}\frac{dp}{i} \ e^{-4\pi s p} \tilde{A}(P_a;P_\beta|p)\\
\tilde{A}(P_a;P_\beta|p)&:=2\pi \left(\mu \pi \gamma(b^2)b^{2-2b^2}\right)^{-\frac{Q/2+i(2P_a+P_{\beta})}{2b}}\frac{\Gamma_b(\frac{Q}{2}-iP_{\beta})^3\Gamma_b(\frac{Q}{2}-2 i P_a-i P_\beta )\Gamma_b(\frac{Q}{2}+2 i P_a-i P_\beta)}{\Gamma_b(Q)\Gamma_b(\frac{Q}{2}+iP_{\beta})\Gamma_b(-2iP_{\beta})\Gamma_b(Q+2iP_a)\Gamma_b(-2iP_a)}\\
& \ \ \ \ \ \ \times \frac{S_{b}\left(p+s_1\right)S_{b}\left(p+s_2\right)}{S_b(p+Q-s_1)S_b(p+Q-s_2)},
\end{aligned}
\end{equation}
where $\mu$ is the bulk cosmological constant, $\gamma(x)=\frac{\Gamma(x)}{\Gamma(1-x)}$, and
$
s_1\equiv \frac{Q}{4}+ i P_a+\frac{i P_\beta }{2}, s_2\equiv \frac{Q}{4}- i P_a+\frac{i P_\beta }{2}.
$
We will find it convenient to introduce the $b-$\textit{deformed hypergeometric function} which is defined as (see e.g. appendix C of \cite{Ponsot:2003ju}):
\begin{equation}
\begin{aligned}
F_b\left(\alpha,\beta;\gamma;-ix\right)=\frac{S_b(\gamma)}{S_b(\alpha)S_b(\beta)}\int_{-i\infty}^{i\infty}\frac{dp}{i}e^{2\pi p x}\frac{S_b(p+\alpha)S_b(p+\beta)}{S_b(p+\gamma)S_b(p+Q)}.
\end{aligned}
\end{equation}
Going back to (\ref{hosom}), we can change variables of integration as $\tilde{p}\equiv p-s_2$ and then re-express the result compactly as
\begin{equation}\label{aliouv}
\begin{aligned}
A(P_a;P_\beta|s)
&=2\pi \left(\mu \pi \gamma(b^2)b^{2-2b^2}\right)^{-\frac{Q/2+i(2P_a+P_{\beta})}{2b}} \times \frac{\Gamma_b(2iP_a)\Gamma_b(\frac{Q}{2}-iP_{\beta})^2\Gamma_b(\frac{Q}{2}-2 i P_a-i P_\beta )\Gamma_b(\frac{Q}{2}-2iP_a+iP_\beta)}{\Gamma_b(Q)\Gamma_b(Q+2iP_a)\Gamma_b(Q-2iP_a)\Gamma_b(-2iP_a)\Gamma_b(-2iP_{\beta})}\\
& \ \ \ \ \ \  \times e^{-4\pi s_2s} \ F_b\left(s_1+s_2,2s_2;Q+s_2-s_1;2is\right).
\end{aligned}
\end{equation}
The modular kernel that relates torus one-point blocks was obtained in \cite{Teschner:2003at}. If we consider an external operator on the torus with $P_0=P_{\beta}$ and the internal operators with $P=P_a$ and $P'=s$, the modular kernel $\mathbb{S}_{PP'}[P_0]$ which we wrote in (\ref{eq:explicitModularS}) can be written in terms of the $b$-deformed hypergeometric function as
\begin{equation}
\begin{aligned}
\mathbb{S}_{P_a \ s}[P_\beta]&=
\sqrt{2}\frac{\Gamma_b\left(Q+2is\right)\Gamma_b\left(Q-2is\right)\Gamma_b(\frac{Q}{2}-iP_{\beta})\Gamma_b\left(\frac{Q}{2}+2iP_a-iP_\beta\right)\Gamma_b\left(\frac{Q}{2}-2iP_a-iP_\beta\right)}{\Gamma_b\left(\frac{Q}{2}+2is-iP_\beta\right)\Gamma_b\left(\frac{Q}{2}-2is-iP_\beta\right)\Gamma_b(\frac{Q}{2}+iP_{\beta})\Gamma_b(2iP_a)\Gamma_b(-2iP_a)} \frac{}{}\\
&  \ \ \ \ \ \ \times \frac{S_b(s_1+s_2)S_b(2s_2)}{S_b(Q+s_2-s_1)} e^{-4\pi s_2s} \ F_b\left(s_1+s_2,2s_2;Q+s_2-s_1;2is\right).
\end{aligned}
\end{equation}
Comparing this with the bulk-to-boundary structure constant of Liouville (\ref{aliouv}), we first get the following relation
\begin{equation}\label{relation}
\begin{aligned}
\frac{\mathbb{S}_{P_a \ s}[P_\beta]}{A(P_a;P_\beta|s)}&= \frac{\left( \pi\mu \gamma(b^2)b^{2-2b^2}\right)^{\frac{Q/2+iP_{\beta}}{2b}}}{2^{\frac{1}{2}}\pi}\left(\frac{\Gamma_b(Q)\Gamma_b\left(Q+2is\right)\Gamma_b\left(Q-2is\right)\Gamma_b(-2iP_{\beta})}{\Gamma_b\left(\frac{Q}{2}+2is-iP_\beta\right)\Gamma_b\left(\frac{Q}{2}-2is-iP_\beta\right)\Gamma_b(\frac{Q}{2}-iP_{\beta})^2}\right)\\
&\times \left( \pi\mu \gamma(b^2)b^{2-2b^2}\right)^{\frac{iP_a}{b}}\left(\frac{\Gamma_b(Q+2iP_a)}{\Gamma_b(2iP_a)}\right)
\end{aligned}
\end{equation}
The term in the second line of (\ref{relation}) can be written in terms of the (1,1) ZZ brane:
\begin{equation}
\begin{aligned}
\left( \pi\mu \gamma(b^2)b^{2-2b^2}\right)^{\frac{iP_a}{b}}\left(\frac{\Gamma_b(Q+2iP_a)}{\Gamma_b(2iP_a)}\right)=-2^{1/4}\Psi_{ZZ}(1,1|-P_a)
\end{aligned}
\end{equation}
where
\begin{equation}
\begin{aligned}
\Psi_{ZZ}(1,1|P):=\left(\pi \mu \gamma(b^2)\right)^{-iP/b}\frac{2^{3/4}2\pi i P}{\Gamma(1-2ibP)\Gamma(1-2iP/b)}.
\end{aligned}
\end{equation}
Hence, we can write
\begin{equation}\label{s1}
\begin{aligned}
\frac{\mathbb{S}_{P_a \ s}[P_\beta]}{\Psi_{ZZ}(1,1|-P_a)A(P_a;P_\beta|s)}&=-\frac{1}{2^{\frac{1}{4}}\pi}\left( \pi\mu \gamma(b^2)b^{2-2b^2}\right)^{\frac{Q/2+iP_{\beta}}{2b}}\\
&\ \ \  \times \frac{\Gamma_b(Q)\Gamma_b\left(Q+2is\right)\Gamma_b\left(Q-2is\right)\Gamma_b(-2iP_{\beta})}{\Gamma_b\left(\frac{Q}{2}+2is-iP_\beta\right)\Gamma_b\left(\frac{Q}{2}-2is-iP_\beta\right)\Gamma_b(\frac{Q}{2}-iP_{\beta})^2}.
\end{aligned}
\end{equation}
Now the RHS of (\ref{s1}) is equal to the ``$g$ function'' of Ponsot and Teschner that enters in the definition of the boundary three-point function of Liouville theory \cite{Ponsot:2001ng}. For generic values of the arguments it is defined as
\begin{equation}
\begin{aligned}
g_{\beta}^{\sigma_2\sigma_1}:=\left(\pi \mu\gamma(b^2)b^{2-2b^2}\right)^{\beta/2b}\frac{\Gamma_b(Q)\Gamma_b(Q-2\beta)\Gamma_b(2\sigma_1)\Gamma_b(2Q-2\sigma_2)}{\Gamma_b(2Q-\beta-\sigma_1-\sigma_2)\Gamma_b(\sigma_1+\sigma_2-\beta)\Gamma_b(Q-\beta+\sigma_1-\sigma_2)\Gamma_b(Q-\beta+\sigma_2-\sigma_1)}.
\end{aligned}
\end{equation}
Choosing $\sigma_1=\sigma_2=\frac{Q}{2}+is$ and $\beta=\frac{Q}{2}+iP_{\beta}$, it is straightforward to see that one can reproduce the RHS of (\ref{s1}). Therefore we obtain the final relation:
\begin{equation}
\begin{aligned}
\mathbb{S}_{P P'}[P_0]=-\frac{1}{2^{\frac{1}{4}}\pi}\Psi_{ZZ}(1,1|-P) \ g^{P'P'}_{P_0} \ A(P;P_0|P').
\end{aligned}
\end{equation}

\end{appendix}

\bibliographystyle{JHEP}
\bibliography{asformulaebcft}
\end{document}